\documentclass{aa}  

\usepackage[varg]{txfonts}

\usepackage{natbib}
\bibpunct{(}{)}{;}{a}{}{,} 

\usepackage[colorlinks,citecolor=blue]{hyperref}
\usepackage{graphicx}
\usepackage{booktabs}
\usepackage{multirow}
\usepackage{enumitem}
\usepackage{color}

\usepackage{txfonts}
\usepackage[T1]{fontenc}
\usepackage{ae,aecompl}
\usepackage{url}

\begin{document}

\def\icarus{Icarus}
\def\apj{ApJ}
\def\apss{Asron. Space Sci}
\def\araa{ARA\&A}
\def\apjl{ApJL}
\def\apjs{ApJS}
\def\aap{A\&A}
\def\mnras{MNRAS}
\def\pasp{PASP}
\def\nar{New Astron. Reviews}
\def\nat{Nature}
\def\aj{AJ}
\def\azh{ARep}
\def\jcph{J. Comput. Phys}
\def\pasj{PASJ}
\def\bain{Bull. of the Astr. Inst. of the Netherlands}
\def\aaps{A\&ASS}
\def\ooo{$\sim$}
\def\na{New Astronomy}
\def\prd{Phys. Rew. D}

\def\Oo {\displaystyle}
\def\kmpskpc{km~s$^{-1}$~kpc$^{-1}$\,}
\def\kmps{km~s$^{-1}$\,}
\def\Bunit{\mu G\,}
\def\Msun{M\ensuremath{_\odot}\,}
\newcommand{\Msunpc}{M\ensuremath{_\odot}~pc\ensuremath{^{-2}}\,}

\title{Global enhancement and structure formation of the magnetic field in spiral galaxies}
\titlerunning{Magnetic field structure in spiral galaxies}

\author{Sergey A. Khoperskov\inst{1,2}\thanks{sergey.khoperskov@obspm.fr}, Sergey S. Khrapov\inst{3}}
\authorrunning{S. Khoperskov}

\institute{GEPI, Observatoire de Paris, CNRS, 5 place Jules Janssen, 92190 Meudon, France \and Institute of Astronomy, Russian Academy of Sciences, Pyatnitskaya st., 48, 119017 Moscow, Russia \and Volgograd State University, Universitetsky pr., 100, 400062, Volgograd, Russia}
 
\abstract{ In this paper we study numerically large-scale magnetic field evolution and its enhancement in gaseous disks of spiral galaxies. We consider a set of models with the various spiral pattern parameters and the initial magnetic field strength with taking into account gas self-gravity and cooling/heating processes. In agreement with previous studies we find out that galactic magnetic field is mostly aligned with gaseous structures, however small-scale gaseous structures~(spurs and clumps) are more chaotic than the magnetic field structure. In spiral arms magnetic field strongly coexists with the gas distribution, in the inter-arm region we see filamentary magnetic field structure. These filaments connect several isolated gaseous clumps. Simulations reveal the presence of the small-scale irregularities of the magnetic field as well as the reversal of magnetic field at the outer edge of the large-scale spurs. We provide evidences that the magnetic field in the spiral arms has a stronger mean-field component, and there is a clear inverse correlation between gas density and plasma-beta parameter, compared to the rest of the disk with a more turbulent component of the field and an absence of correlation between gas density and plasma-beta. We show the mean field growth up to $3-10~\Bunit$ in the cold gas during several rotation periods~($500-800$~Myr), whereas ratio between azimuthal and radial field is equal to $4/1$. We find an enhancement of random and ordered components of the magnetic field. Mean field strength increases by a factor of $1.5-2.5$ for models with various spiral pattern parameters. Random magnetic field component can reach up to 25\% from the total strength. By making an analysis of the time-depended evolution of radial Poynting flux we point out that the magnetic field strength is enhanced stronger at the galactic outskirts which is due to the radial transfer of magnetic energy by the spiral arms pushing the magnetic field outward. Our results also support the presence of sufficient conditions for the development of magnetorotational instability~(MRI) at distances $ > 11$~kpc after $300$~Myr of evolution.}

\maketitle

\section{Introduction}\label{sec::intro}
It has long been recognized that numerous nearby spiral galaxies contain a large-scale magnetic field~\citep{2009ASTRA...5...43B,2016A&ARv..24....4B}. Giant galaxies generate their regular magnetic fields already at $z=3-4$~\citep{2008Natur.454..302B,2009A&A...494...21A}. General properties of magnetic fields and their variation as a function of resolution, seed field and baryon physics in cosmological simulations have been intensively studied by~\cite{2014ApJ...783L..20P,2015MNRAS.453.3999M,2016MNRAS.456L..69M}. \cite{2016MNRAS.457.1722R} suggest that weak initial seed fields were first amplified by a small-scale dynamo during a violent feedback-dominated early phase in the galaxy formation history.

In the disks of galaxies magnetic field pressure appears to be comparable with thermal pressure of the ISM in the current epoch~\citep{1996ARA&A..34..155B,1999ApJ...520..706C}. Magnetic field strength is also related to the galactic star formation rate~\citep[see, e.g.][]{2009ASTRA...5...43B,2010ApJ...721..975O,2012AstL...38..543M}. Indeed, total magnetic field strength grows nonlinearly with star formation rate (SFR), while the regular magnetic field strength does not seem to depend on SFR~\citep{2009RMxAC..36...25K}. Magnetic field structure in Milky Way~\citep{2012ApJ...757...14J} and its role in the gas kinematics are still under debates~\citep[see  models of the Milky Way rotation][]{2013MNRAS.433.2172S,2014A&A...568A.104E}. 

Some grand design spiral galaxies demonstrate the presence of prominent magnetic arms, situated in the inter-arm regions, between the material arms~\citep{1998MNRAS.299L..21S}. The origin of this feature remains unclear and possible mechanisms have been discussed in a several works by~\cite{2013A&A...556A.147M,2013MNRAS.428.3569C,2015MNRAS.446L...6C,2015A&A...578A..94M}. Polarized synchrotron emission and Faraday rotation measurements suggest that in the a number of disk galaxies regular magnetic field is generally aligned with the global gas motions~\citep[see e.g.][]{1988A&A...192...66G,2013pss5.book..641B,2014A&A...571A..61G} and molecular gas structures~\citep{2011Natur.479..499L}.  There are observational evidences that magnetic field can control the flow of the diffuse interstellar gas at kiloparsec scales~\citep{2005A&A...444..739B}. There are galaxies where bisymmetric magnetic structure dominates~\citep[e.g., M~81, M~51 in ][]{2000RSPTA.358..777B}. At the same time in some galaxies the magnetic field is represented by axisymmetic spiral structures~\citep{1998A&A...335..789K}. The mean magnetic field strength for a sample of spiral galaxies observed by~\cite{1995PhDT........57N} is on average $9\pm 3~\Bunit$, higher values can be found in spiral arms, up to $20~\Bunit$~\citep{1991A&A...251...15B}. For barred galaxies strength of the total magnetic field can be even larger. For instance in the knots of NGC~1097 it is about $60~\Bunit$~\citep{2005A&A...444..739B}.

The growth of magnetic field strength is clearly demonstrated in dynamo theory simulations of the interstellar medium in spiral galaxies~\citep{2016A&A...585A..21F}, barred galaxies~\citep{1994A&A...292..409B} and ring galaxies~\citep{2016A&A...592A..44M}. Numerical models of evolving galaxies show fast field amplification and ordering due to differential rotation \citep{2009MNRAS.397..733K,2009ApJ...696...96W}, possibly supported by the magneto-rotational instability \citep{2013A&A...560A..93G,2013MNRAS.432..176P}. In these models the ordered magnetic field forms spiral arm segments. \cite{2006ApJ...647..997S} also established that the magnetic field plays an important role in the morphology of gas in spiral galaxies. They also claimed the magnetic field distortion due to spiral shocks evolution. By doing SPH simulations \cite{2008MNRAS.383..497D} showed that the magnetic field is able to inhibit the formation of structures in the disk, but small scale features still appear even for highly magnetized gas~($\beta\leq 0.1$).  Through the paper we define the standard plasma beta parameter:
\begin{equation}
\beta = \frac{8\pi p}{B^2}\,, 
\end{equation}
which measures the ratio of the thermal to the magnetic pressure.

Except the large scale magnetic field and random fields there are several specific phenomena, related to the galactic magnetic field structure, for example, magnetic field reversals~\citep[see e.g.][]{1996A&A...308..433V}. Magnetic field reversals from the early phases may survive until recent epoch~\citep{2012A&A...537A..68M}. In simulations large-scale field reversals are common feature of the models by~\citep{1993MNRAS.264..285P, 2012A&A...537A..68M,2013GApFD.107..497M}. Magnetic field reversal is studied in SPMHD simulations of the Milky Way disk~\citep{2016MNRAS.461.4482D} in detail. It was shown that the magnetic field reversals occur when the velocity jump across the spiral shock is above $\approx 20$~\kmps. Reversals also occur at corotation and at the inner Lindblad resonance.

Believed that spurs in spiral galaxies are formed due to Kelvin-Helmholtz-type instability in the vicinity of galactic shock~\citep{1980ApJ...242..528E,2004MNRAS.349..270W}. From another point of view, wiggle instability is originating from the generation of potential vorticity at a deformed shock front~\citep{2014ApJ...789...68K,2017MNRAS.471.2932S}. \cite{2015ApJ...809...33K} showed that magnetic fields suppress wiggle instability, but not completely, at least for $\beta \geq 1$. The stabilizing role of magnetic fields is not from the perturbed fields but directly from the background unperturbed fields that tend to reduce the shock compression factor by exerting magnetic pressure. Despite these studies, magnetic field structure in the vicinity of galactic spurs has never been described in detail. 

Current high resolution galactic scale simulations include detailed molecular chemistry of ISM~\citep{2009ApJ...697...55G,2012MNRAS.425.3058C,2014MNRAS.441.1628S}, as well as star formation and feedback~\citep{2009ApJ...700..358T,2013MNRAS.436.1836R,2015MNRAS.453.3082P,2016ApJ...822...11P,2017arXiv170310421I}. These results suggest that hydrodynamic effects and the self-gravity of the gas play some important roles regarding the gas structures in a spiral~\citep{2006MNRAS.367..873D,2006MNRAS.371.1663D,2008MNRAS.389.1097D,2011MNRAS.413.2935D,2013MNRAS.436.1836R} and bar potential~\citep{2014MNRAS.439..936F,2016MNRAS.461.1684F}. MHD turbulence impact on the  molecular clouds formation and evolution has been recently stressed by \citep{2009ApJ...704..161I,2015MNRAS.451.3340K}. Morphology and evolution of molecular clouds in MHD converging flow simulations also have been studied by~\cite{2009MNRAS.398.1082B,2016MNRAS.460.2110F}. 

The effects of self-gravity and magnetic fields highly depend on the gaseous temperatures and phases~\citep{2009MNRAS.398.1082B,2012ApJ...761..156F}. However, it is not clear how the magnetic fields affect the stability of the multiphase ISM in a spiral potential. For instance, \cite{2008MNRAS.383..497D} studied gas dynamics in a spiral potential, taking into account magnetic field evolution. However, in their non self-gravitating SMHD simulations, they do not solve an energy equation with realistic cooling and heating processes where warm and cold components are treated separately as isothermal gases without phase exchange. In this paper we aim to explore the role of magnetic field on structure of galactic gaseous disks as well as enhancement of various components of the magnetic field with taking into account gas selfgravity and multiphase ISM in disks of spiral galaxies. We focus our analysis on the magnetic field structure in the simulated spiral galaxies with different morphology and pattern kinematics. Indeed star formation and feedback are very important for the structure and dynamics of the ISM, but instead of including all of the relevant detailed physics, here we focus on improving the aspect of initial parameters of magnetic field~(spatial structure and strength) on the small and large scale structures formation. The major interest of this work is the possibility of probing physical quantities and structures at in a consistent galactic context. Rather than studying the amplification of a primordial field, we start our calculations with magnetic field strengths closer to the present day values. However, we also study the simulation with initial turbulent magnetic field which already have a significant strength. The paper is organized as follows. In Sect.~\ref{sec::model} we refer our numerical models and introduce basic definitions. In Section~\ref{sec::results} we present our results. In Section~\ref{sec::concl} we summarize and draw out some concluding remarks.

\section{Model}\label{sec::model}
\subsection{Basic equations}
The simulations have been run using the unspilt TVD MUSCL (Total Variation Diminishing Multi Upstream Scheme for Conservation Laws) code, which includes standard physics modules previously described in~\cite{2013MNRAS.428.2311K,2014JPhCS.510a2011K,2016MNRAS.455.1782K}. For divergence cleaning we apply constrained transport technique for magnetic field transport through the computational domain~\citep{1988ApJ...332..659E}. In this approach magnetic field strength is defined at faces of a cell, while other gas dynamical variables are defined at the center of a cells. We numerically investigated the gas dynamics in 3D in a computational box of the size $12\times12\times4$ kpc, with a spatial resolution of $1600\times1600\times256$ grid zones in the $x$, $y$, and $z$ directions, respectively and the corresponding grid spacing is $15$~pc. In the simulations we also implement gas self-gravity and the network of cooling/heating processes for solar metallicity ISM~\citep[see details in][]{2013MNRAS.428.2311K}. The equations considered in our simulations are:
\begin{equation}
 \frac{\partial \rho }{\partial t} + \nabla \cdot \left( \rho {\bf u} \right) = 0\,,
\end{equation}
\begin{equation}
 \frac{\partial (\rho {\bf u})}{\partial t}  + \nabla\cdot\left[\rho {\bf u}\times{\bf u} - \frac{1}{4\pi}{\bf B}\times{\bf B} \right] = - \nabla p - \rho \nabla \Psi\,,
\end{equation}
\begin{equation}
 \rho \frac{\partial e}{\partial t} +\nabla \cdot \left[ (e + p){\bf u} - \frac{1}{4\pi} ({\bf B}\cdot{\bf u}){\bf B} \right] = \rho\Lambda\,,
\end{equation}
\begin{equation}
\frac{\partial {\bf B}}{\partial t} - \nabla \times ({\bf u}\times {\bf B}) = 0\,,
\end{equation}
where $\rho$ is the gas density, ${\bf u}$ and  ${\bf B}$ are the vectors of the gas velocity and magnetic field correspondingly, $p$ is the total gas pressure, $e$ is the total gas energy, $\Psi = \Psi_{\rm ext} + \Psi_{\rm gas}$ is the sum of external gravitational potential $\Psi_{\rm ext}$ and gas self-gravity $\Psi_{\rm gas}$, $\Lambda$ is the loss function, including heating and cooling terms~\citep[see details in][]{2013MNRAS.428.2311K}. Boundary conditions were chosen to allow for free outflow \citep[see e.g.,][]{2010A&A...523A..72D}. In our models magnetic field at the outer border of computational domain is much weaker than the average and the boundary conditions do not contribute significantly to the overall magnetic energy evolution.

\begin{table}
\caption{Table showing the parameters used in the calculations: $m$ is the number of arms of the gravitational potential perturbation, $\varepsilon$ is the relative amplitude of the adopted perturbation, $i$ is the pitch angle, $\Omega_{\rm p}$ is the pattern speed, $\beta$ is the initial plasma beta. Note, that each model differs from the fiducial run by a single parameter, which is mentioned in the model name. All models are characterized by the purely toroidal initial magnetic field except the model marked by astreics~(rf0*) where we impose the initial turbulent field~(see Sect.~\ref{sec::results7}).}
\begin{center}
\begin{tabular}{lcccccccccccccccc}
\hline
Model  & m & $\varepsilon$ & i & $\Omega_{\rm p}$ & $\beta$ \\
& & & & \kmpskpc & \\
 \hline
 \hline
 fiducial & 2 & 0.1 &  20  & 20 & 10 \\
 m3		  & 3 & 0.1 &  20  & 20 & 10    \\
 m4  	  & 4 & 0.1 &  20  & 20 & 10    \\
 e0		  & - & 0 &  -  & - & 10 \\
 i10		  & 2 & 0.1 &  10  & 20 & 10   \\   
 i30		  & 2 & 0.1 &  30  & 20 & 10   \\
 O2		  & 2 & 0.1 &  20  & 40 & 10    \\
 O05 	  & 2 & 0.1 &  20  & 10 & 10    \\
 b1		  & 2 & 0.1 &  20  & 20 & 1 \\
 b03		  & 2 & 0.1 &  20  & 20 & 0.3\\
rf0*		  & 2 & 0.1 &  20  & 20 & 10 \\
  \hline
\end{tabular}\label{tab::tabular1}
\end{center}
\end{table}

We consider the dynamics of gaseous disk in external potential of spherical isothermal halo $\Psi_{\rm h}$, Plummer model of spherical bulge $\Psi_{\rm b}$ and flat disk component in Myamoto-Nagai form~$\Psi_{\rm d}$~\citep{1975PASJ...27..533M}. Combination of the disk, bulge and halo parameters let us study the dynamics of gas in the isolated Milky Way-type disk, with a flat rotation curve and circular velocity of $\approx 220$~\kmps at $8$~kpc from the center. We investigate gas response on the presence of spiral potential described by the logarithmic spiral perturbation $\Psi_{\rm sp}$ of the axisymmetric potential of stellar disk, and then total external potential is following:
\begin{equation}
\Oo \Psi_{\rm tot} = \Psi_{\rm h} +  \Psi_{\rm b} + \Psi_{\rm d} (1 + \zeta \Psi_{\rm sp}).\label{eq::potential}
\end{equation}
Amplitude of spiral perturbation is adopted in the form:
\begin{equation}
\Oo \Psi_{\rm sp} = \frac{(r/r_c)^2}{((r/r_c)^2 + 1)^{3/4}}  \cos{\left(m\left(\varphi - \Omega_{\rm p} t - \frac{\cos{i}}{\sin{i} }\log{\left(r/h_c\right)}\right)\right)}\\,\label{eq::spirals}
\end{equation}
where $r$, $\varphi$ are the radial and azimuthal coordinates respectively, $i$ is the spiral pitch angle, $\Omega_{\rm p}$ is the pattern speed, $r_c$ is the spiral scale length, $h_c$ is the spiral width.
Amplitude of perturbation $\zeta$ is a function of time until $t_1$ and it is used in the form proposed by~\cite{2000AJ....119..800D}:
\begin{equation}
\zeta = \varepsilon \left( \frac{3}{16}\phi^5 -\frac{5}{8}\phi^3 + \frac{15}{16}\phi + \frac{1}{2} \right),\,\, {\rm where} \,\, \phi = 2 \frac{t}{t_1} -1\,.
\end{equation}
After $t_1$ amplitude of perturbation is constant and equals~to~$\varepsilon$. Such approach guarantees  a smooth transition from the axisymmetric potential to the spiral one during the time interval $t_1$. In our simulations we adopted $t_1 = 100$~Myr which let us diminish initial transient effects.

To avoid an  initial magnetic field divergence in the computational domain we introduce the magnetic field by using vector potential~${\bf A}$:
\begin{equation}
{\bf B} = - {\bf \bigtriangledown} \times {\bf A}\, \label{eq::ini_magn_f}
\end{equation}
where we set up ${\bf A} = \{0,0,\exp(-r/r_{\rm m})\}$ for a purely toroidal initial magnetic field structure. Such approach let us set up values for the magnetic field on the faces of the computational cells, because these values are actually used in computations and divergence cleaning procedures.

We set up the gaseous disk equilibrium state according to the radial balance of the gas rotation versus radial gradient of gas pressure, gravitational forces (external potential and self-gravity) and magnetic field pressure. For gaseous disk profile we choose an exponential law with the radial scale length equals to $5$~kpc.  Our models are listed in Table~\ref{tab::tabular1}, where we  vary initial magnetic field strength, morphology of spiral potential and its pattern speed. We also simulated three special models: without magnetic field (pure hydrodynamics), without spiral pattern (axisymmetric external potential, or $\varepsilon=0$) and with initial turbulent magnetic field structure. Our simulation was evolved for about $0.8-1$~Gyr.

\section{Results}\label{sec::results}
\begin{figure*}
\includegraphics[height=0.24\hsize]{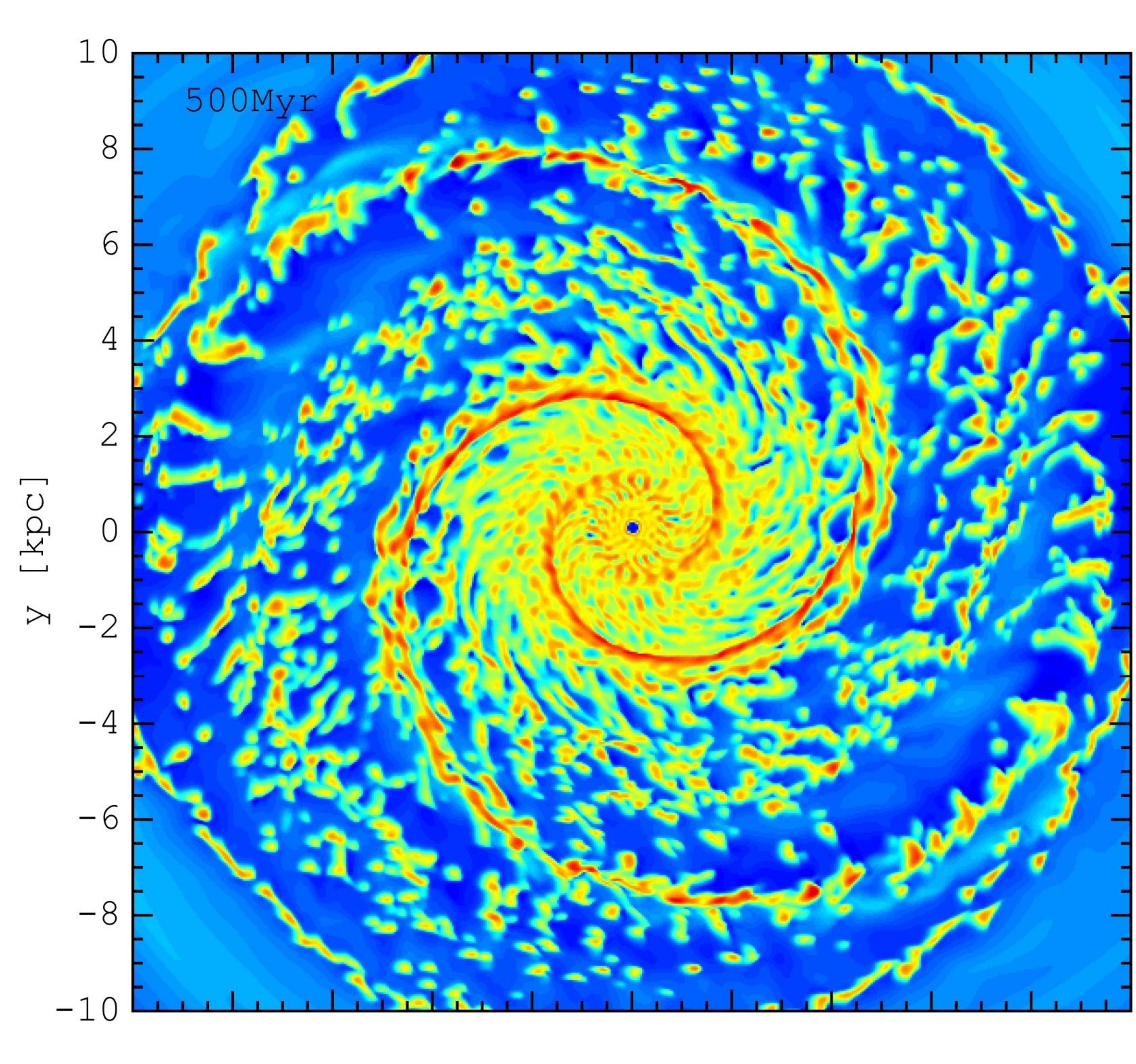}
\includegraphics[height=0.24\hsize]{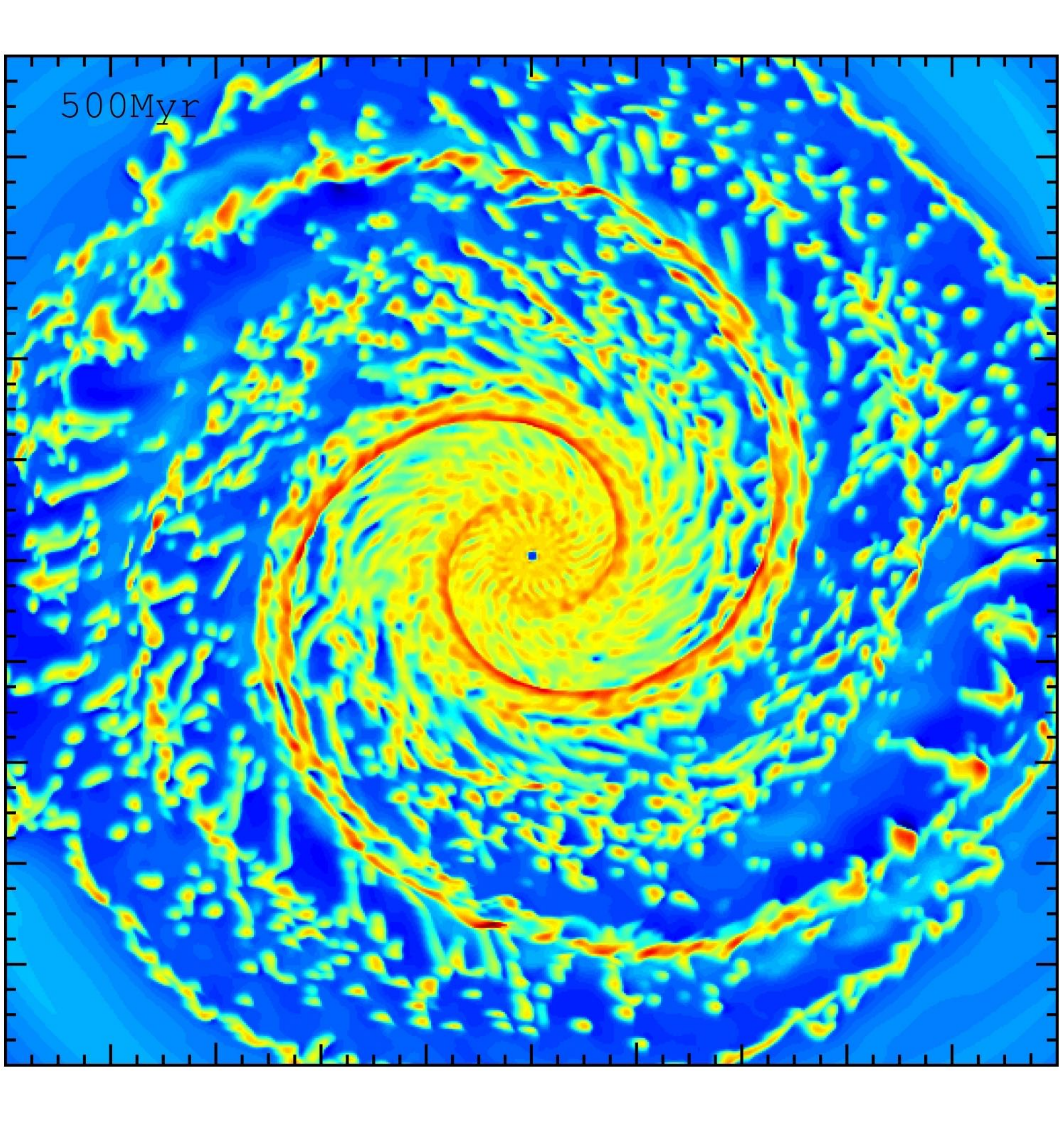}
\includegraphics[height=0.24\hsize]{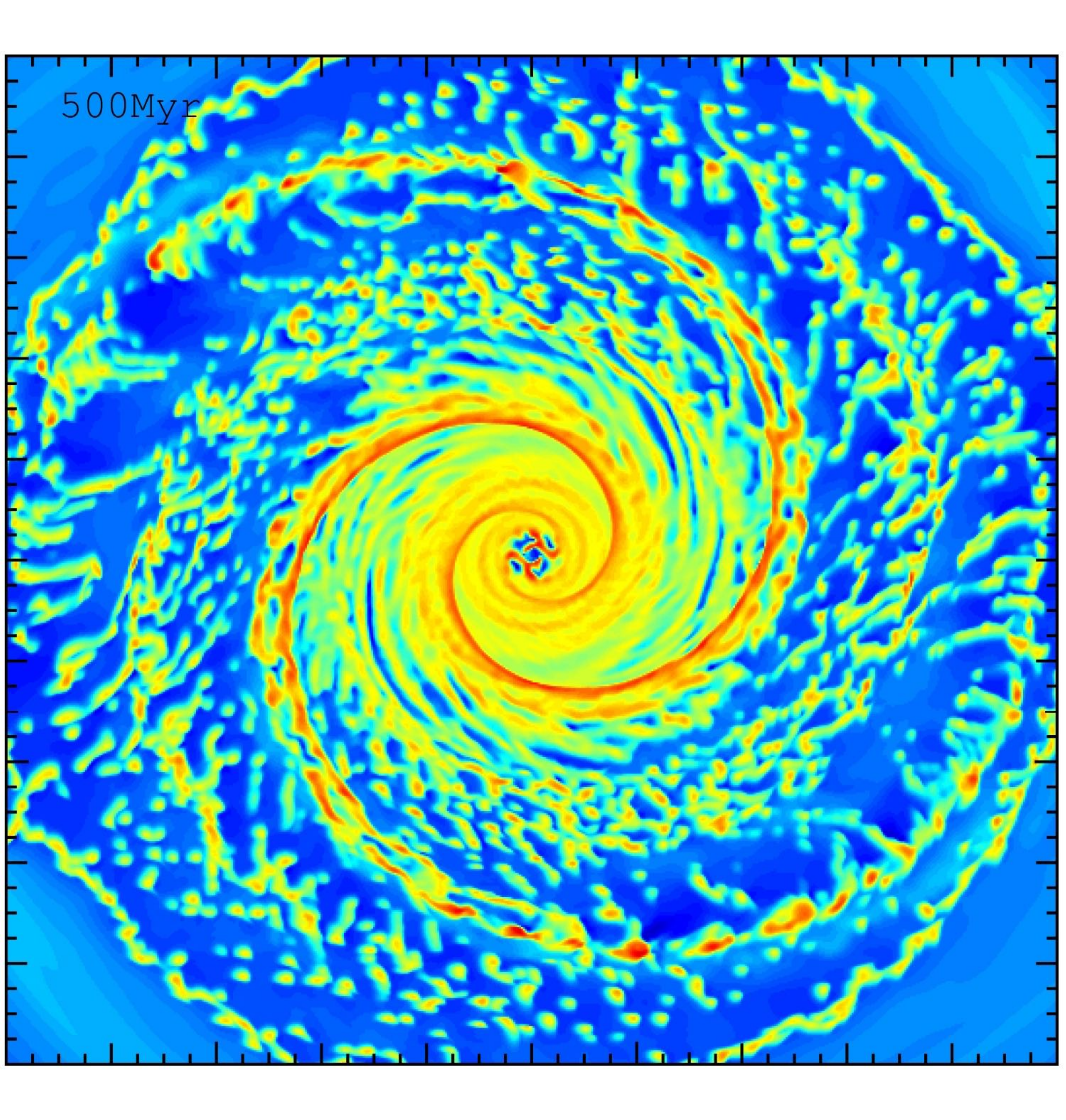}
\includegraphics[height=0.24\hsize]{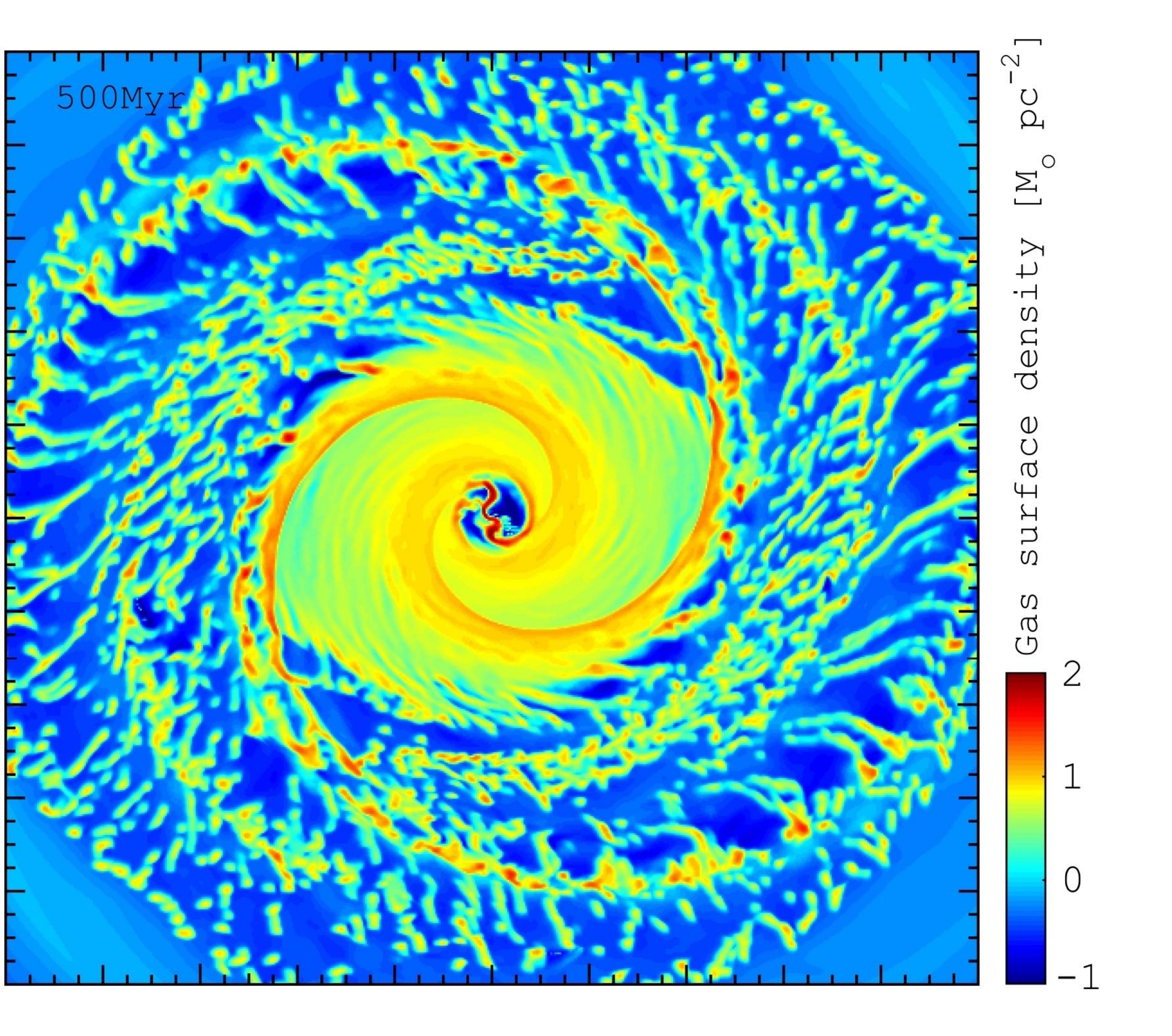}
\caption{The face-on view of the surface density in the whole galactic disk at $500$~Myr for three models with different initial magnetic field strength, from left to right: no magnetic field, $\beta=10$, $\beta=1$, $\beta=0.3$. Rotation is clockwise.}\label{fig::different_beta}
\end{figure*}

First, we present the analysis of evolution of both gas and magnetic field morphology in our fiducial simulation in Section~\ref{sec::results1} where we consider the response of slightly magnetized medium $\beta=10$ on the two-armed spiral potential perturbation representing the stellar spiral arms of a galaxy. We discuss the morphology of the magnetic field, and the magnetic field strengths of the spiral arm and inter-arm regions in Sections~\ref{sec::results3},~\ref{sec::results4}. Evolution of various magnetic field components is considered in Section~\ref{sec::results5}. We also examine the growth of magnetic  field strength in models with various spiral perturbation~(Sections~\ref{sec::results6}-\ref{sec::results6a}). Finally, turbulent magnetic field evolution and its impact on the gas dynamics is presented in Section~\ref{sec::results7}.

\subsection{Magnetic field impact on the structures formation and gas kinematics}\label{sec::results1}
\begin{figure}
\includegraphics[width=1\hsize]{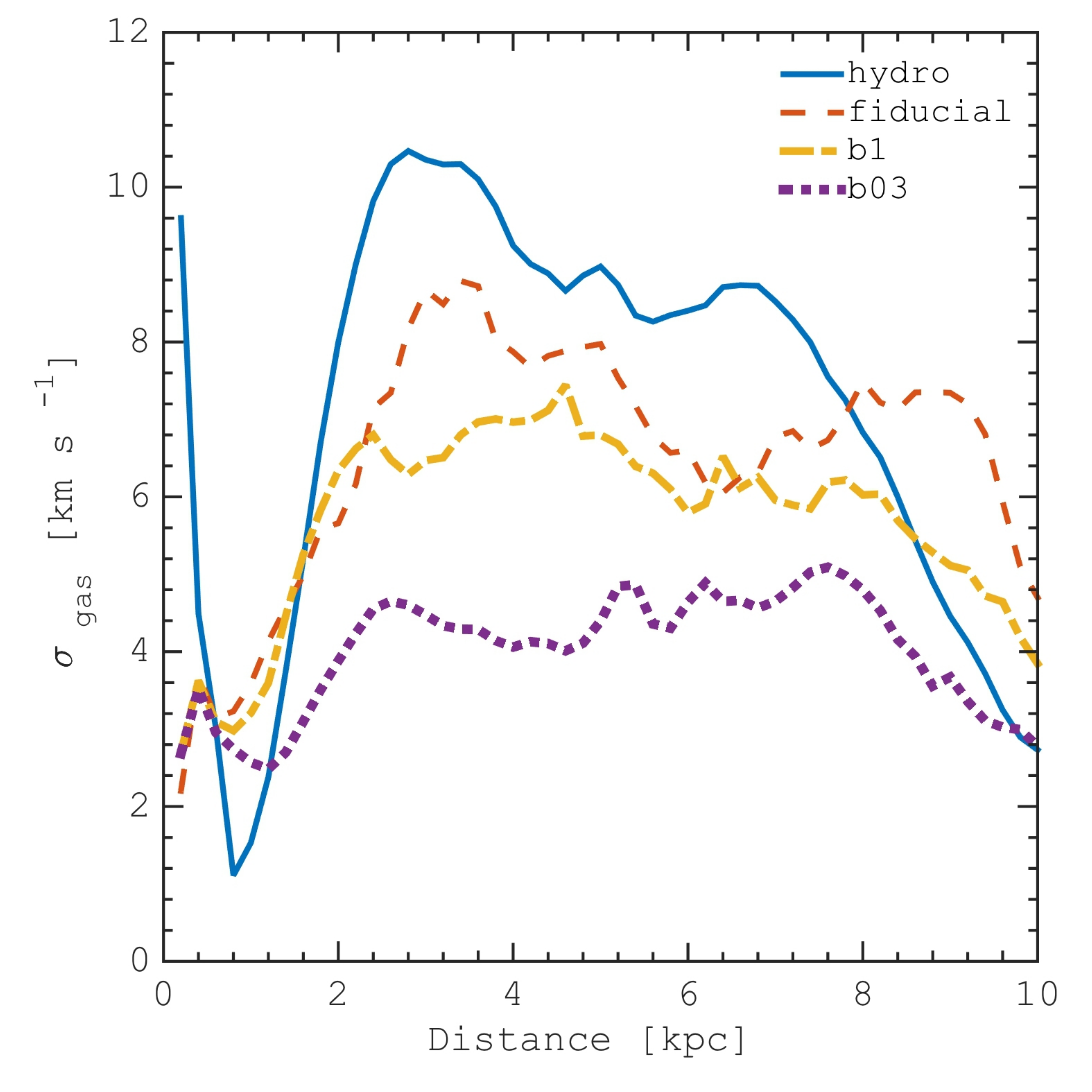}
\caption{Radial dependence of the gas velocity dispersion $\sigma_{\rm gas}$~(see Eq.~\ref{eq::velocity_dispersion}) at $t=500$~Myr  in models with different initial magnetic field strength: no magnetic field~(solid blue line), $\beta=10$~(red dashed line), $\beta=1$~(yellow dash-dotted line), $\beta=0.3$~(purple dotted line). }\label{fig::velocity_dispersion} 
\end{figure}

Large scale disk galaxy morphology with magnetic fields looks similar to models without it. For small initial magnetic field strength ($\beta = 10$) we see a very small difference from the pure hydro simulation~(see Fig.~\ref{fig::different_beta}). In both cases we observe the formation of spiral structure which is driven by the external potential perturbation. In the inter-arm region there is a number of small scale isolated clumps. Various small scale gas features, such as spurs, feathering and clumps are formed due to gravitational, thermal and shear flow instabilities~\citep{2006MNRAS.367..873D,2008MNRAS.389.1097D,2008ApJ...675..188W}. The clumps are connected to large filaments which build up in the intersection region of the colliding flows. For larger initial magnetic field~($\beta\leq 1$), difference in gas density distributions is seen well. In the very center small scale features formation is strongly suppressed where we see only global spiral shocks. At the same time, filaments are dominant type of structures in the outer disk  instead of isolated clumps seen in models with much lower initial magnetic field strength. These filaments appear aligned with the ambient magnetic field, suggesting they are magnetically dominated. The degree of structure in the disc is reduced as the magnetic field strength increases. Comparison of our models with and without magnetic fields suggests that magnetic fields have only a minor effect on the global disc structure, merely smoothing out substructures in the disc similar to an extra pressure term. Here our models demonstrate a good agreement  with the previous studies by~\cite{2006ApJ...647..997S} and \cite{2008MNRAS.383..497D}.  

We demonstrated that substructures formation in the gaseous is reduced when we include magnetic fields in our models. However, magnetic field also affect the gas kinematics. To estimate the impact we introduce the velocity dispersion of the gas  $\sigma_{gas}$ as following:
\begin{equation}
\sigma_{\rm gas}^2 = \frac{1}{3}\left(\sigma_{\rm r}^2 + \sigma_{\rm \phi}^2 + \sigma_{\rm z}^2 \right)\,, \label{eq::velocity_dispersion}
\end{equation}
where $\sigma_{\rm r}$, $\sigma_{\rm \phi}$, and $\sigma_{\rm z}$ are the mass-weighted gas velocity dispersion components in radial, azimuthal and vertical directions respectively. The gas velocity dispersion is calculated over an annulus of width 100 pc located at the specified radius. In Fig.~\ref{fig::velocity_dispersion} we demonstrate radial profiles of the gas velocity dispersion computed in models with different initial magnetic field strength. An emerging trend for these simulations is that rate of turbulence in the galactic disk decreases with increasing the magnetic field strength. Difference between purely hydrodynamical model~($\beta=\infty$) and our fiducial model~($\beta=10$) are not particularly drastic, however, and the global decrease of the gas velocity dispersion is consistent.

\subsection{Basic model description}\label{sec::results2}

\begin{figure*}
\includegraphics[height=0.19\hsize]{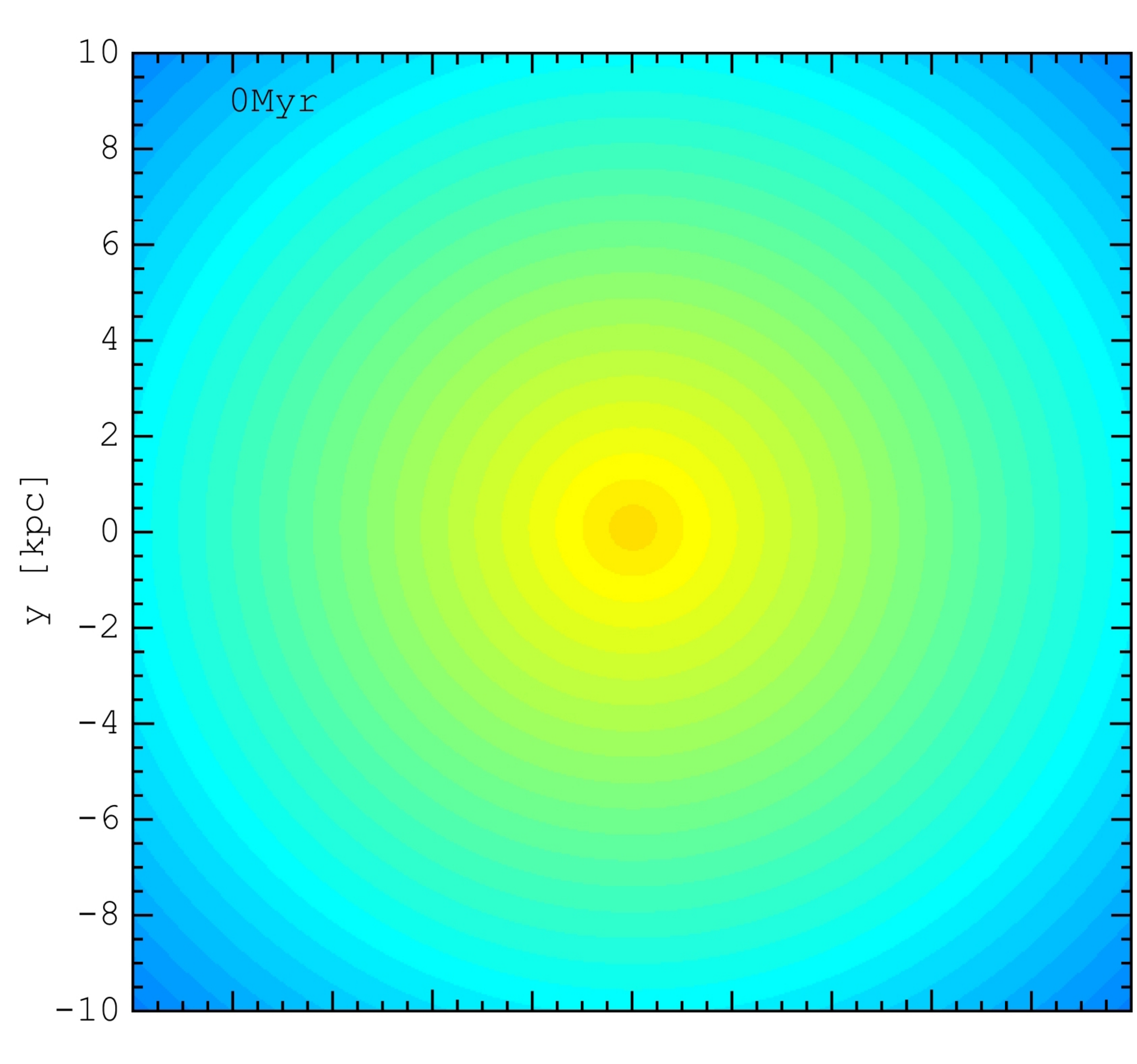}
\includegraphics[height=0.19\hsize]{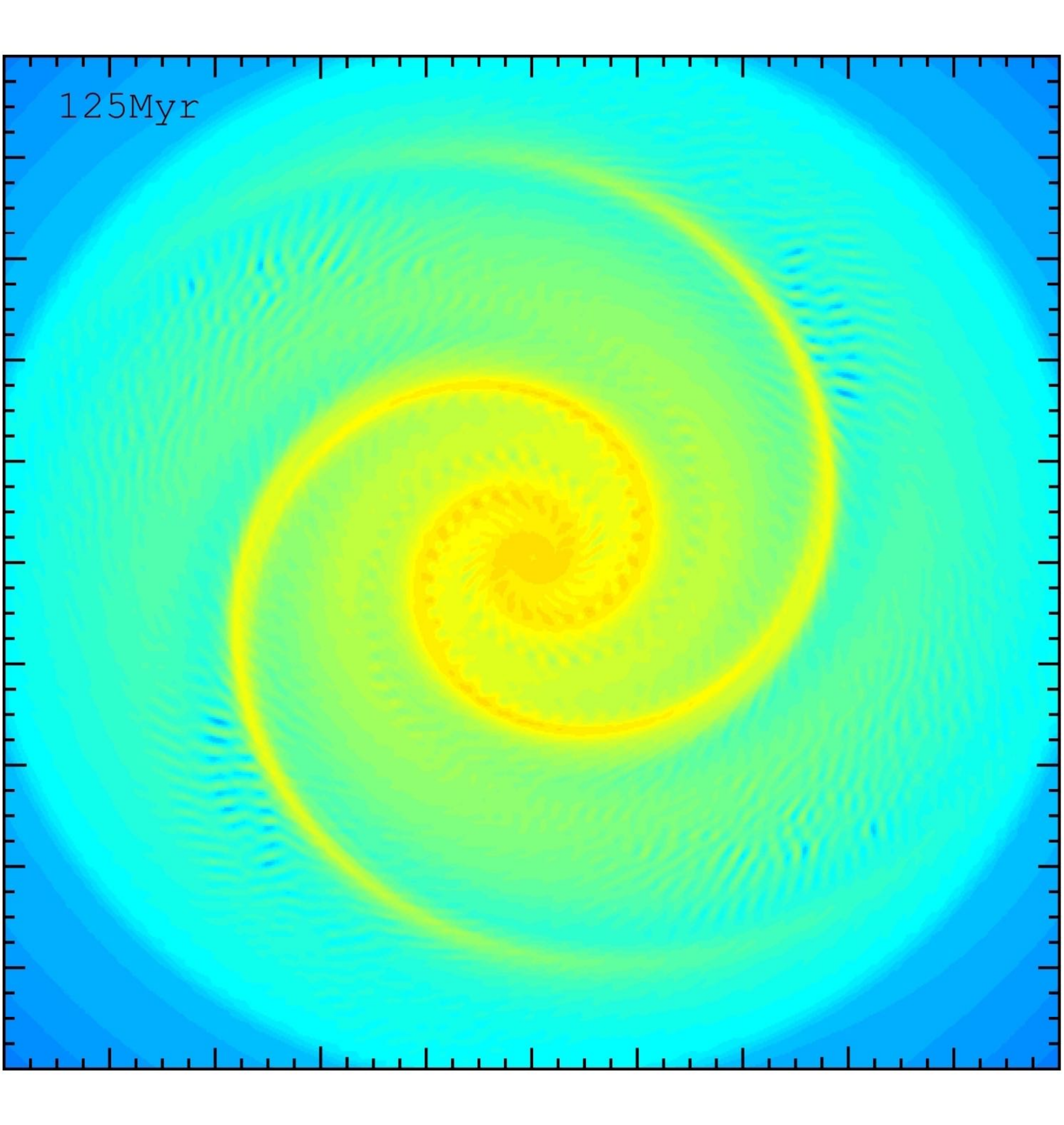}
\includegraphics[height=0.19\hsize]{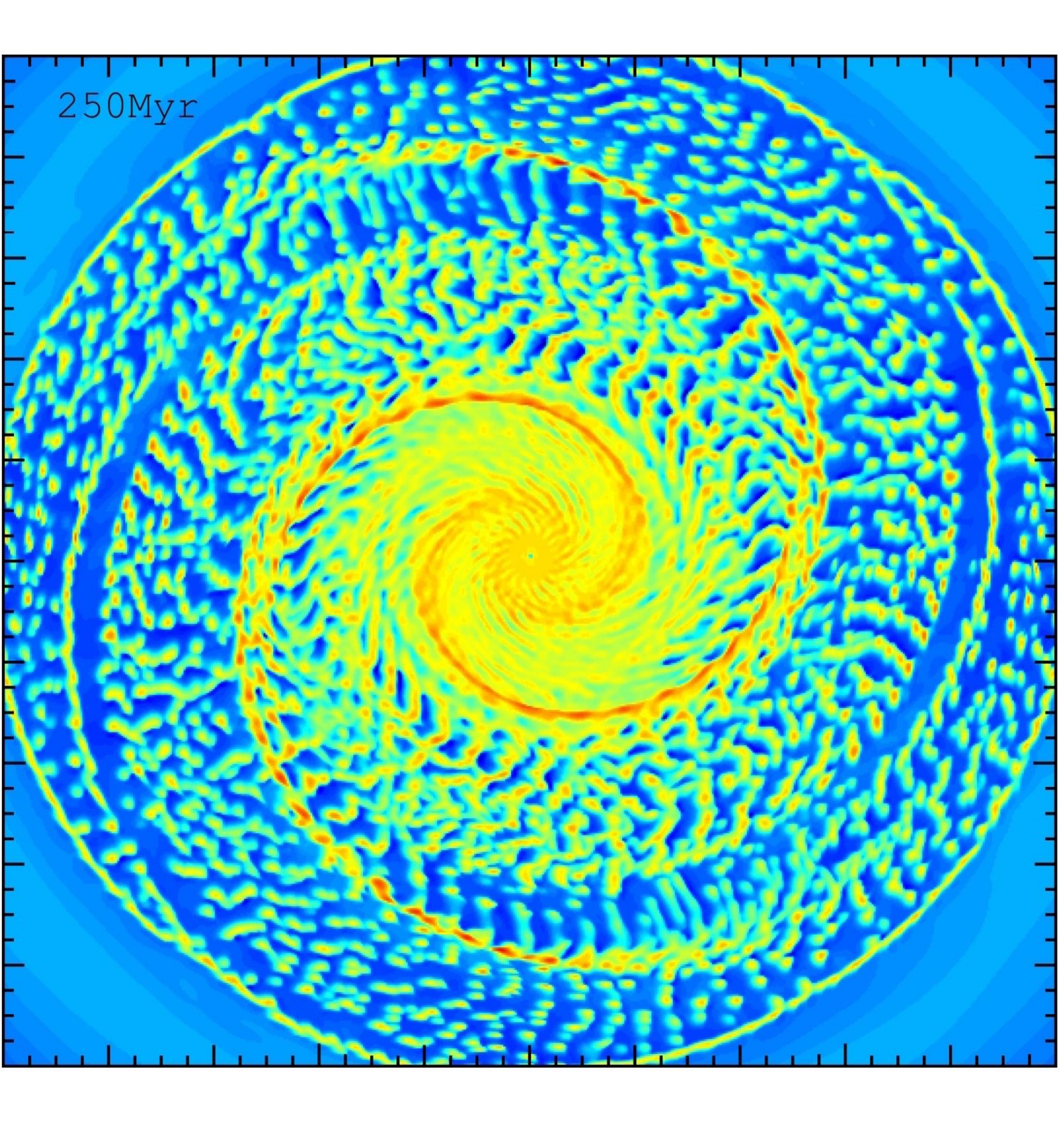}
\includegraphics[height=0.19\hsize]{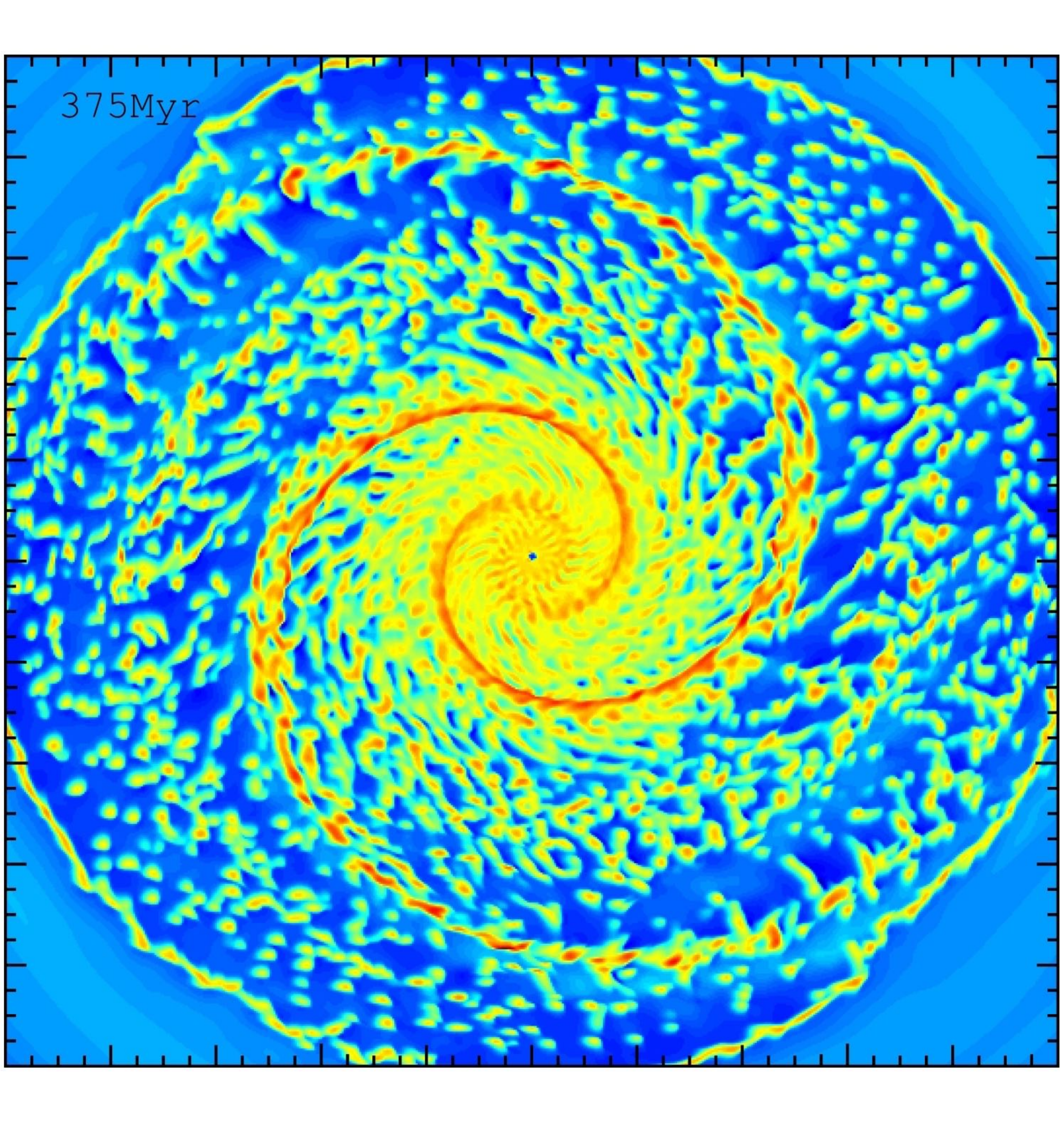}
\includegraphics[height=0.19\hsize]{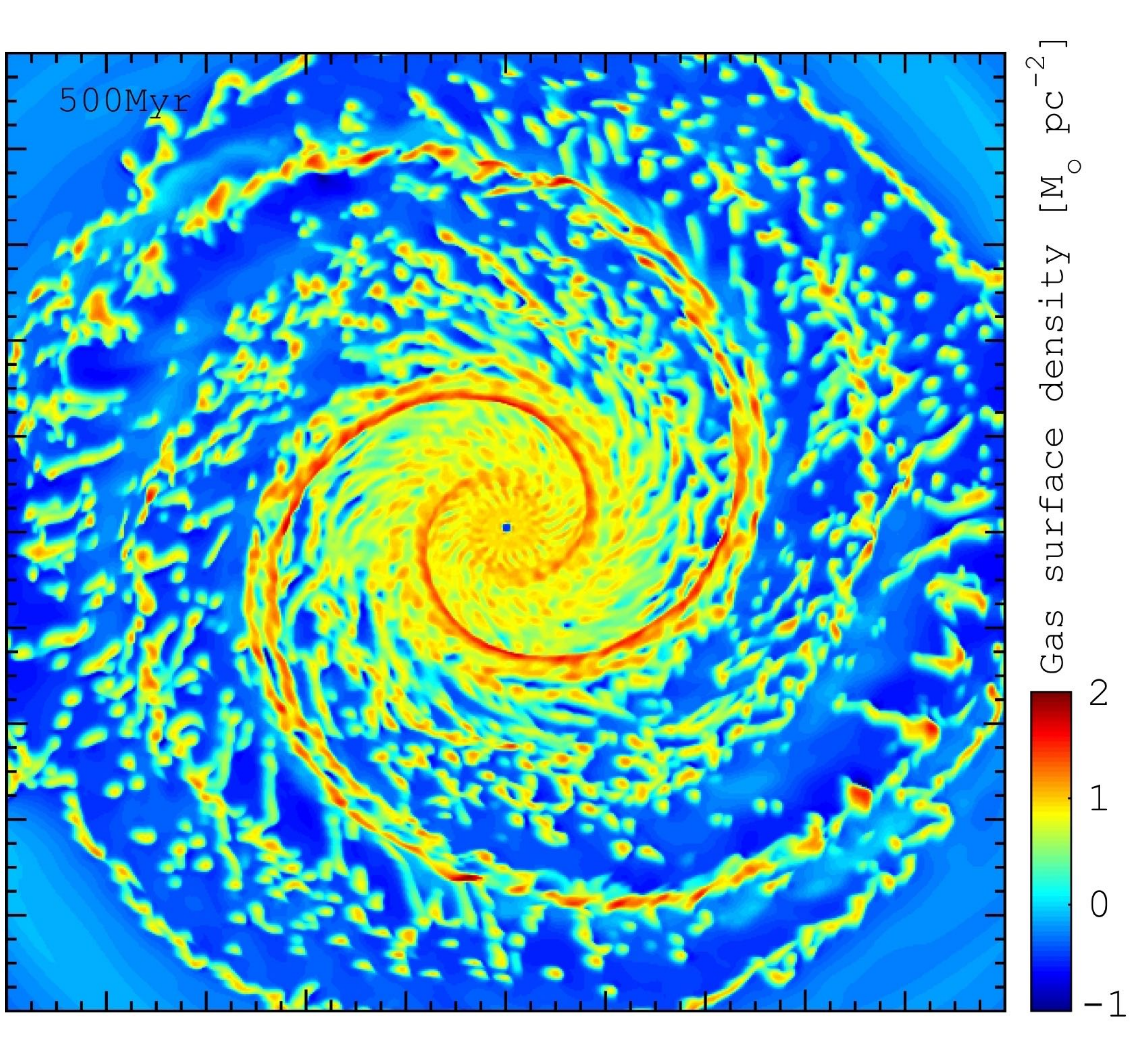}\\
\includegraphics[height=0.19\hsize]{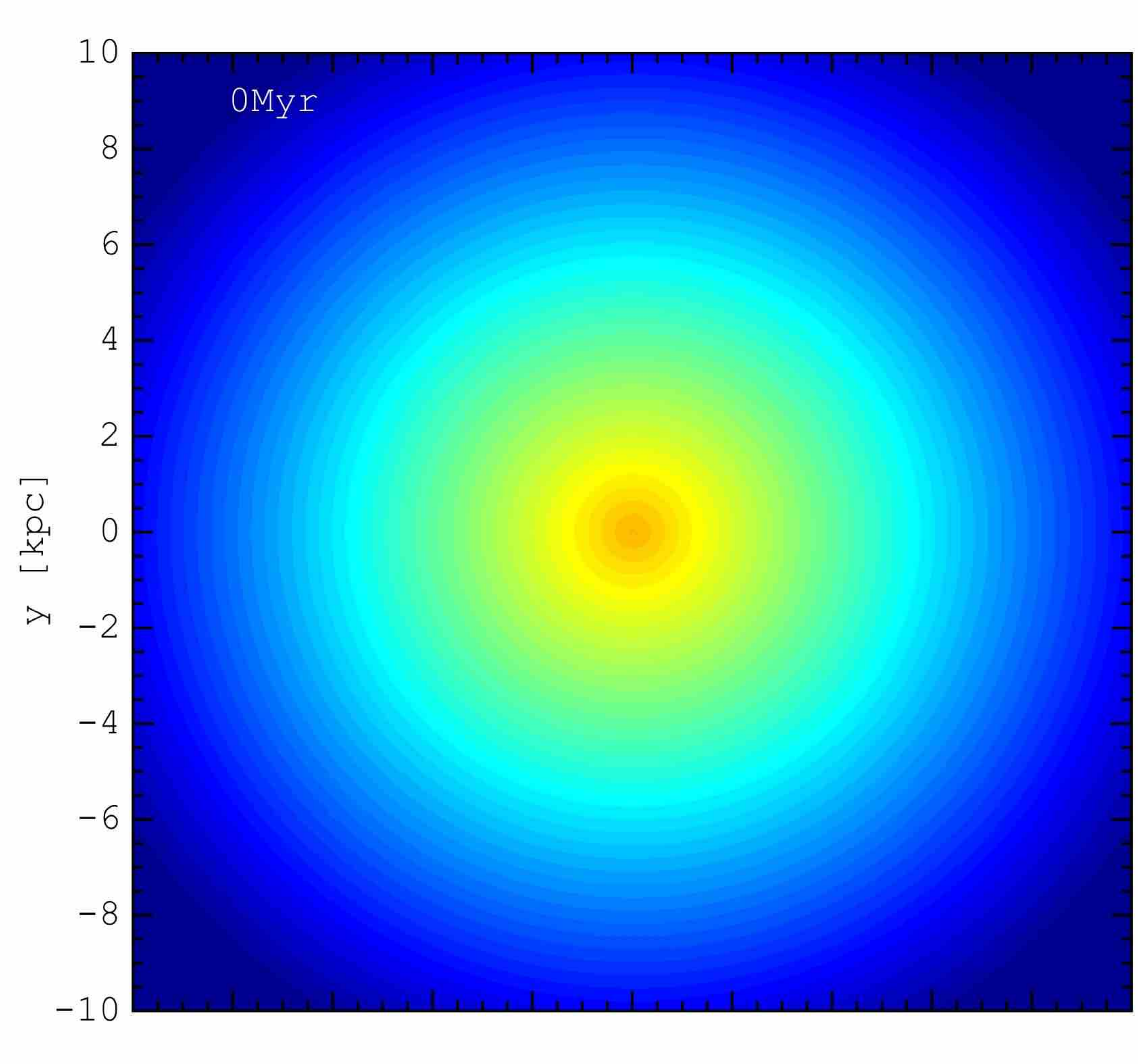}
\includegraphics[height=0.19\hsize]{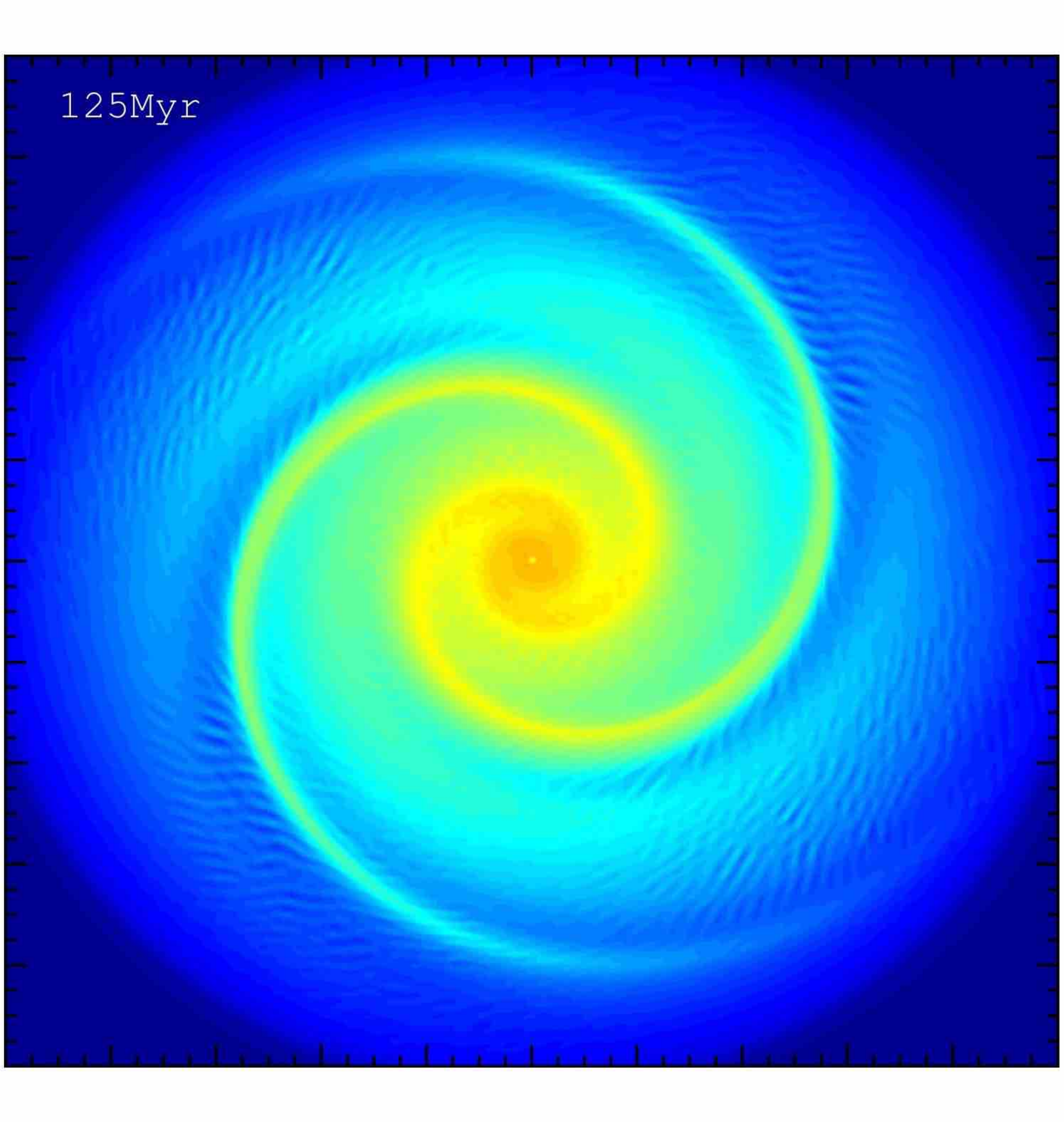}
\includegraphics[height=0.19\hsize]{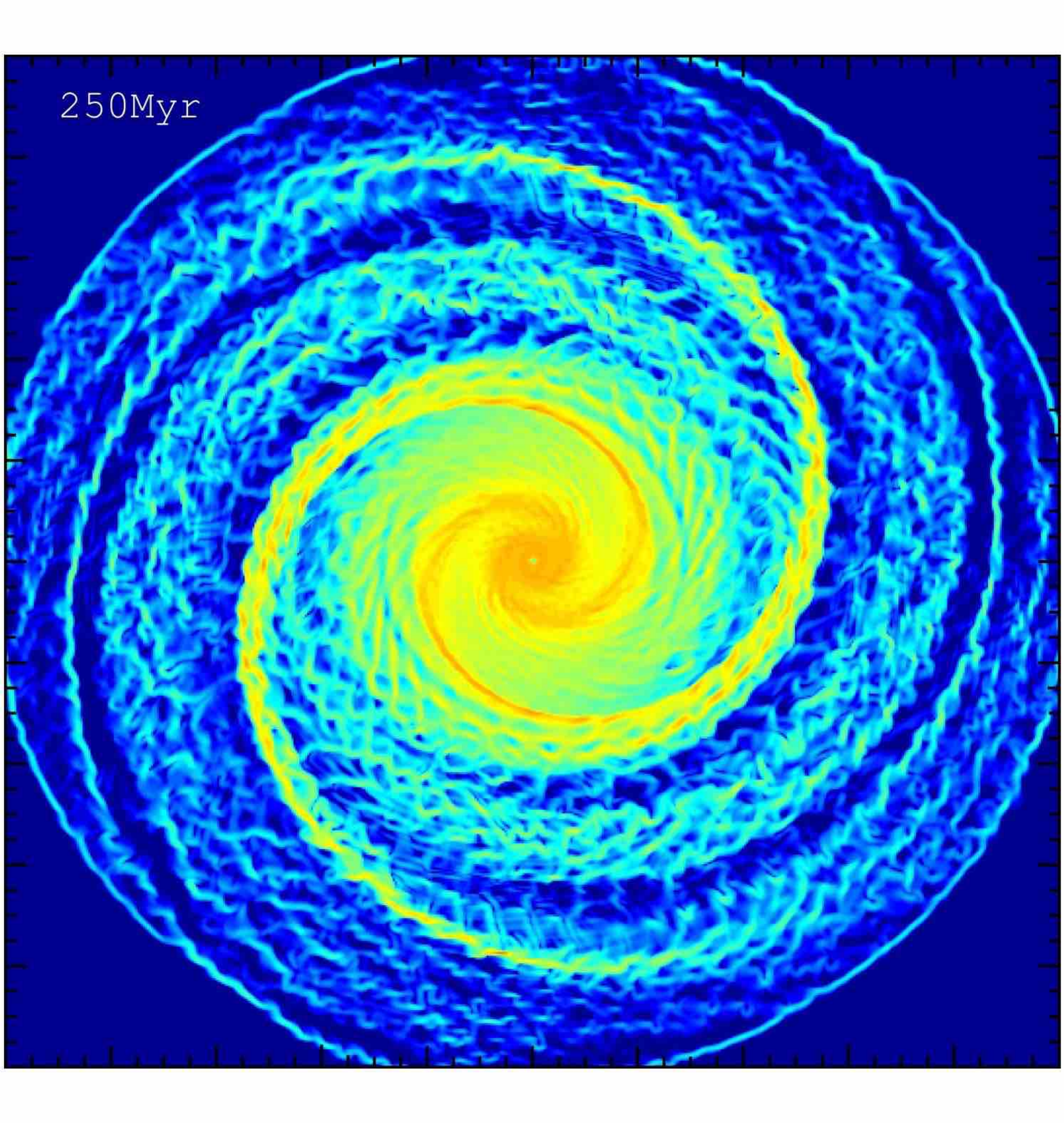}
\includegraphics[height=0.19\hsize]{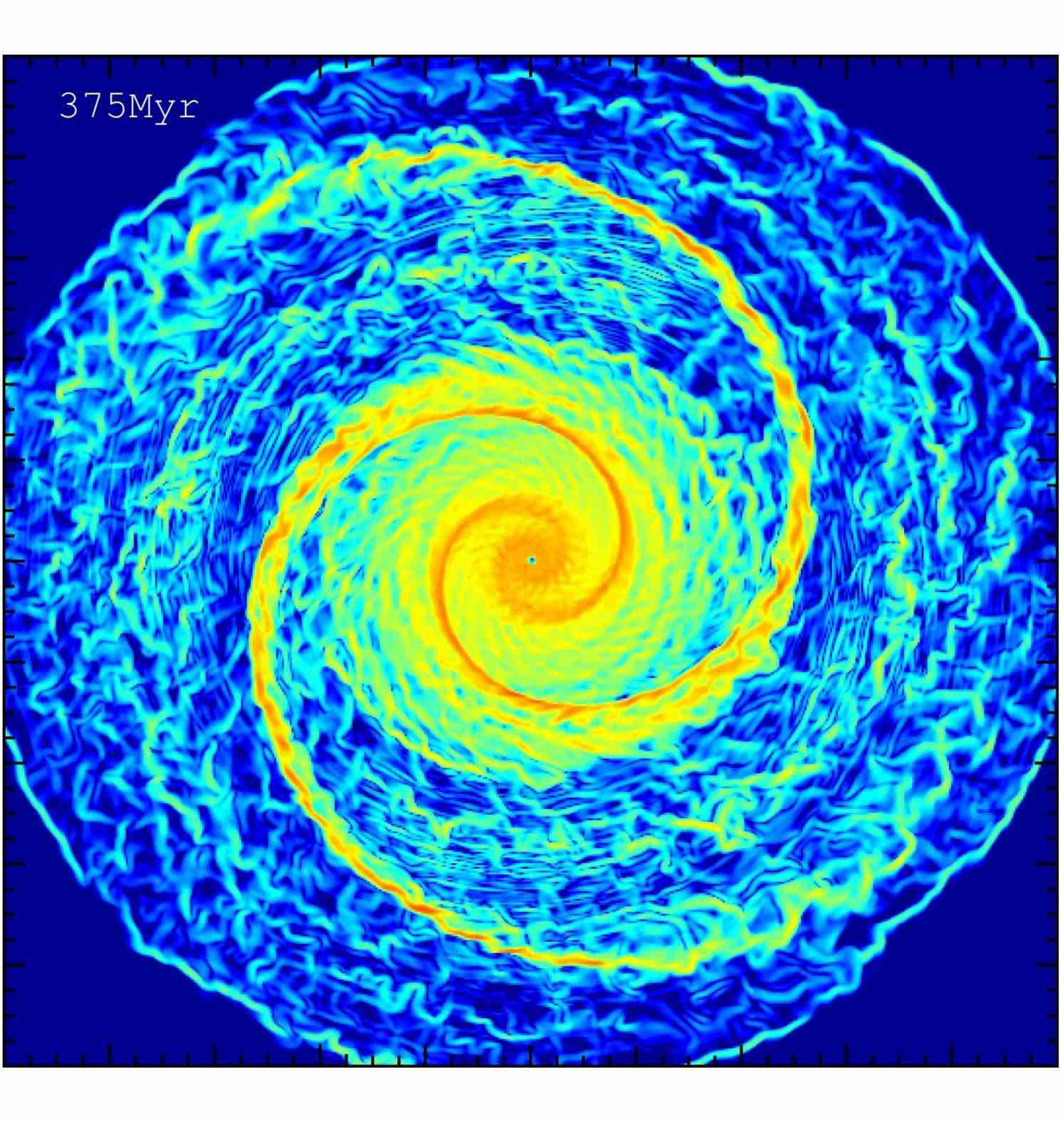}
\includegraphics[height=0.19\hsize]{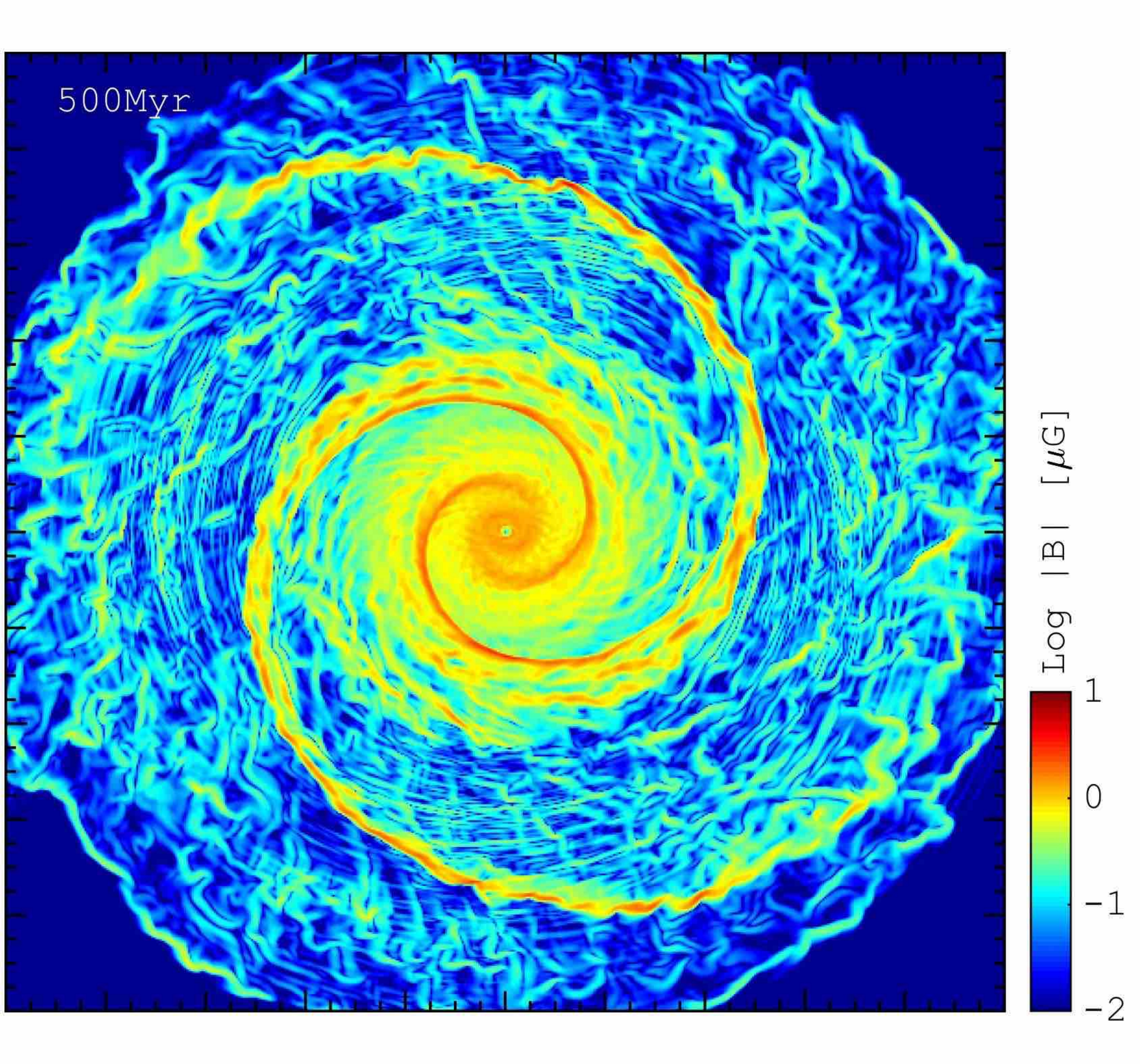}\\
\caption{Time depended evolution of the gas surface density (top) and the magnetic field strength $|B_{\rm tot}|$ in the disk plane (bottom) in the fiducial model. Rotation is clockwise.}\label{fig::evolution_fid}
\end{figure*}

First, we analyze our basic model evolution. Global gas dynamics is mostly driven by the action of the large scale spiral pattern~\citep[e.g., see also][]{2002ApJ...580..235G}. Global spiral shock amplitude increases rapidly in the simulation. It is initially very thick, but it starts to be narrower then the amplitude of spiral perturbation reaches the saturation level $\varepsilon$ at $t_1 = 100$~Myr. Shortly after $t_1$ small scale structures start forming due to shear instability~\citep[see e.g.][]{2004MNRAS.349..270W, 2006ApJ...647..997S}. These basic stages of the gas density evolution can be seen in Fig.~\ref{fig::evolution_fid}~(top row).

At early stages, the magnetic field structure mainly follows the gas distribution and it is characterized by the two-arm spiral structure~(see Fig.~\ref{fig::evolution_fid}, bottom row). Similar to~\cite{2008MNRAS.383..497D} we find that magnetic field is compressed by the spiral shocks. The relative increase in the magnetic field strength is greater where the density wave is stronger.  However, then small scale magnetic field structure starts to be different from the gas density distribution ($t>250$~Myr). In the inter-arm region magnetic field has a filamentary structure while a magnetic filament connects numerous substructures in the gas. Magnetic filaments look like curved structures, but they are mostly orientated along the azimuthal coordinate because of the galactic rotation. Radial scales lengths of the filaments do not exceed $1$~kpc.

Basic run starts from $1$~$\Bunit$ in the center and $0.1\Bunit$ at the outskirts~($R\approx 10$~kpc). At initial stages of evolution ($<200$~Myr) we mostly see redistribution of magnetic field which increases in spirals and decreases in the inter-arm region. In the inter-arm region higher plasma-beta decreases the tension force acting on the flow, supporting the formation of higher density structures and corresponding field, for slightly higher efficient winding. Nevertheless, density of filaments in the inter-arms is lower that that is in the spiral arms, and these high-density filaments tend to be parallel to the local magnetic field~\citep[see e.g.,][]{2012A&A...541A..63P,2013A&A...550A..38P}. After that, in the filaments magnetic field strength vary in the range $0.1-3~\Bunit$. In the very center and in spiral arms magnetic field can reach up to $10~\Bunit$. Large fluctuations in magnitude for different clumps, suggesting that it can be distorted by the random motions of the dense gas clumps.

\subsection{Spurs structure}\label{sec::results3}

\begin{figure*}
\includegraphics[width=1\hsize]{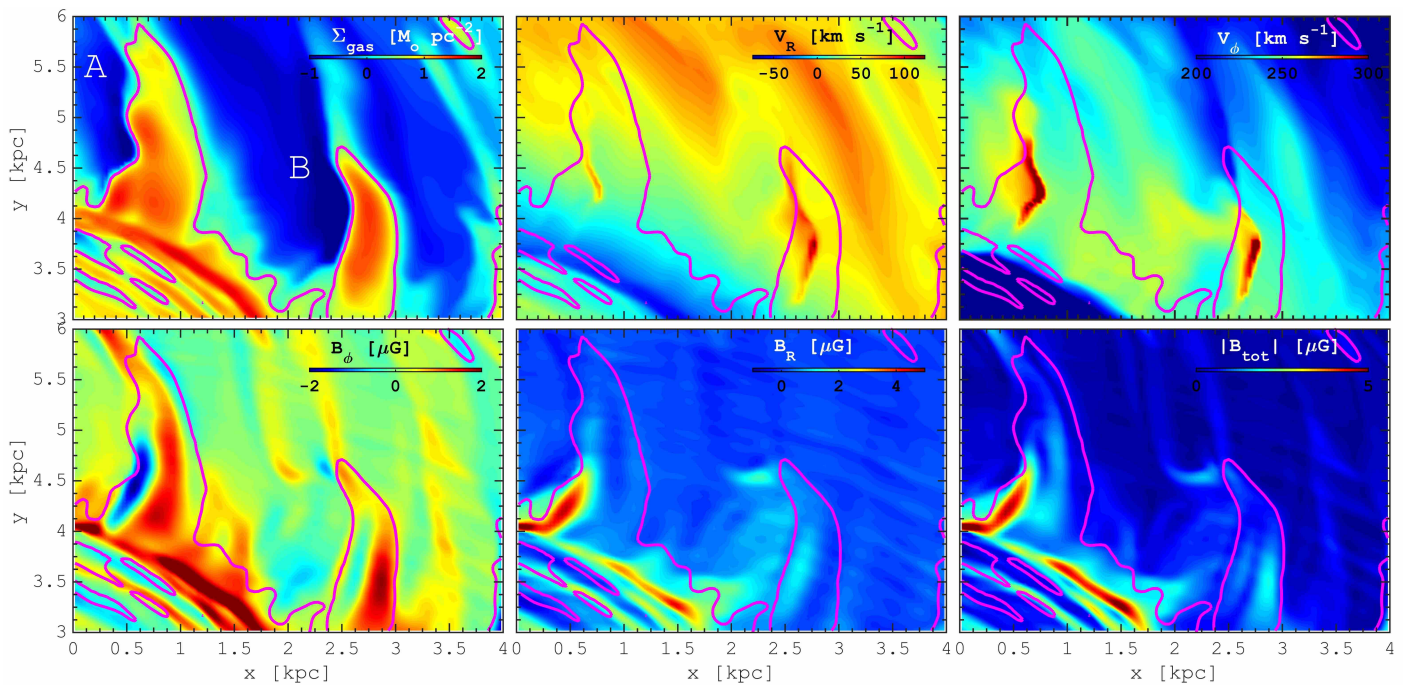}\caption{Face on distributions of the gas surface density~(top left), radial velocity component~(top center), azimuthal velocity component~(top right), azimuthal magnetic field component~(bottom left), radial magnetic field component~(bottom center) and total field stength~(bottom right) in the part of the disk at $t=500$~Myr for model O05~(with slow pattern speed rotation). We have labelled the spurs with letters A and B for easier reference. Magenta solid line shows the same surface density level in all frames. Rotation is clockwise.}\label{fig::shpurs}
\end{figure*}

Spurs are very well known features of spiral galaxies~\citep{1967MNRAS.137..157B,1980ApJ...242..528E}, which are clearly seen in a number of galaxies. From a theoretical point of view excitation of spur-like structures in spiral arms and the effect of magnetic fields on the wiggle instability have been investigated in MHD simulations by many authors~\citep[][]{2002ApJ...570..132K,2004JKAS...37..243K,2006ApJ...646..213K}.  In our basic model spiral arms fragmentation is very strong and it is not easy to identify large scale spur-like structures. However in model O05~(with the slower pattern speed rotation, see Table~\ref{tab::tabular1}) we find a several very extended spurs which we analyze below in a more precise way.  

In our simulations gas cools and becomes dense in vicinity of galactic spiral arms, leading to the agglomeration of clumps into the large-scale structures which shear off the spiral arms to become spurs. In Fig.~\ref{fig::shpurs} we plot the gas column density, velocity and magnetic field components. Two large-scale spurs can be easily found and we marked these patterns as A~(left) and B~(right). Particular attention is paid to these structures because they are prominent, which facilitate our analysis and we are able to extract and more clearly demonstrate the key features that we want to identify, namely the magnetic field structure across the spurs. Here we define spur as a structure above the certain column density threshold~$5$~\Msunpc~(marked by magenta line in Fig.~\ref{fig::shpurs}). As it can be seen, spurs are not homogeneous structures. For instance, there are several substructures inside the left one~(see A in Fig.~\ref{fig::shpurs}). Inside the spurs velocity components have a jump across the spurs with an amplitude of $\approx 20-30$~\kmps. Magnetic field structure looks very different from the density distribution. At the front edge of both spurs magnetic field is strongly positive and its strength is $1-3$~$\Bunit$ which is much higher than in the surrounding medium. At the outer edge of spurs azimuthal magnetic field component changes the direction and becomes negative. The positive and negative $B_\phi$ regions of enhanced magnetic field strength are separated by a strip of almost zero field along the spurs. Radial magnetic field component is maximal in a region where azimuthal component is negative. Thus, there is a reversal of azimuthal magnetic field across the spurs.   Formation of reversal field within the spurs in our models is consistent with the result by~\cite{2016MNRAS.461.4482D} because they found that the magnetic field reversals occur when the velocity jump across the spiral shock is above $\approx 20$~\kmps. Note that similar small-amplitude reversals of the magnetic field are also seen for more tiny structures in the inter-arm region. These small scale reversals can be consistent with polarization observations of pulsars in the disk of the Milky Way which claimed the evidence for the field reversal near the Perseus-arm as well as three reversals found in the solar circle~\citep{1999MNRAS.306..371H}.

\subsection{$\beta$ - gas density relation}\label{sec::results4}
\begin{figure*}
\begin{center}
\includegraphics[width=0.4\hsize]{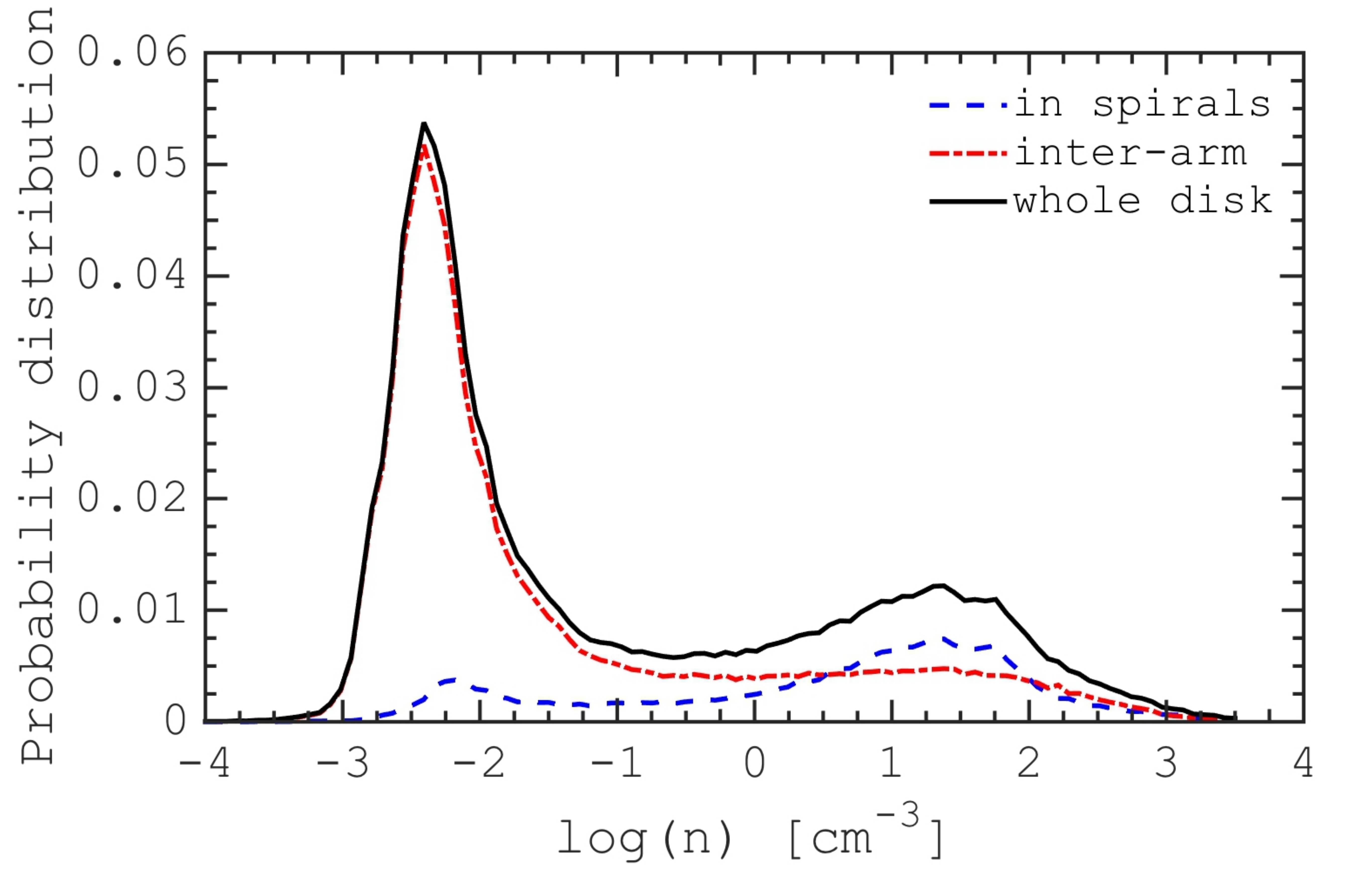}\includegraphics[width=0.4\hsize]{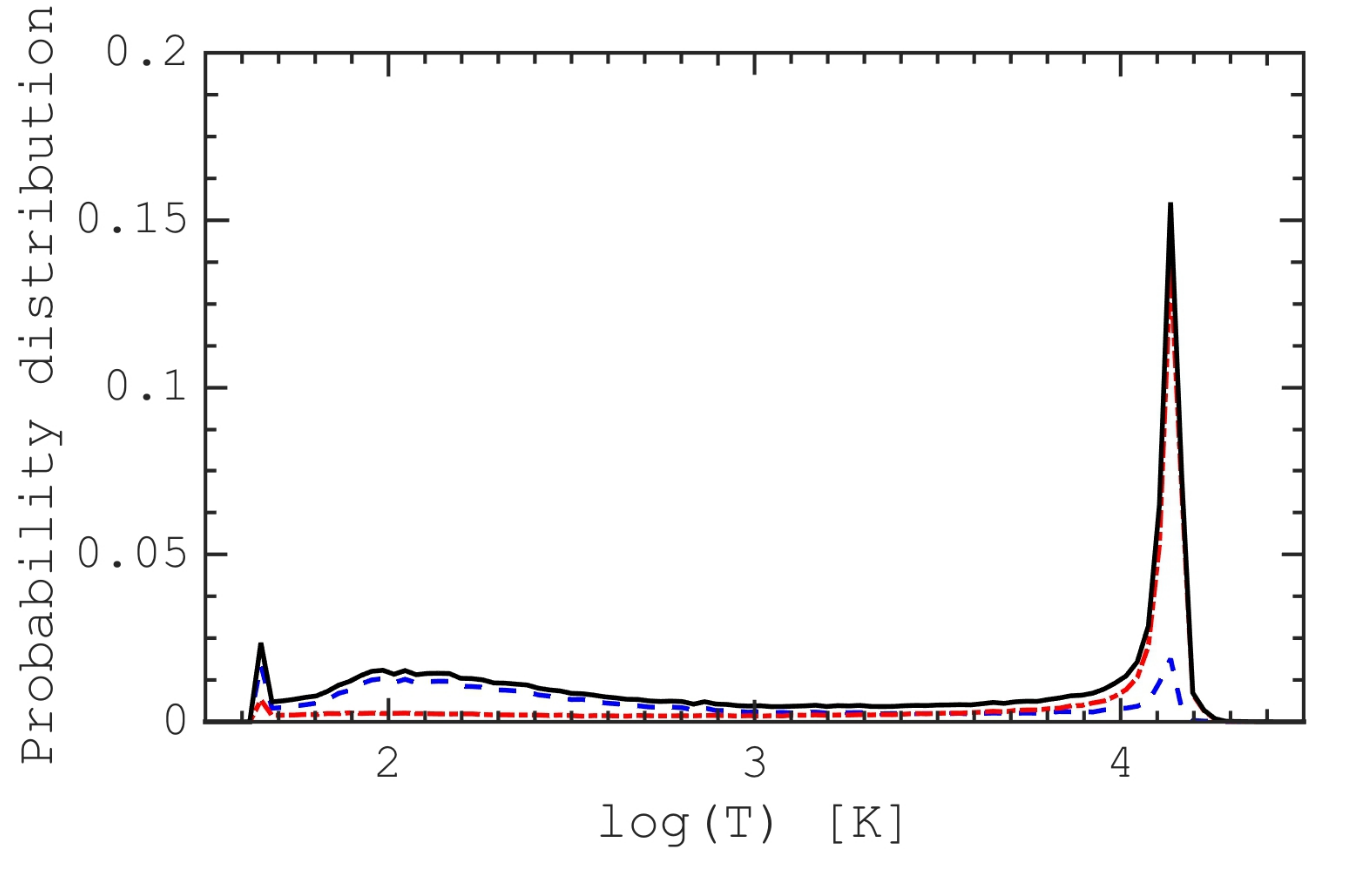}\\\includegraphics[width=0.4\hsize]{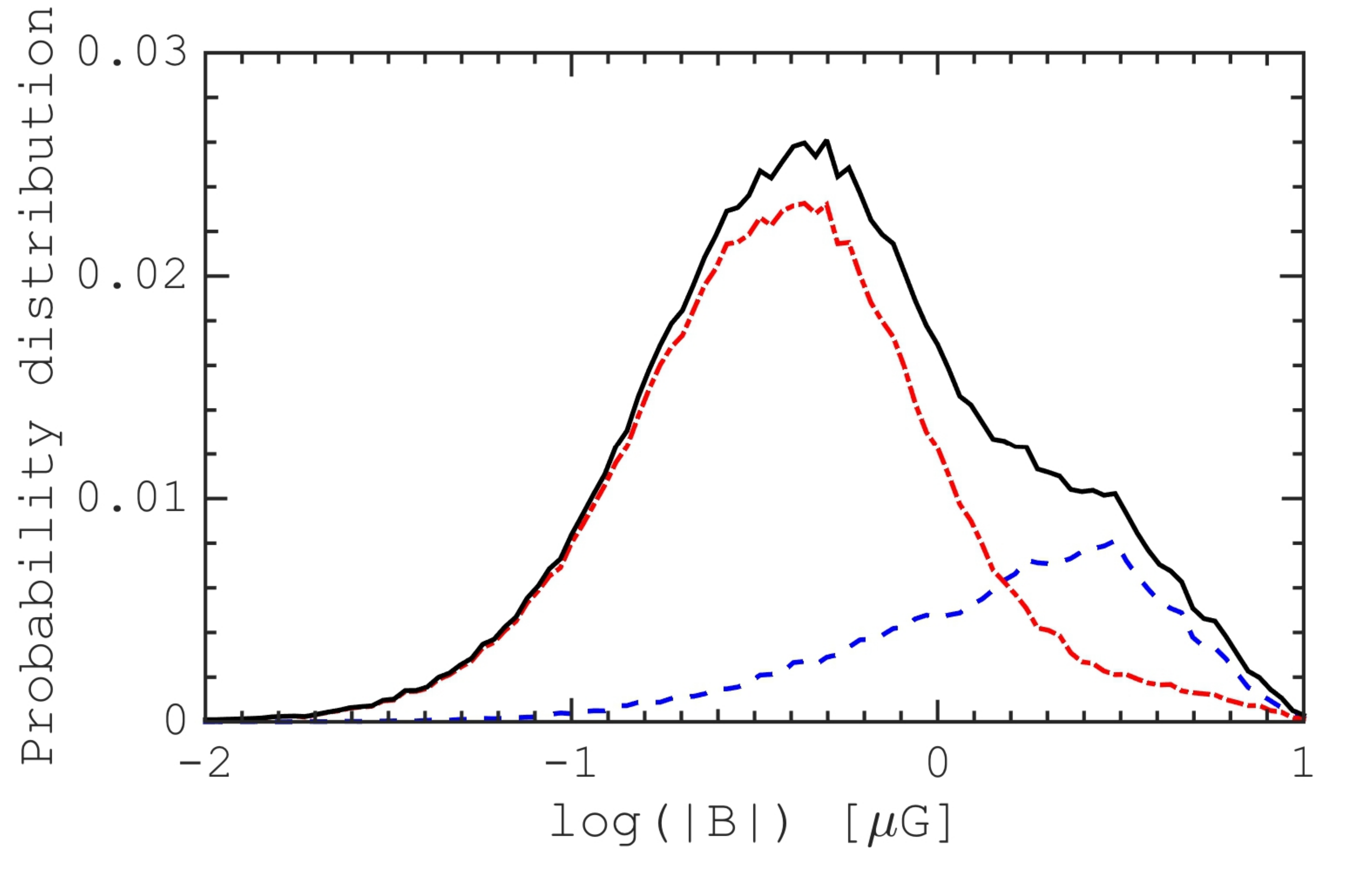}\includegraphics[width=0.4\hsize]{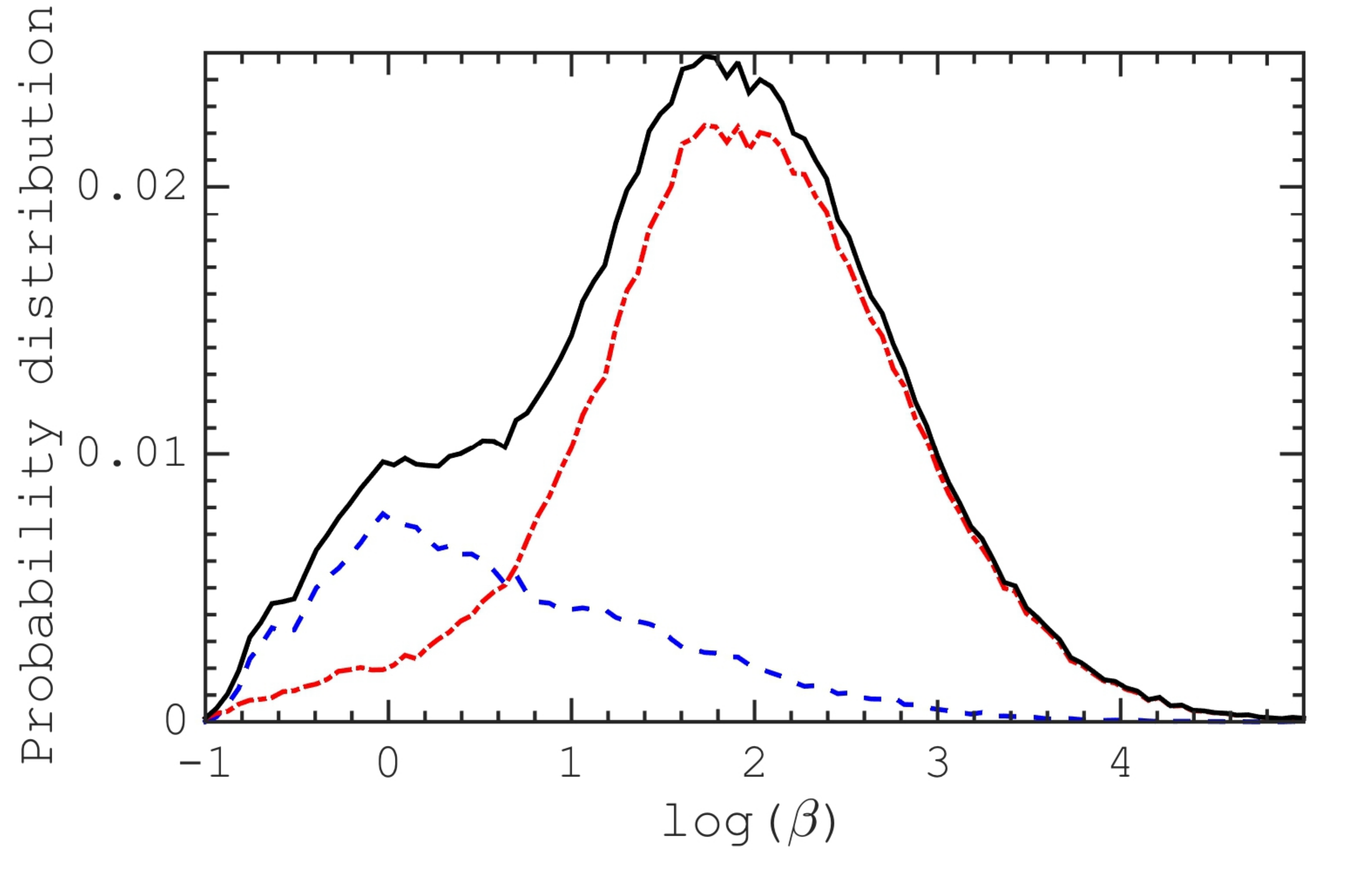}
\caption{Probability density functions for logarithms of gas concentration~(top left), temperature~(top right), magnetic field strength~(bottom left) and plasma beta~(bottom right) at 500~Myr of the fiducial model. Distributions are plotted for the whole computational domain~(solid black lines), for spiral arms region~(dashed blue lines) and for inter-arm region~(dash-dot lines).}\label{fig::ism_statistics}
\end{center}
\end{figure*}

In this section, we focus on the properties of magnetic field in different parts of the disk at late epoch ($t>500$~Myr) when the grand design spiral pattern is strong enough. ISM in the galactic disk rapidly segregates into two phases, as can be seen from Fig.~\ref{fig::ism_statistics}, where we plot probability density functions for gas density, temperature, mean magnetic field strength and plasma $\beta$. There is a warm phase, with typical temperature $10^4$~K and concentration $\approx 0.1$ cm$^{-3}$, most of which is concentrated in between of spiral arms. The mean density of the cold gas~($\approx 10^2$~K) is typically of the order of $100$~cm$^{-3}$, high enough to allow gaseous clouds formation on a timescale of $10-30$~Myr. Note also that all quantities the distribution functions have a well-pronounced double-peaked shape.

\begin{figure*}
\includegraphics[width=0.33\hsize]{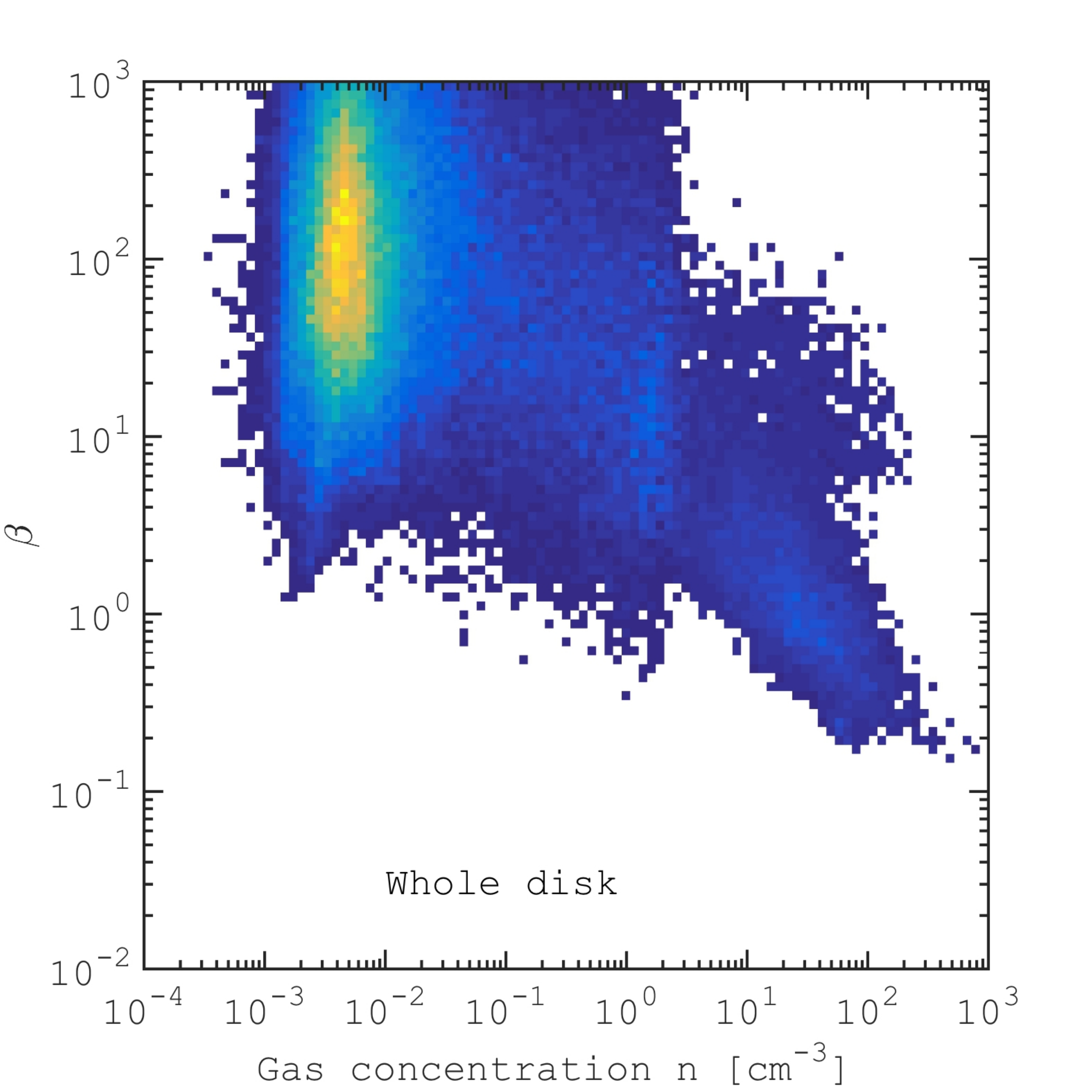}
\includegraphics[width=0.33\hsize]{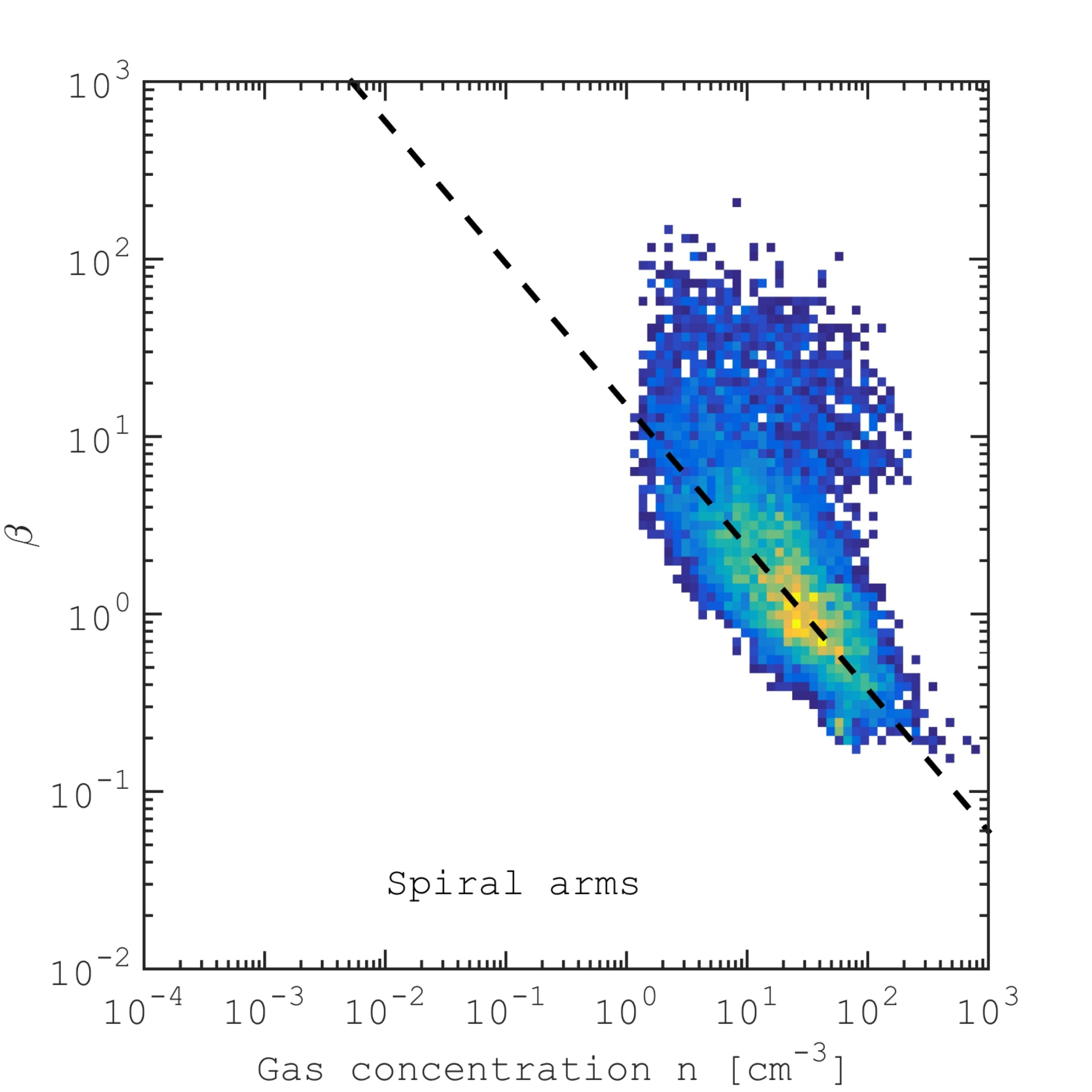}
\includegraphics[width=0.33\hsize]{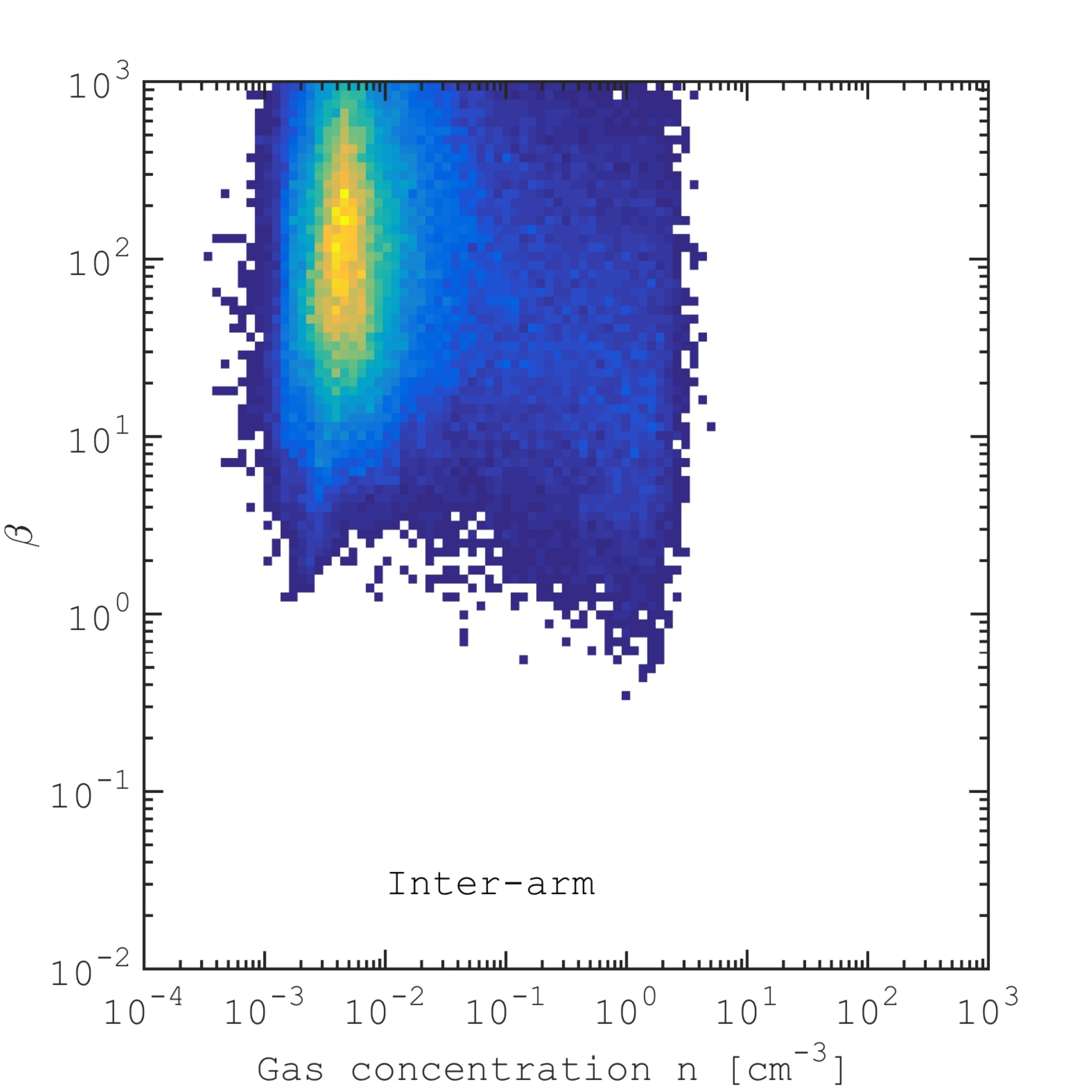}
\caption{Relation between plasma beta~($\beta$) and the gas concentration (cm$^{-3}$) for the whole disk~(left), in spirals arms~(center) and in the inter-arm regions~(right). Dashed line in the center frame is the distribution fit: $\beta \propto n^{-0.8}$}\label{fig::beta_dens}
\end{figure*}

Since we introduce the gas cooling, our simulations argue that, plasma parameter $\beta$ can vary with the gas parameters~(see Fig.~\ref{fig::beta_dens}). Plasma beta appears to be decreasing with the gas concentration $n$ and it varies in a wide range: from $10^{-3}-10^{-2}$ cm$^{-3}$ for the low density medium to $10-100$ cm$^{-3}$ in the densest parts of the spiral arms. Whereas we underline that the dense gas is much more magnetized than the low dense gas.  However, as it can be seen in the figure, for the whole galactic disk there is no clear relation $\beta-n$. Various galactic disk structures can develop the high-density contrasts. High-density regions in the galactic centre or in spirals created by gravitational collapse and gas compression due to spiral potential co-exist with a low-density fields at galactic outskirts and in the inter-arm region. Since, ISM physical conditions in spiral galaxies vary rapidly, the strength of magnetic field also changes on short time-scale and small spatial scales. 

In the inter-arm regions magnetic field strength is significantly lower $0.1 \Bunit$, in the arms magnetic field strength $1-10 \Bunit$~(see Figs.~\ref{fig::evolution_fid},~\ref{fig::ism_statistics}). It is clear that the inter-arm region is filled by the warm $\sim10^4$~K medium, meanwhile in spirals gas mostly appears in the dense phase with temperature $\leq 100$~K. Thus, it is natural to expect that magnetic filed parameters would be different in these regions~\citep[see e.g.][]{2008MNRAS.383..497D}. To check this point, we split computational domain on spirals and inter-arm regions. We assume that gas is located in spirals if the gravitational potential of spiral perturbation~(see.~Eq.~\ref{eq::spirals}) is deeper than $0.5$, otherwise gas is located in the inter-arm region.  Potential threshold value $0.5$ is chosen with the aim to have the gaseous arms located in the spiral region because of the possible spatial offset between minimum of the gravitational potential and the spiral shock~\citep[see e.g.,][]{2011AstL...37..563K}. In general the field tends to be slightly more ordered, and stronger in the spiral arms, and more random in the inter-arm region. For spiral arms region we find a relation $\beta \propto n^{-0.8}$~(see middle frame in Fig.~\ref{fig::beta_dens}). In the spiral region such relation is rather clear because of the close coupling of magnetic fields to gaseous spirals. From another side, there are numerous dense clumps or filaments in the inter-arm region~(see Fig.~\ref{fig::evolution_fid}) which are supported by a high mean magnetic pressure and consequently they are characterized by a lower plasma $\beta$. However, filling factor of such clumps and filaments is very small and the warm, low dense and less magnetized medium is  a dominant in the inter-arm region. Moreover, as we mentioned above, magnetic filaments can be spatially decoupled from the gaseous structures exactly in the inter-arm region~(see Fig.~\ref{fig::evolution_fid}). Such structures decoupling and a wide range of physical condition in the inter-arm region make the relation $\beta - n$ insignificant. 

\subsection{Global field growth and evolution of various components}\label{sec::results5}

\begin{figure*}
\includegraphics[width=0.33\hsize]{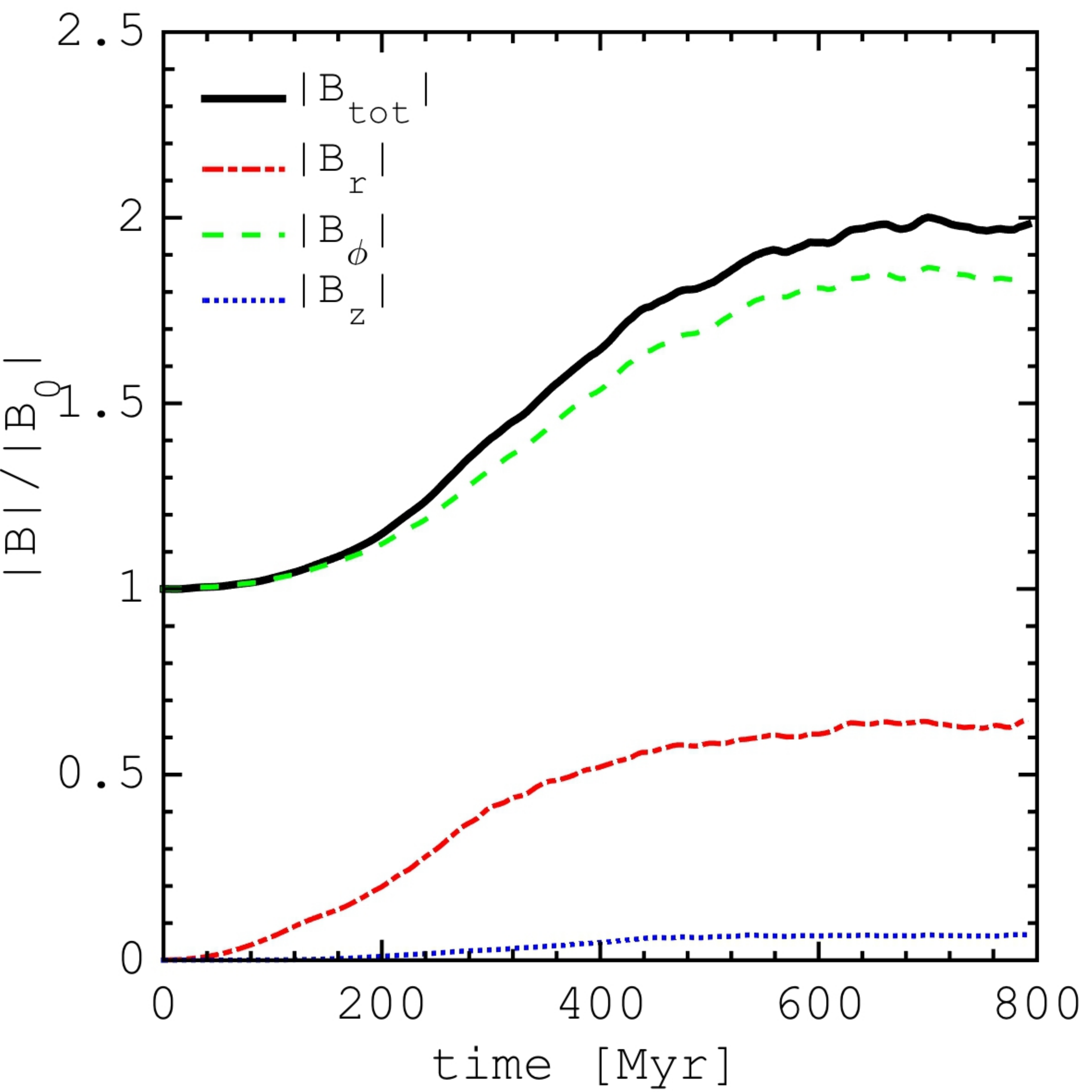}
\includegraphics[width=0.33\hsize]{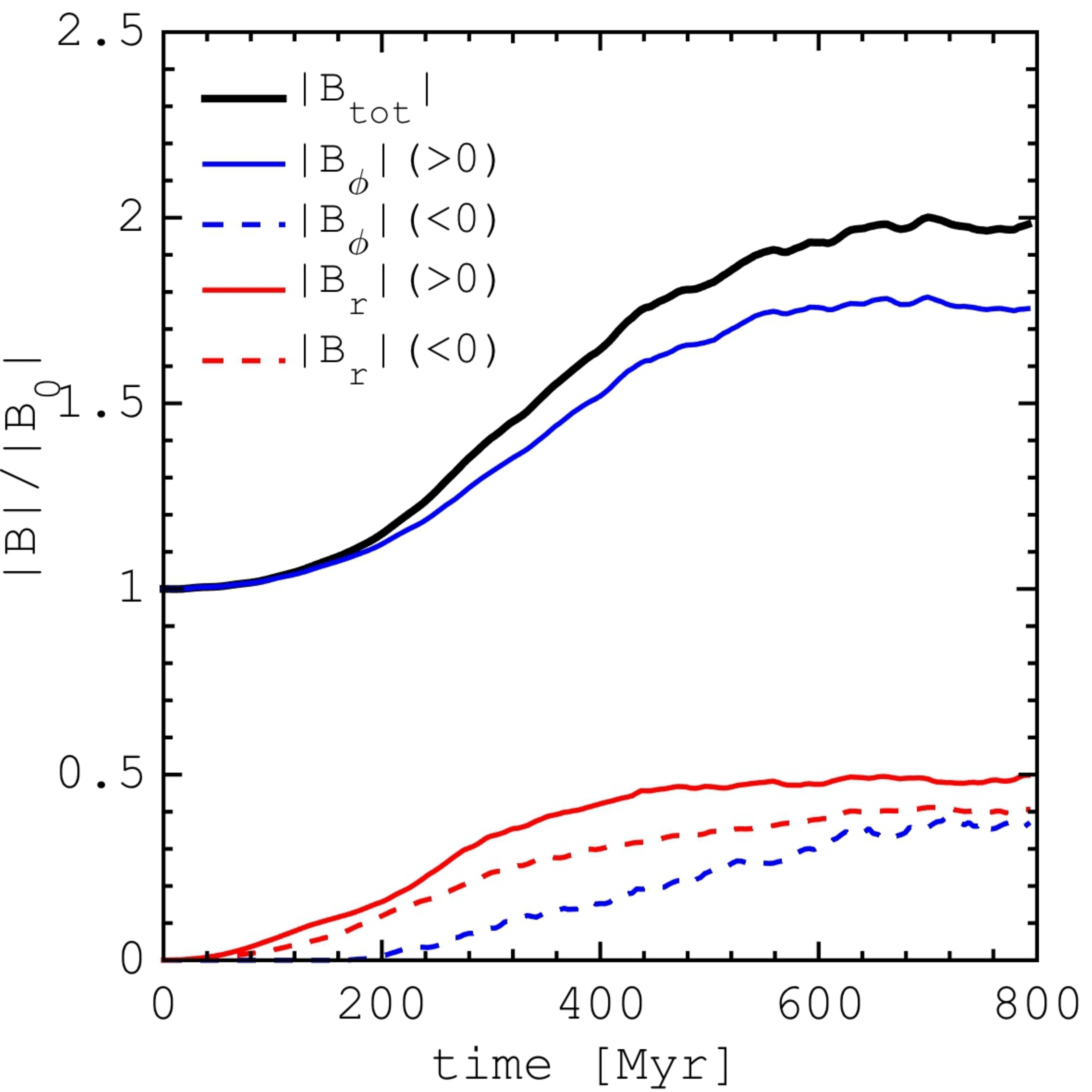}
\includegraphics[width=0.33\hsize]{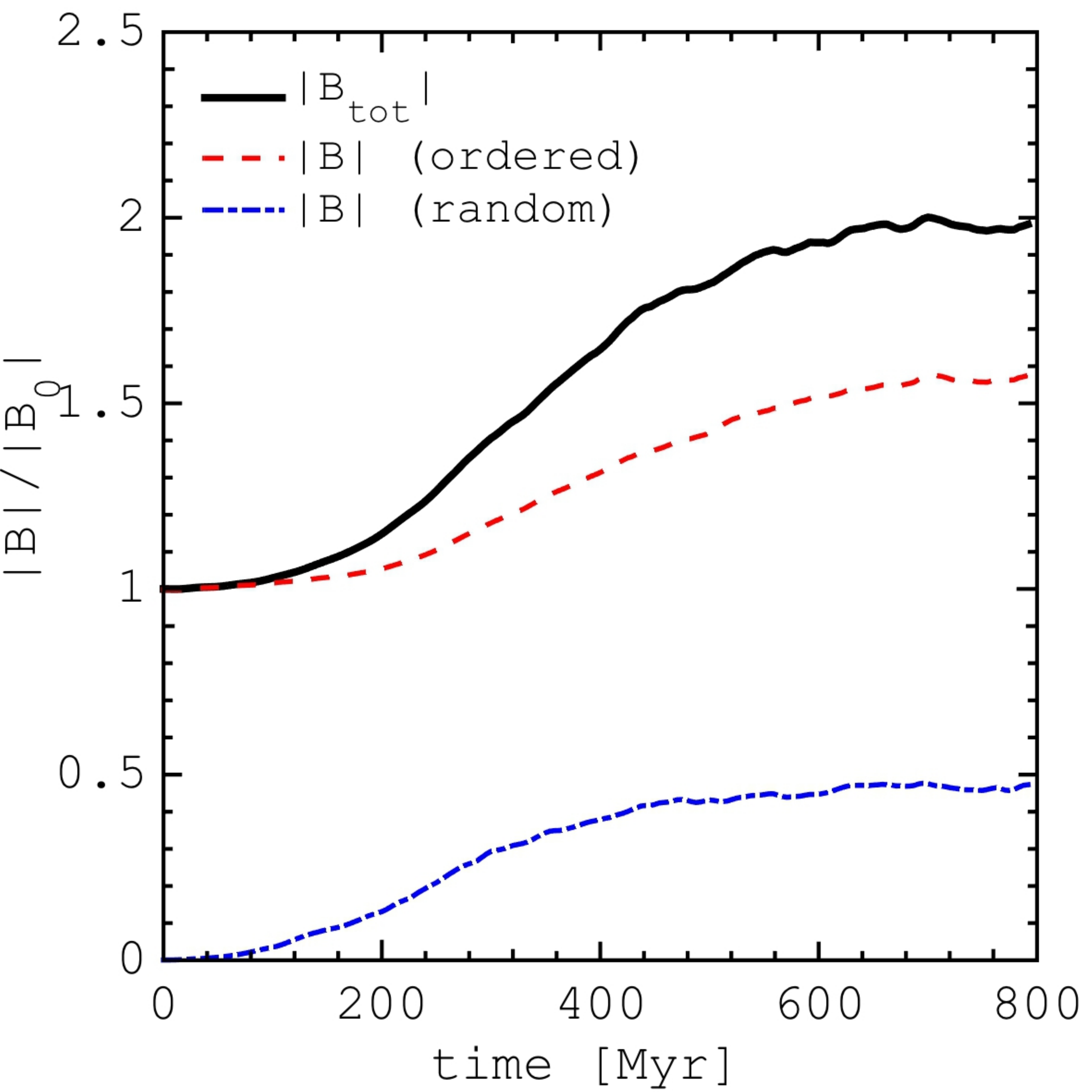}
\caption{Evolution of the magnetic field components. In all frames black line shows the total disk-averaged $|B|$ value evolution. In the left frame evolution of $|B_r|$, $|B_\phi|$ and $|B_{\rm z}|$ is shown. In the center we show modules of the positive and negative components of azimuthal and radial magnetic field. In the  right we show ordered and random magnetic field components. }\label{fig::b_evol}
\end{figure*}

Enhancement of various components of magnetic field is discussed in this paragraph. We employ this analysis to judge which component is dominant and has a larger growth rate in our fiducial model with two-arm spiral perturbation.  In Fig.~\ref{fig::b_evol} we show a time depended evolution of magnetic field projections~($B_{\rm r}, B_\varphi, B_{\rm z}$ in left), positive and negative magnetic field components~(center) and both ordered and random magnetic field strength~(right). By initial parameters set up~(see Eq.~\ref{eq::ini_magn_f}) there is a pure ordered, azimuthal field in the disk $B_\varphi>0$.  At latter times large scale field tends to be mostly connected with grand designed spiral perturbation during the whole simulation meanwhile in the inter-arm region the topology of the field is tangled. We find that azimuthal field $B_\phi$ increases by a factor of $2$ during first $500-800$~Myr of evolution since spiral pattern has been imposed. Maximum growth rate is between $200$ and $500$ Myr, and later magnetic field strength reaches saturation level.  We also report about the sensible growth of the radial component of magnetic field which reaches up to 20\% from the mean value of magnetic field. Since spiral structure has a reasonable pitch angle ($20^\circ$), radial field can be partially driven by the spiral arms. In the absence of star formation and corresponded feedback, and magnetic field $B_z$ stays negligible at a level $0.1-1$\% of the total strength.  

In the middle frame of Fig.~\ref{fig::b_evol} we plot the evolution of positive and negative components of radial and azimuthal magnetic field. Since the magnetic field shows a significant level of alignment with the velocity field,
positive azimuthal magnetic field component $B_\phi$ is characterized by the strongest enchansement, up to $1.8$ times from the initial value. Reversal magnetic field component ($B_\varphi<0$) also appears in our simulations~(up to 15\%) and as it has been shown above~(see Section.~\ref{sec::results3}), most of this field is related to spurs and other tiny structures in the inter-arm region. Since these structures have a controversial direction of magnetic field at front and back sides, radial field components~(both negative and positive) are characterized by the same growth rate~(see middle frame in Fig.~\ref{fig::b_evol}).

We also calculate ordered $B_{\rm ord}$ and random $B_{\rm rnd}$ magnetic field components evolution. We introduce these quantities similar to ~\cite{2008MNRAS.383..497D}:
\begin{equation}
\Oo B_{\rm ord} = \sqrt{\langle B_\varphi \rangle^2 + \langle B_r \rangle^2 + \langle B_z \rangle^2 }\,,
\end{equation}
\begin{equation}
\Oo B_{\rm rnd} = \sqrt{\langle(B_\varphi - \langle B_\varphi \rangle)^2  + (B_r - \langle B_r \rangle)^2 + (B_z - \langle B_z \rangle)^2 \rangle}\,.
\end{equation}
In the right frame of Fig.~\ref{fig::b_evol} evolution of these components is shown. Ordered magnetic field is the dominant component at all the time and it reaches up to 75\% from the final mean field. Random magnetic field is mostly driven by the wiggle instability in the inter-arm region where the gas motions are turbulent and the largest density fluctuations are seen on the small scales. However, relative increase (difference between final at $800$~Myr and initial field) of ordered and random field is very similar and it equals to 0.25\% from the final mean field strength. Thus, we claim that we observe growth of both components, but not a transformation of ordered component to the random field.

MHD studies of SN driven turbulence have had considerable success in predicting the magnetic field growth rate. The main interest of such works is to simulate a turbulent galactic dynamo. For instance, \cite{1999A&A...350..230K} used a resolution of $10$~pc for a local patch of the galactic disk with taking into account galactic shear. Their initial field $0.1~\Bunit$ rapidly increases to $1.3~\Bunit$ strength. Various models suggest that supernova-driven turbulence in conjunction with shear leads to an exponential increase of the mean magnetic field with the time scales $0.1-0.25$~Gyr~\citep{2008A&A...486L..35G, 2013MNRAS.430L..40G}. By means of cosmic ray driven dynamo \cite{2009ApJ...706L.155H} also found that timescale of the large-scale magnetic field component growth is close to the disk rotation period. 

\cite{2008A&A...486L..35G} claimed that the rotation frequency is the critical parameter allowing the dynamo to operate because for models without shear \cite{2008AN....329..619G} found no amplification of the mean field. Hence, we can compare magnetic field enchantment rate found by \cite{2008A&A...486L..35G} with our model e0~(without spiral pattern) where the field radial migration is negligible. In all our models we adopted the same rotation curve and it matches the condition established by \cite{2008A&A...486L..35G} $\Omega>25~$~\kmpskpc for galactocentric radii $r<8$~kpc. We estimated total magnetic field growth rate as a function of angular velocity $\Omega(r)$ for model without spiral perturbation. Since our model does not extend long beyond the apparent saturation of the magnetic field growth, we are able to detect a linear increase of regular field strength with the doubling times of $\approx 250$, $150$, and $80$~Myr for rotation frequencies~$\Omega=50$, $100$, $200$~\kmpskpc, respectively. Meanwhile \cite{2008A&A...486L..35G} found the exponential time scales  $147$, $102$ and $52$~Myr for the same set of rotation rates. Thus in the absence of dynamo, we obtained much lower field growth rate in comparison to local dynamo models. Note also that the magnetic field enchantment we found is smaller by a factor of a few than \cite{2016MNRAS.461.4482D} found in similar galactic scale SPMHD simulations.

\subsection{Comparison of different models with various spiral pattern parameters}\label{sec::results6}
\begin{figure*}
\includegraphics[width=0.24\hsize]{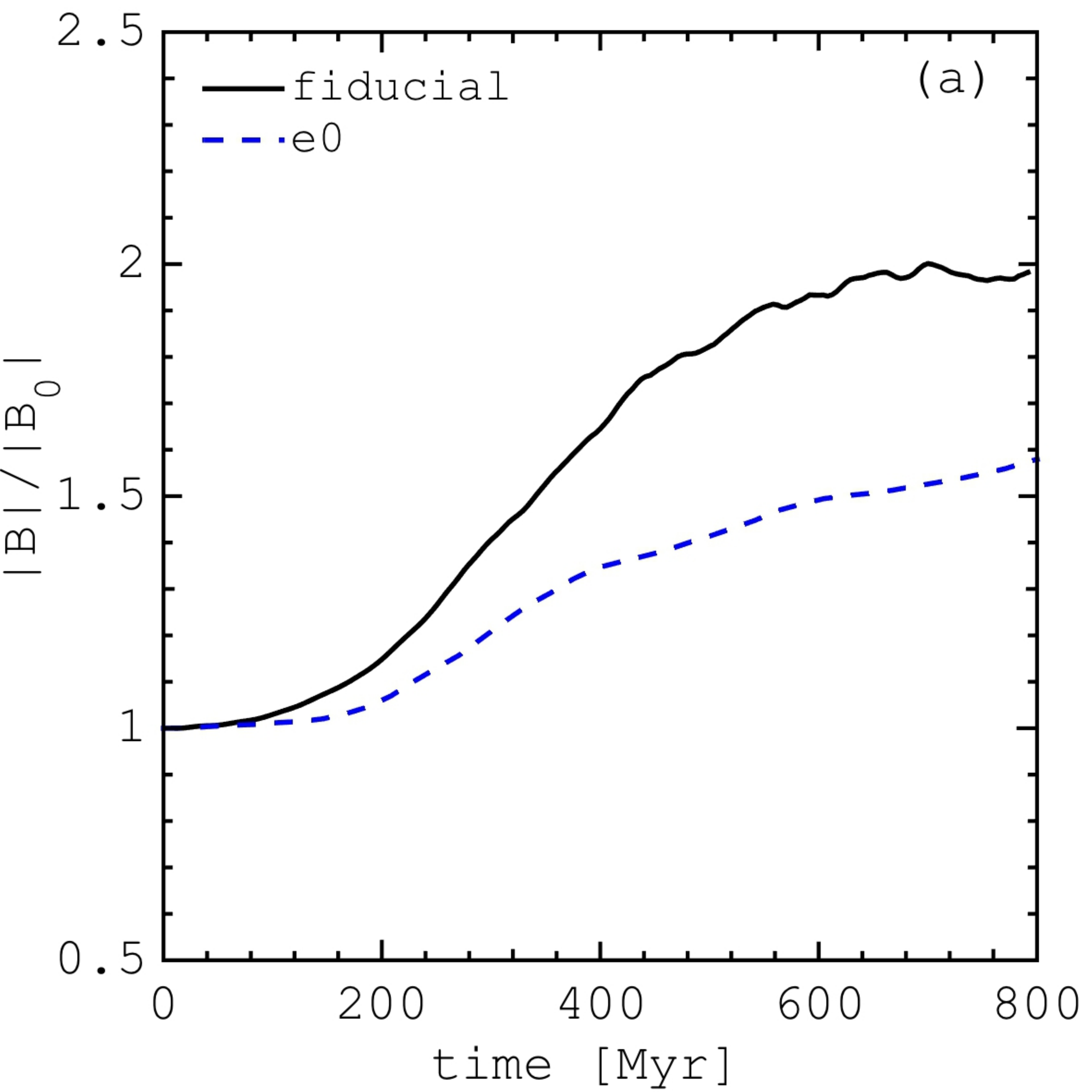}
\includegraphics[width=0.24\hsize]{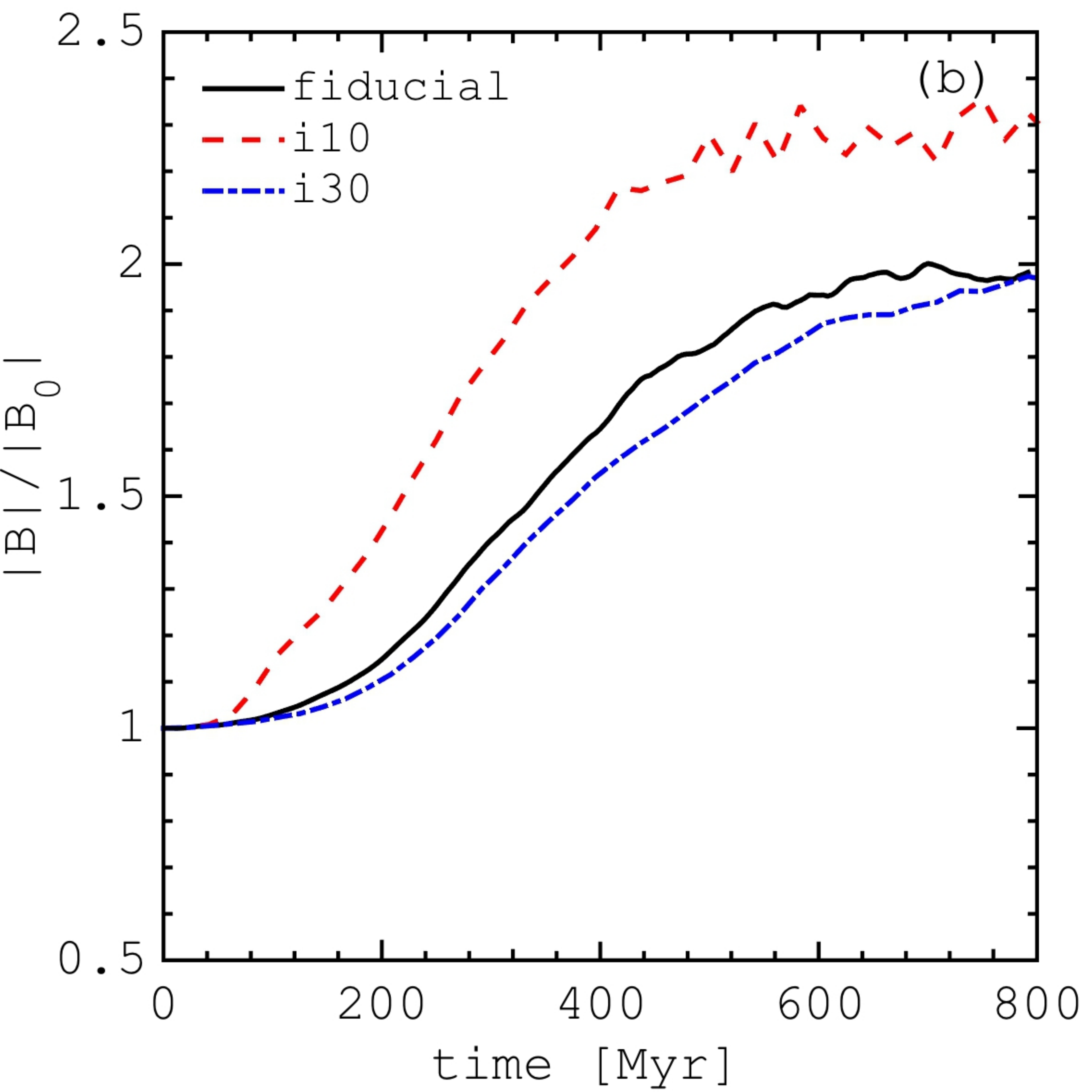}
\includegraphics[width=0.24\hsize]{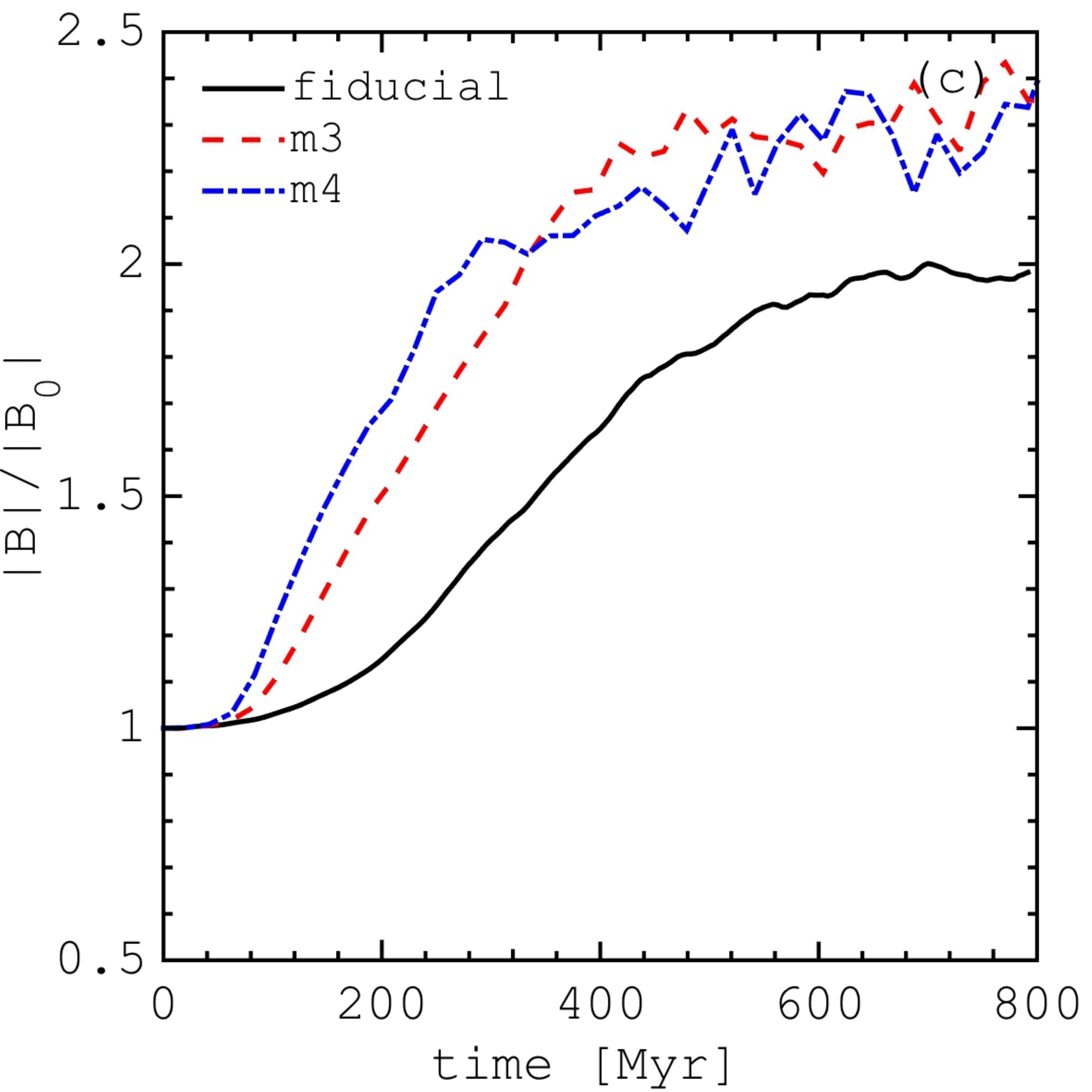}
\includegraphics[width=0.24\hsize]{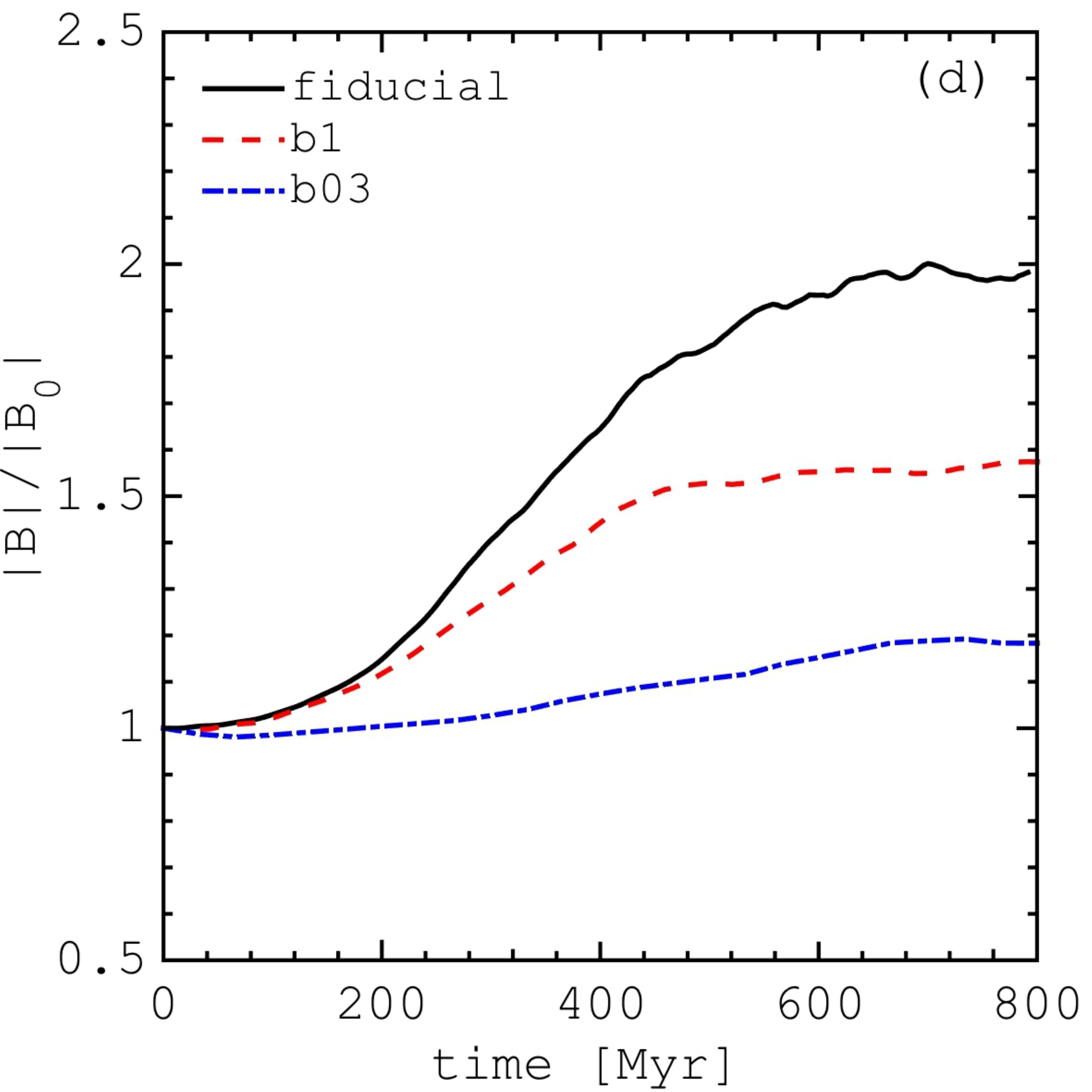}
\caption{Mean magnetic field strength as a function of time in various models. In all frames fiducial model is shown by black solid line, another models are the following: (a) blue dashed line, model without imposed spirals; (b) models with a different pitch angle $10^\circ$~(red dashed line) and $30^\circ$~(blue dash-dot line); (c) models with a different arm number $m=3$~(red dashed line) and $m=4$~(blue dash-dot line); (d) models with a different initial magnetic field $\beta=1$~(red dashed line) and $\beta=0.3$~(blue dash-dot line). }\label{fig::evolution_different_models}
\end{figure*}

In this paragraph we discuss the magnetic field evolution in models with a various spiral structure parameters. First we demonstrate the role of spiral arms. The density structure of the ISM in a spiral potential can be affected by three free parameters here: the pattern speed, strength of the spiral potential and the pitch angle. In Fig~\ref{fig::evolution_different_models}a we compare our fiducial run with model without any spiral perturbation. Absence of global  structures in the disk let enhance the mean magnetic field by a factor of $1.5$ which is 25\% lower than in models with $m=2$. Spiral arms play a significant role in the field generation, however the impact is comparable to the field growth driven by the gas random motions and differential rotation of the disk. In our fiducial model the increase of gas density in spiral arms~(in comparison to axisymmetric model, e0) increased rate of winding of the field, hence the field strength is higher.

\begin{figure}
\includegraphics[height=1\hsize]{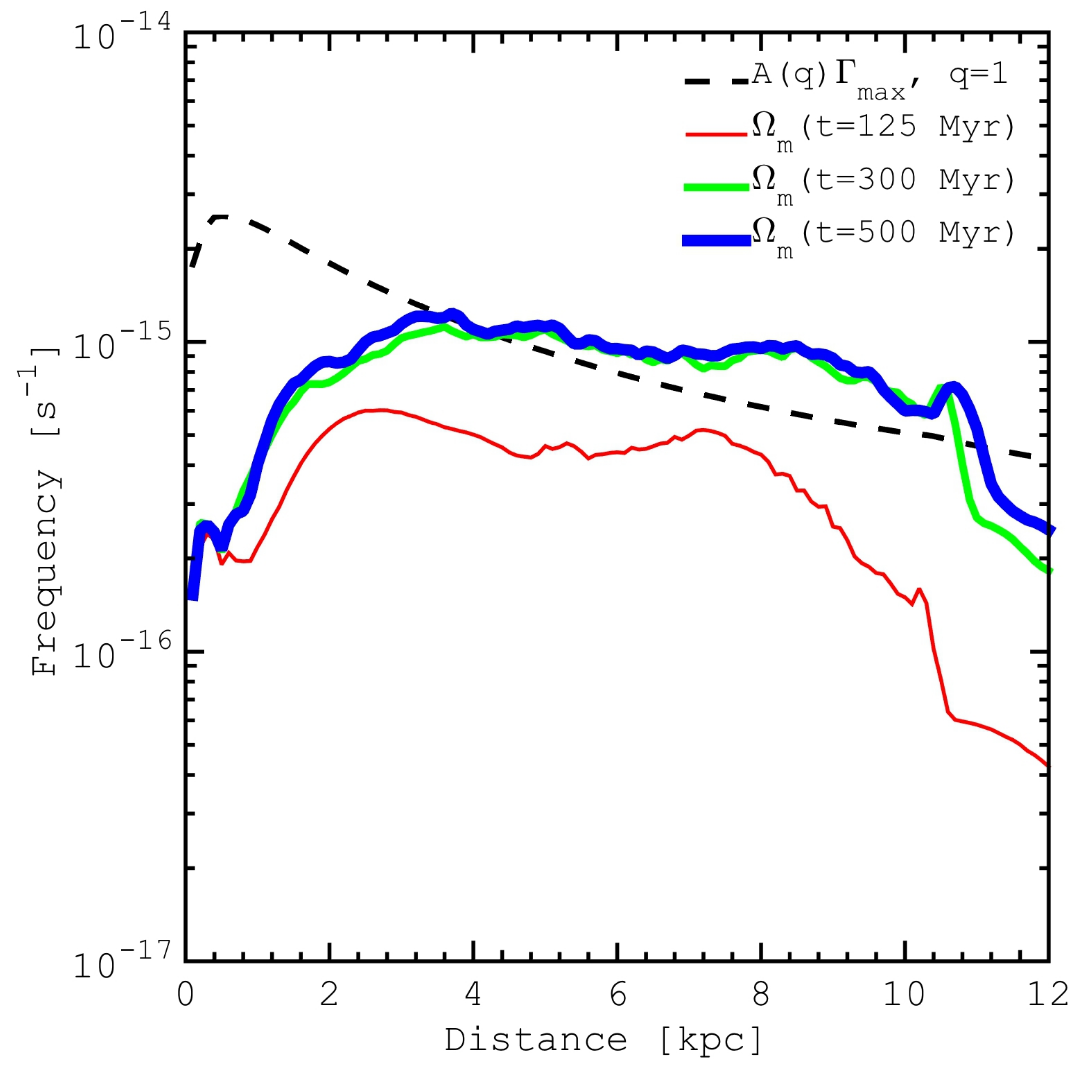}\caption{Comparison of simulated galaxy properties with MRI conditions from~\cite{2010AN....331...34K}: MRI damping curve~(see Eq.~\ref{eq::dampling_curve}) for fiducial model~(dashed line) and expected growth rate of the MRI $\Omega_m$~(see Eq.~\ref{eq::omm}) at different times~(solid lines). MRI is predicted to be damped by gas random motions in the regions where the solid line is above the dashed line.}\label{fig::mri_check}
\end{figure}

Tightly wound spirals generate larger magnetic field in comparison to more opened spirals~(see Fig.~\ref{fig::evolution_different_models}b). For pitch angle $i=10^\circ$, magnetic field strength increases much faster in comparison to models with $i=20^\circ$ and $i=30^\circ$. Saturation level for this model is also by $10-15$\% higher. Multiple spiral arm models~($m=2, 3$) are characterized by a rapid mean magnetic strength growth. Mean magnetic field is higher by a factor of $\approx 2.5$ in comparison to the initial field strength~(see Fig.~\ref{fig::evolution_different_models}c). In case of tightly wound spirals~($i=10^\circ$) and for larger mode number~($m=3,4$) spiral structure action is stronger, and hence the rate of magnetic field winding is higher within a given disk radius. Hence, we detect a faster growth rate of magnetic field strength with a higher equilibrium value.

\subsection{Conditions for MRI}\label{sec::MRI}
\begin{figure*}
\includegraphics[width=0.33\hsize]{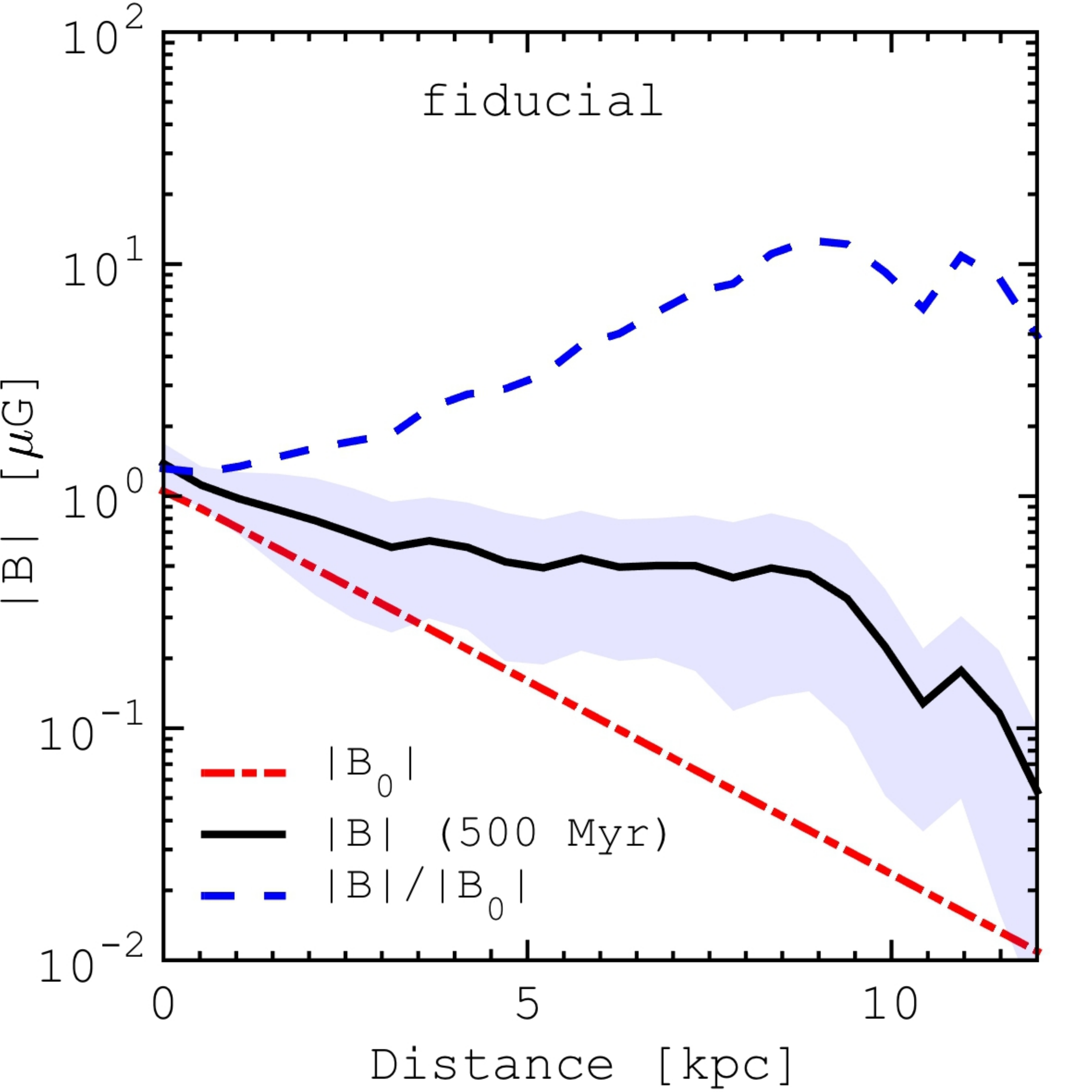}\includegraphics[width=0.33\hsize]{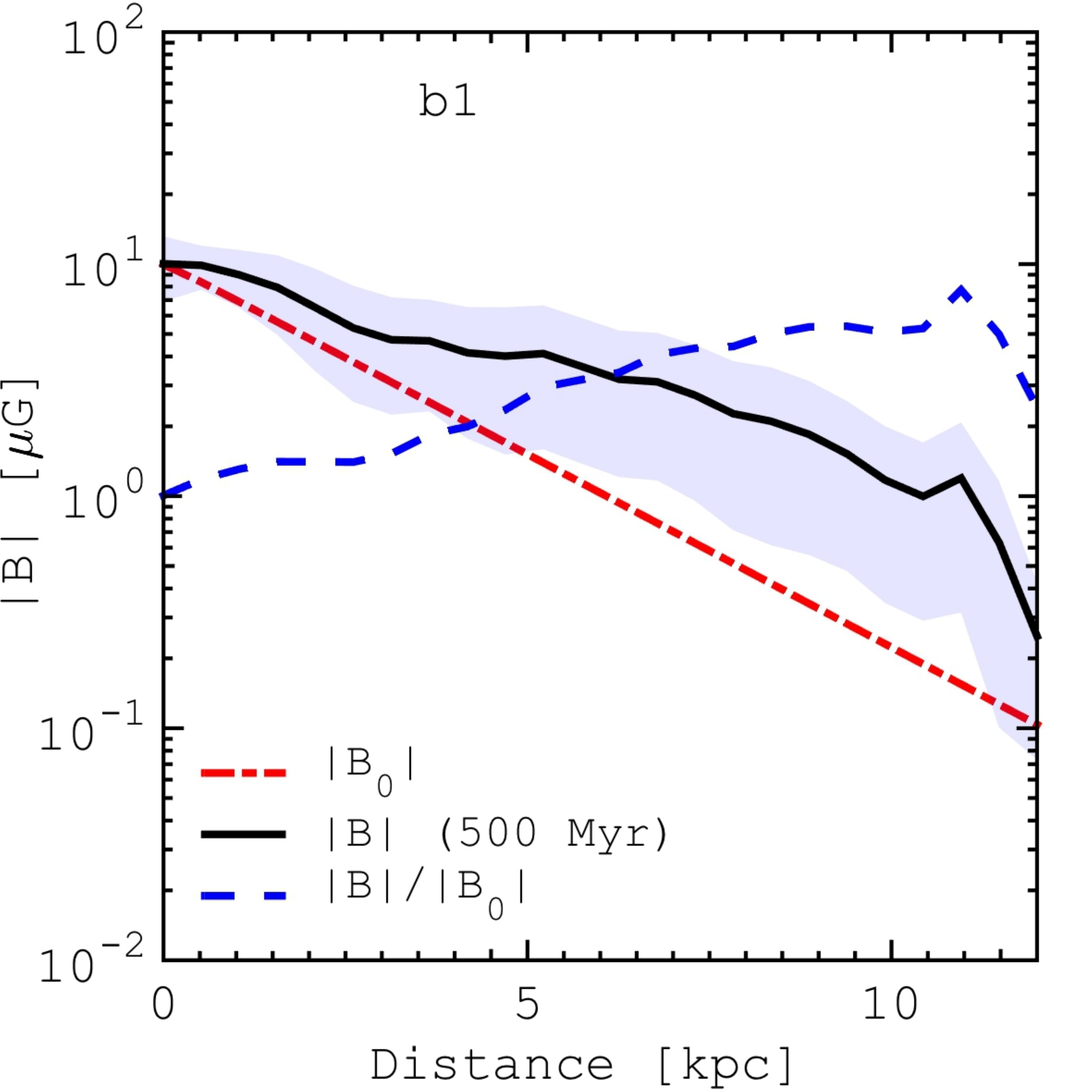}\includegraphics[width=0.33\hsize]{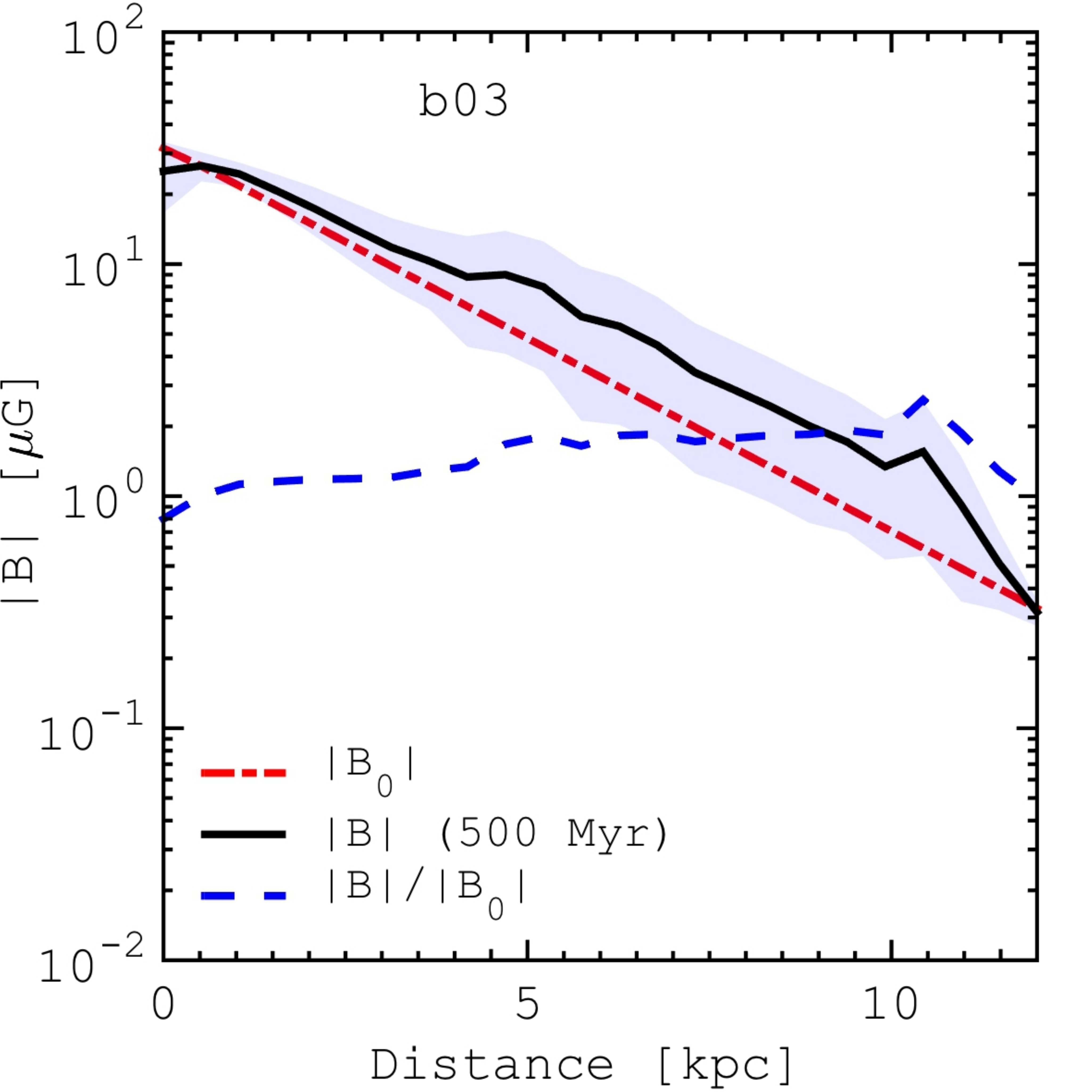}\caption{Radial profiles of the initial~(red dash-dot line) and final (at $t=800$~Myr, solid black line) magnetic field strength in models with a different initial magnetic field strength. Ratio between these quantities is shown by blue dashed line. Filled area around the final magnetic field strength profile indicates the scatter at a given distance.}\label{fig::compare_different_ini_magnetic_radial}
\end{figure*}

SN-driven turbulent ISM dynamics have been studied numerically by numerous authors~\citep[see e.g.][]{2002ApJ...581.1047D,2008A&A...486L..35G}, but MRI has not been found in the inner disk part because it may be suppressed by SN driven turbulence. For instance, \cite{2013A&A...560A..93G} demonstrated that MRI turbulence can be sustained only for distances $R>14$~kpc. Similar result was obtained by~\cite{2010AN....331...34K} who estimated the limiting radius inside of which the SN activity can suppress the MRI for the Milky Way-type galaxy. Indeed, MRI might be important in extended HI disks including outskirts of the Milky Way where the star formation is almost impossible~\citep{1999ApJ...511..660S}. For this reason, absence of star-formation in our models allows us to compare our simulations with conditions for MRI.

By using linear analysis of MHD equations, \cite{2010AN....331...34K} derived the limiting turbulence level and hence the turbulent diffusion needed to damp the MRI. From the linear stability analysis MRI is damped if
\begin{equation}
\Omega_m > A(q) \Gamma_{max} = \frac{qA(q)}{2}\Omega_0\,,\label{eq::dampling_curve}
\end{equation}
where 
\begin{equation}
\Omega_m = \frac{1}{3} \sigma_{\rm gas} k_f\,,\label{eq::omm}
\end{equation}
and $k_f$ is the forcing wavenumber, $A(q)$ varies from $1$ to $3$~\citep[see Fig. 2 in][]{2010AN....331...34K}. All MRI-modes to be damped for $q = 1$~(flat rotation curve) if $A=1.4$. 

In Fig.~\ref{fig::mri_check} we show the corresponding damping curve and estimated turbulence diffusion rate at different times~($125$, $300$ and $500$~Myr). MRI is predicted to be damped by gas random motions in the regions where the solid lines are above the dotted line. For our estimation, we adopt the value of forcing wavelength is $100$~pc, which is the typical spatial scale of small scale structures in our models, but we neglect any variation in space and time. Large gas condensations, clumps and filaments with masses of $10^5-10^6$~\Msun require $\approx 100$~Myr to form via gravitational instabilities and accretion of mass from surrounding ISM~\citep[see e.g.][]{2011MNRAS.413.2935D,2013MNRAS.428.2311K,2013MNRAS.432..653D}. Until this time scale, gas velocity dispersion is not enough to stabilize the disk against MRI growth. In the inner regions of simulated spiral galaxies rapid formation of clumps and further gravity driven random motions are the major source of turbulence. Once small scale structures appear in the gas, velocity dispersion is increases rapidly and MRI to be damped up to the radius of roughly $<11$~kpc at $t>200$~Myr. 

\subsection{Radial profiles of the magnetic field}\label{sec::results6a}
\begin{figure*}
\includegraphics[height=0.25\hsize]{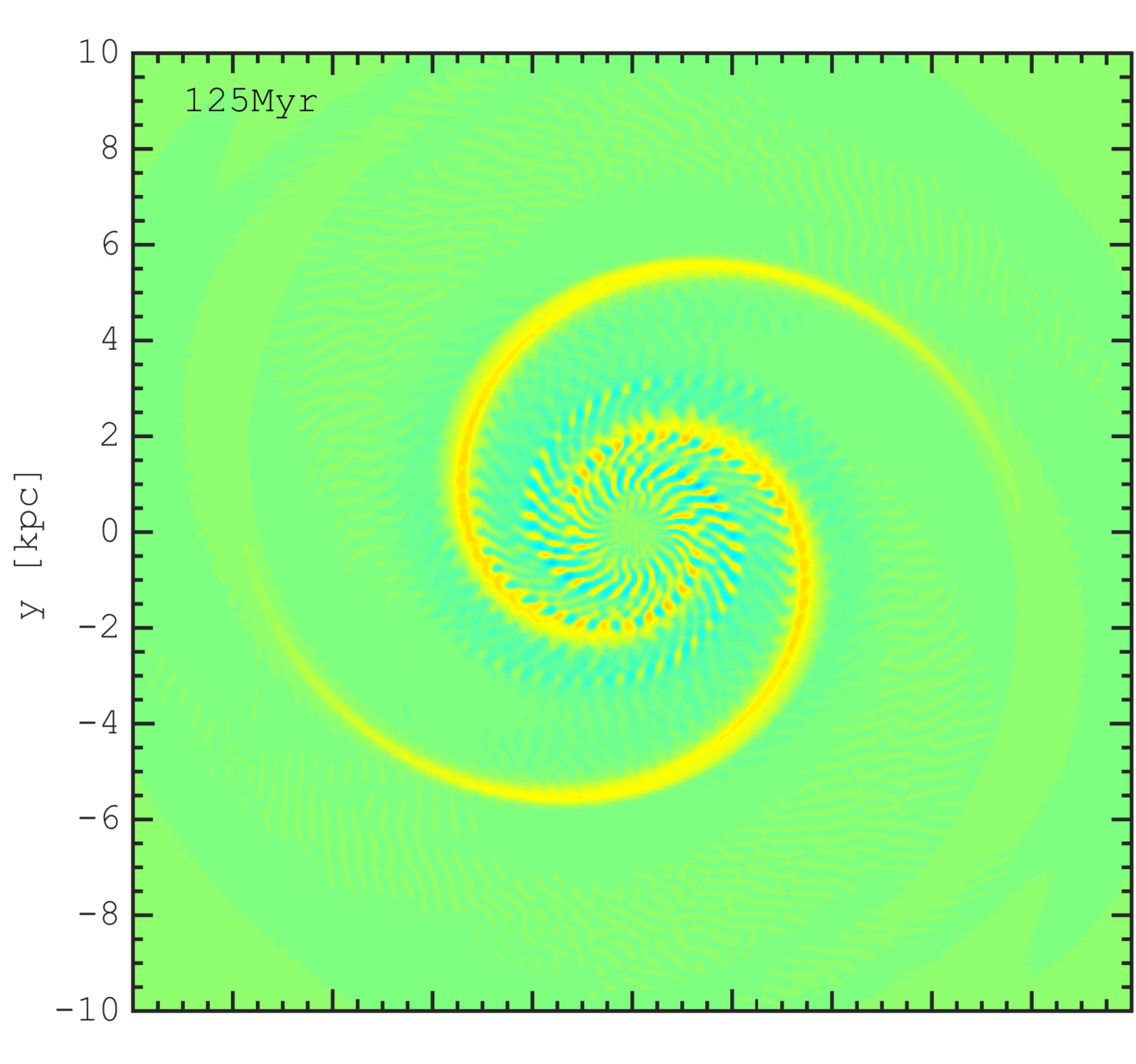}\includegraphics[height=0.25\hsize]{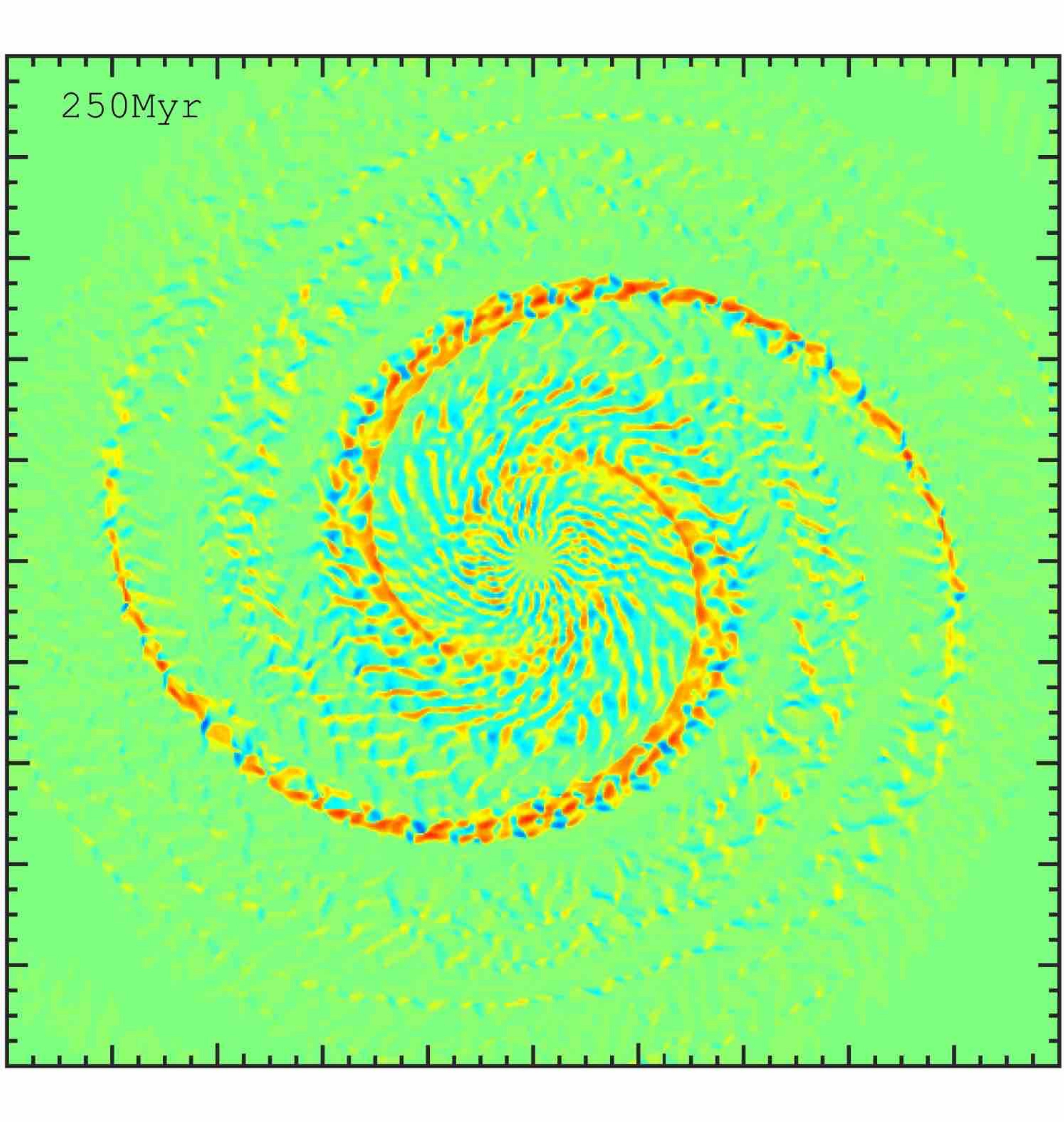}\includegraphics[height=0.25\hsize]{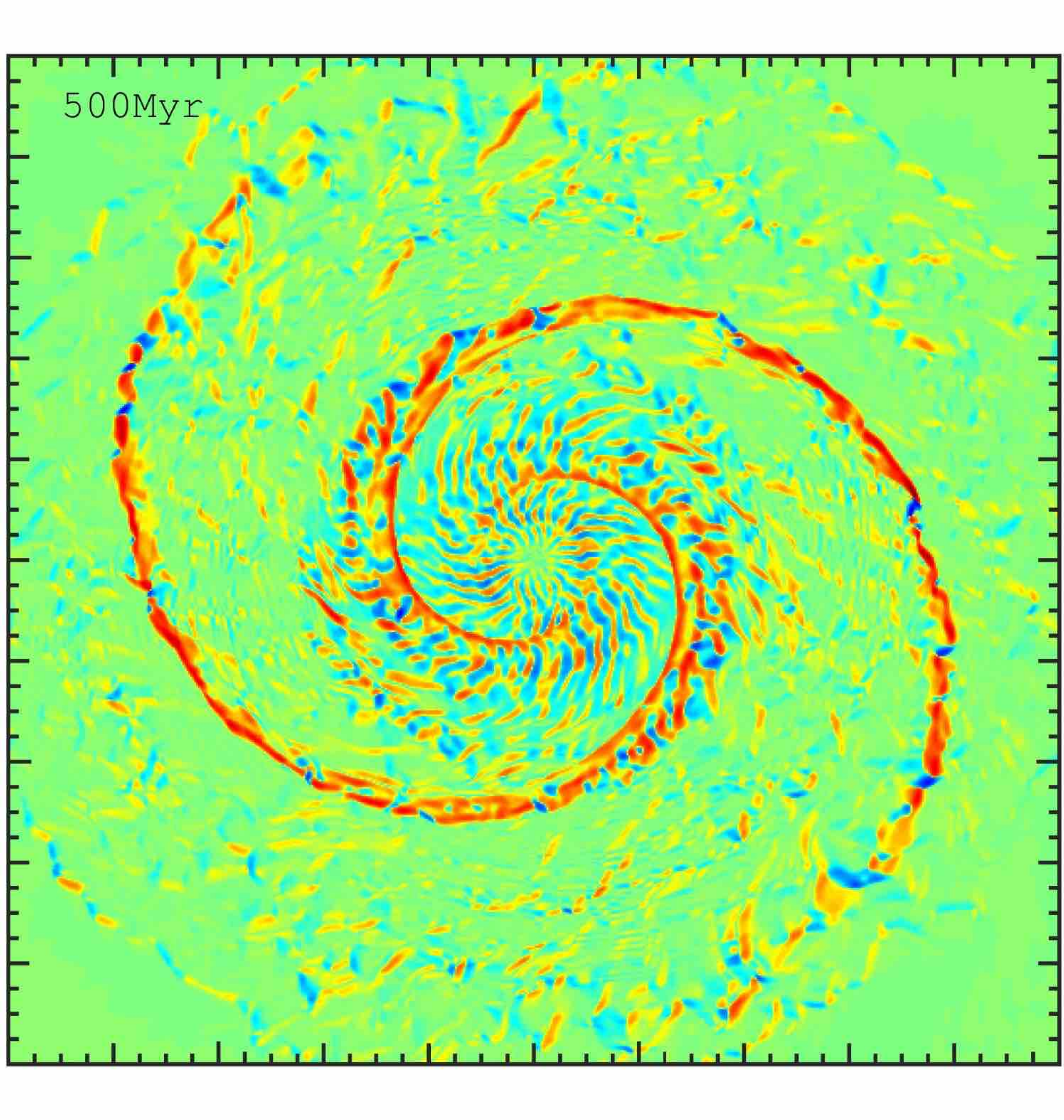}\includegraphics[height=0.25\hsize]{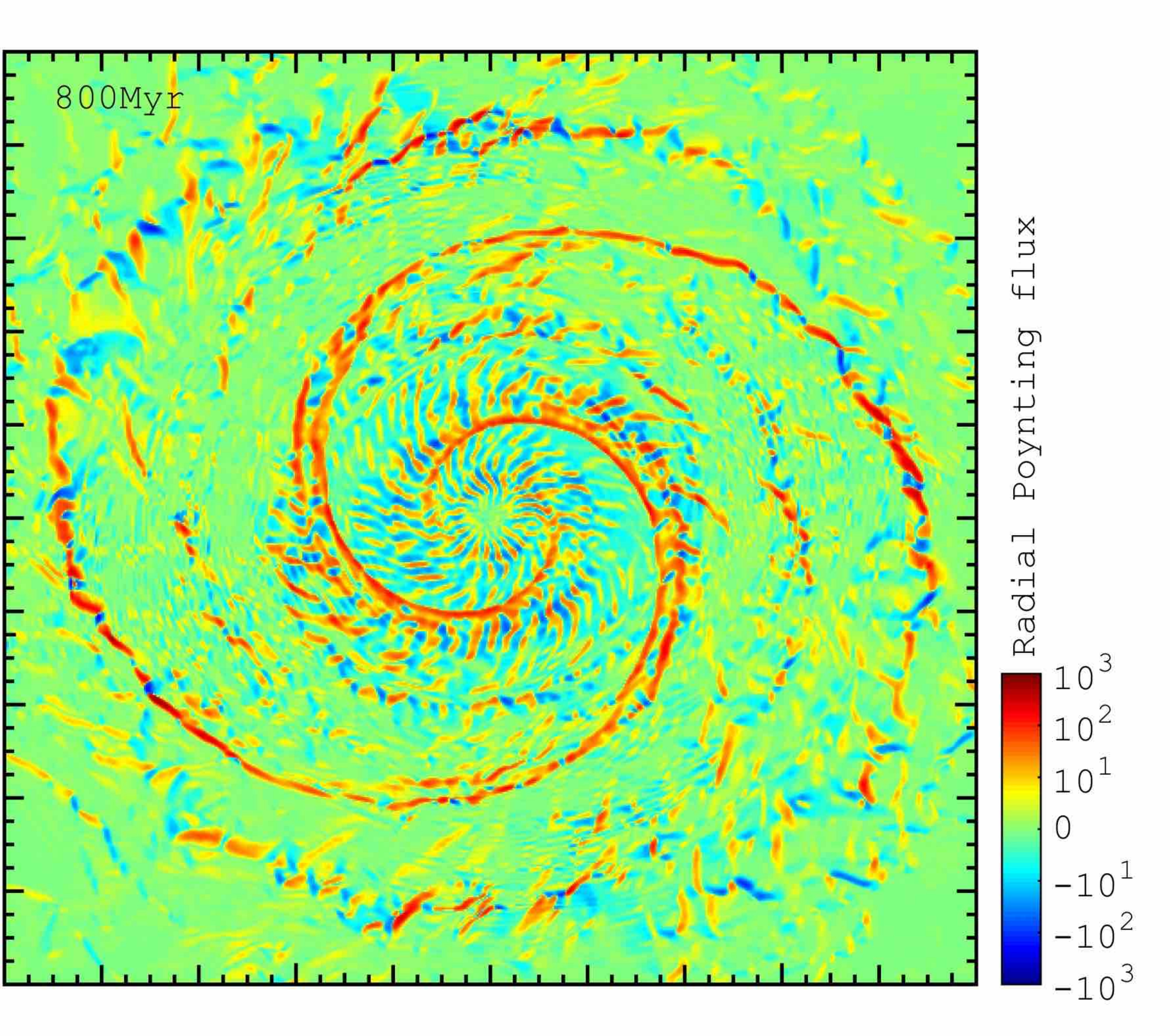}\caption{Time dependence of radial component Poynting flux distribution in fiducial model.}\label{fig::poynting_flux_2d}
\end{figure*}

As we showed above, initial magnetic field strength plays the crucial role in the further disk evolution~(see Fig.~\ref{fig::different_beta}). Indeed, the amount of small structures in gas rapidly decreases for lower values of $\beta$. For these models we have a very different magnetic field strength evolution and relative value of the equilibrium magnetic field strength $|B_{\rm eq}|/|B_0|$ is proportional to  $\beta $~(Fig.~\ref{fig::evolution_different_models}d). We compare initial magnetic field profile with the final state for models with different initial $\beta$~(see Fig.~\ref{fig::compare_different_ini_magnetic_radial}). We found that the magnetic field ratio is a function of radius for all models we analyze. We also claim flattening of mean field radial profile. In the inner part magnetic field amplification is negligible but at the outer edge magnetic field ratio is larger by a factor of $10$~for models with initial $\beta=10$. Hence the magnetic field growth rate increases from the inner to the outer galaxy. Such picture is in consistent with dynamo action models where the regular fields generated in the inner disk could be transported outwards by the joint action of a dynamo~\citep[see e.g.,][]{1998GApFD..89..285M}.

\begin{figure}
\includegraphics[height=1\hsize]{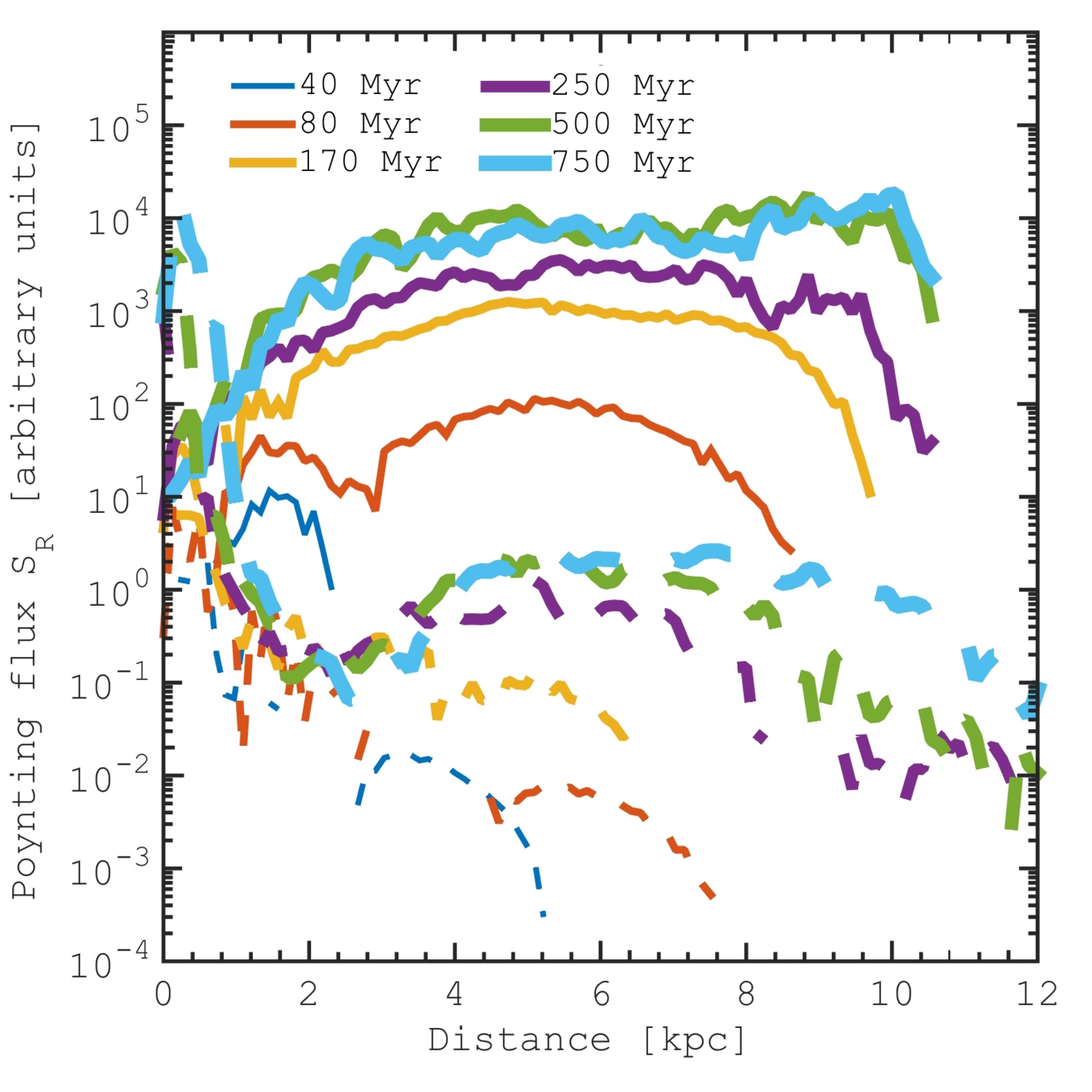}\caption{Time dependence of azimuthally averaged radial component of Poynting flux in fiducial model~(solid lines) and in model without spiral structure~(e0).}\label{fig::poynting_flux_1d}
\end{figure}

The Poynting flux for MHD is interpreted as the rate at which energy is transported into (or out) of a given volume by the electromagnetic fields. To explore the transport of magnetic energy radially in our galactic disks we measure the Poynting flux in the radial direction. In Fig.~\ref{fig::poynting_flux_2d} we show the radial component of Poynting flux at different times in our fiducial model. Positive values imply outward transport and negative -- inward transport of the magnetic field. One can see the positive values of Poynting flux is nicely coincide with the spiral pattern at all the time, which means that the the spiral perturbation is responsible for the radial transport of the magnetic field.

In rotating spiral potential the gas density peaks occur downstream from the minimum of potential. In the outer region~(beyond the corotation radius) the spiral pattern passes through the gas, leaving the shocked wave behind. The shock front itself is always upstream from the density peak and in the outer disk the shock forms on the outer face of the spiral arm~\citep{2011AstL...37..563K,2014ApJ...789...68K}. Across the spiral shock both radial and tangential velocity rapidly increase their amplitude~\citep{2006ApJ...647..997S}. This breaks the symmetry  of the galactic velocity field and as the magnetic field variations are coupled to the velocity field variations, the magnetic field is winded up efficiently by the rotating non-axisymmetric structures. The winding  of fields also enhances the radial diffusion for non-axisymmetric modes. Rapid increase of the gas velocity component implies an increase of the shear. The increase of shear leads to the radial diffusion of fields. Thus we conclude that in our models the non-axisymmetric spiral pattern is the driving force of the magnetic field transfer across the disk~\citep[see also][]{2009MNRAS.397..733K}.  

To better understand how spirals acting the Poynting flux we have checked the location of the Poynting flux peaks in the vicinity of gaseous spirals. At early times~(<$200$~Myr) we found that Poynting flux peaks are located on outer surface of spirals with spatial offset $\approx 50-100$~pc. However later in our models we are not able to find any clear spatial offset between the gas density and Poynting flux peaks. Depending on the region in the disk  we can see that either the Poynting flux has its maximum at the inner edge of spiral structure or we have the opposite configuration or Poynting flux peaks are coexist with the gas density peaks. Probably for well developed spiral pattern we obtained the mixture of these configurations due to a non-steady and nonlinear evolution of gas in a rotating spiral potential.

In Fig.~\ref{fig::poynting_flux_1d} we show azimuthally averaged radial profiles of the mean radial component of the Poynting flux values at different times. {\bf In our fiducial model} at early stages of evolution ($<200$~Myr), Poynting flux profile does not expand out to the outer disk edge, its profile has a break at a given radius where the gaseous spirals significantly decrease their amplitude. However, gaseous spirals increase their amplitude in time and expand over the larger and larger area. Consequently, at later times~($\geq 200$~Myr) radial Poynting flux profile saturates at all radii. In the model without spiral perturbation~(e0) we do not observe significant increase of the Poynting flux in time and space. Thus we can finally summarise that in our simulations spiral waves action is indeed the major driver of the radial transport of magnetic field outward.

\subsection{Model with initial turbulent magnetic field structure}\label{sec::results7}
\begin{figure*}
\includegraphics[width=0.5\hsize]{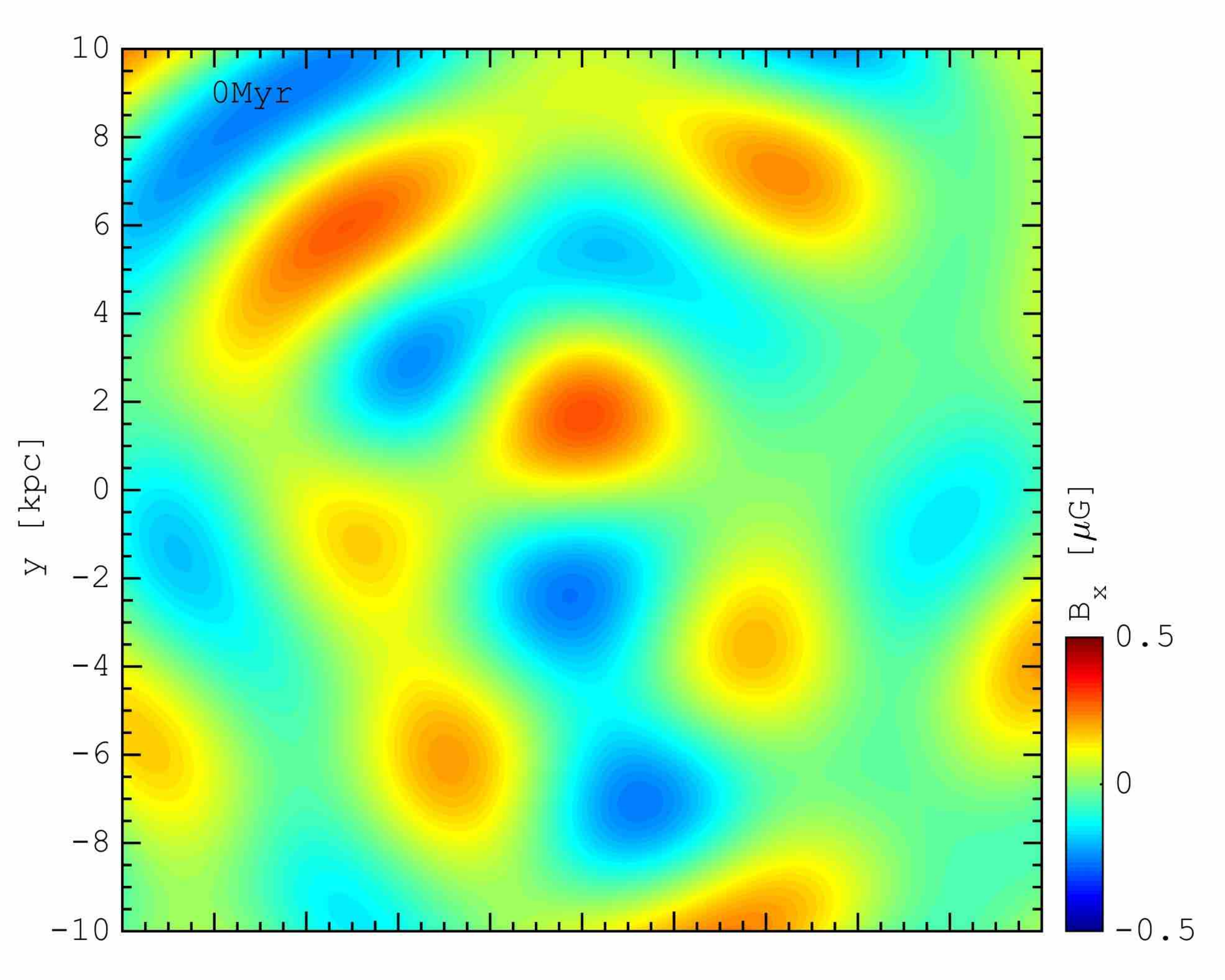}\includegraphics[width=0.5\hsize]{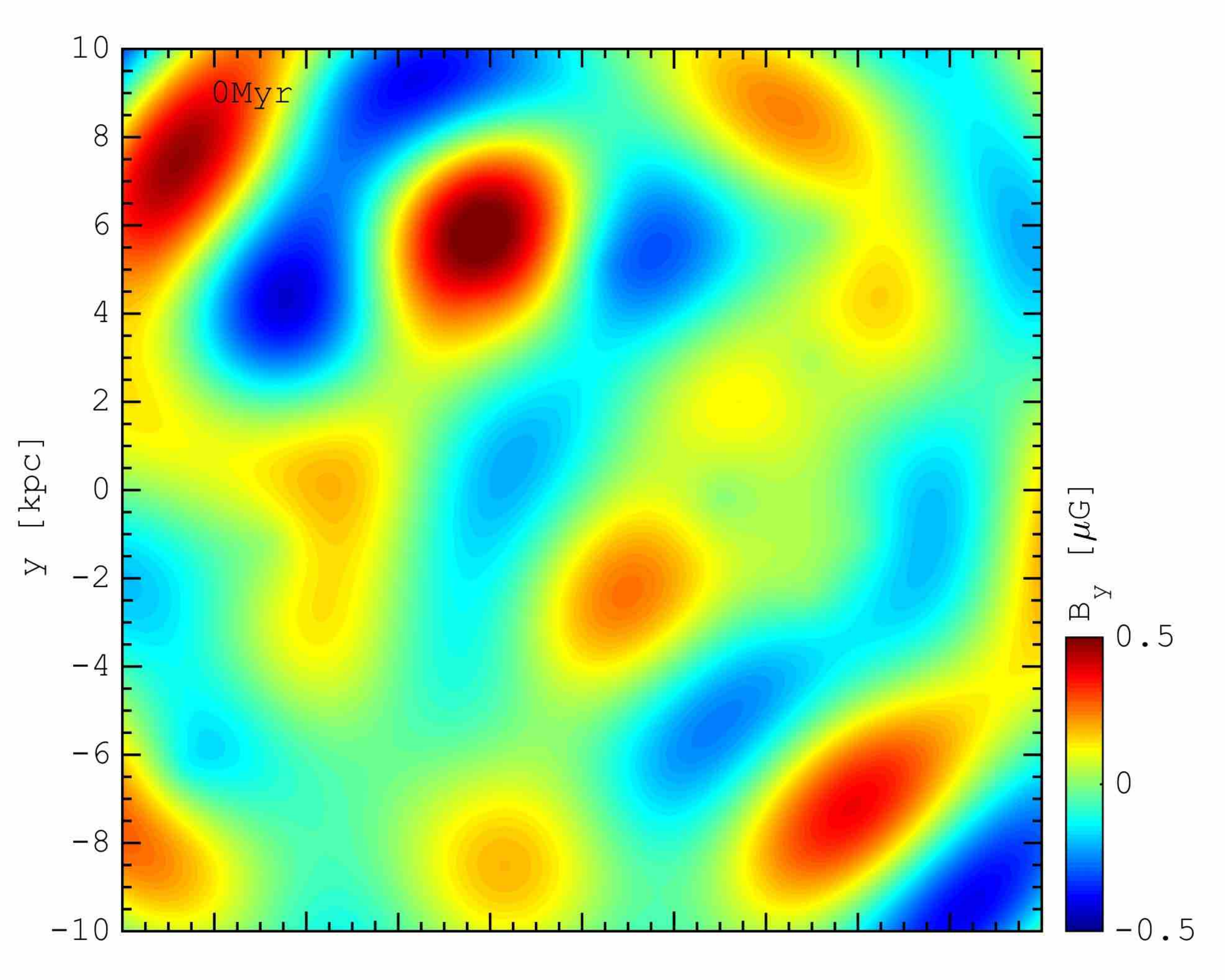} \\ \includegraphics[width=0.5\hsize]{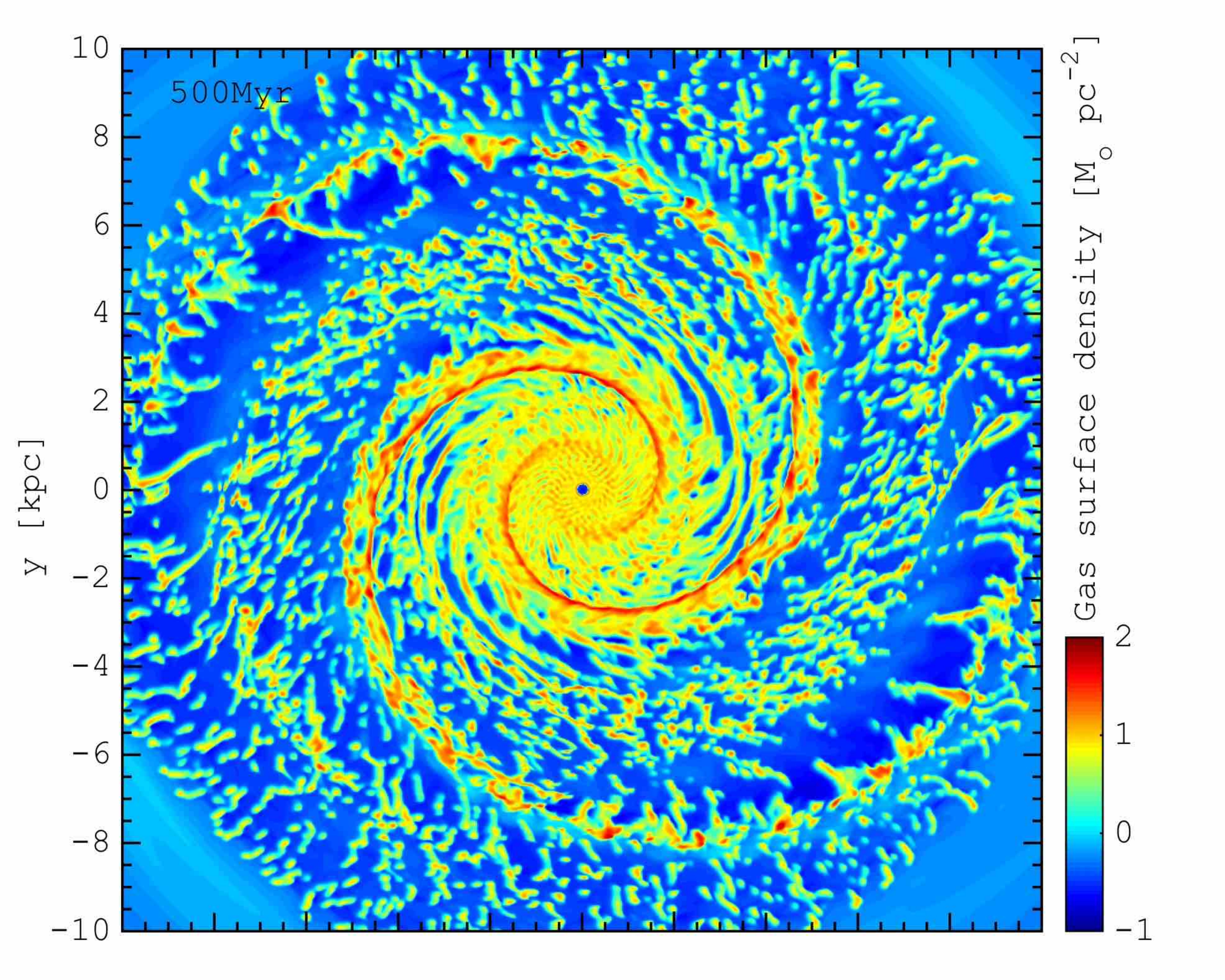}\includegraphics[width=0.5\hsize]{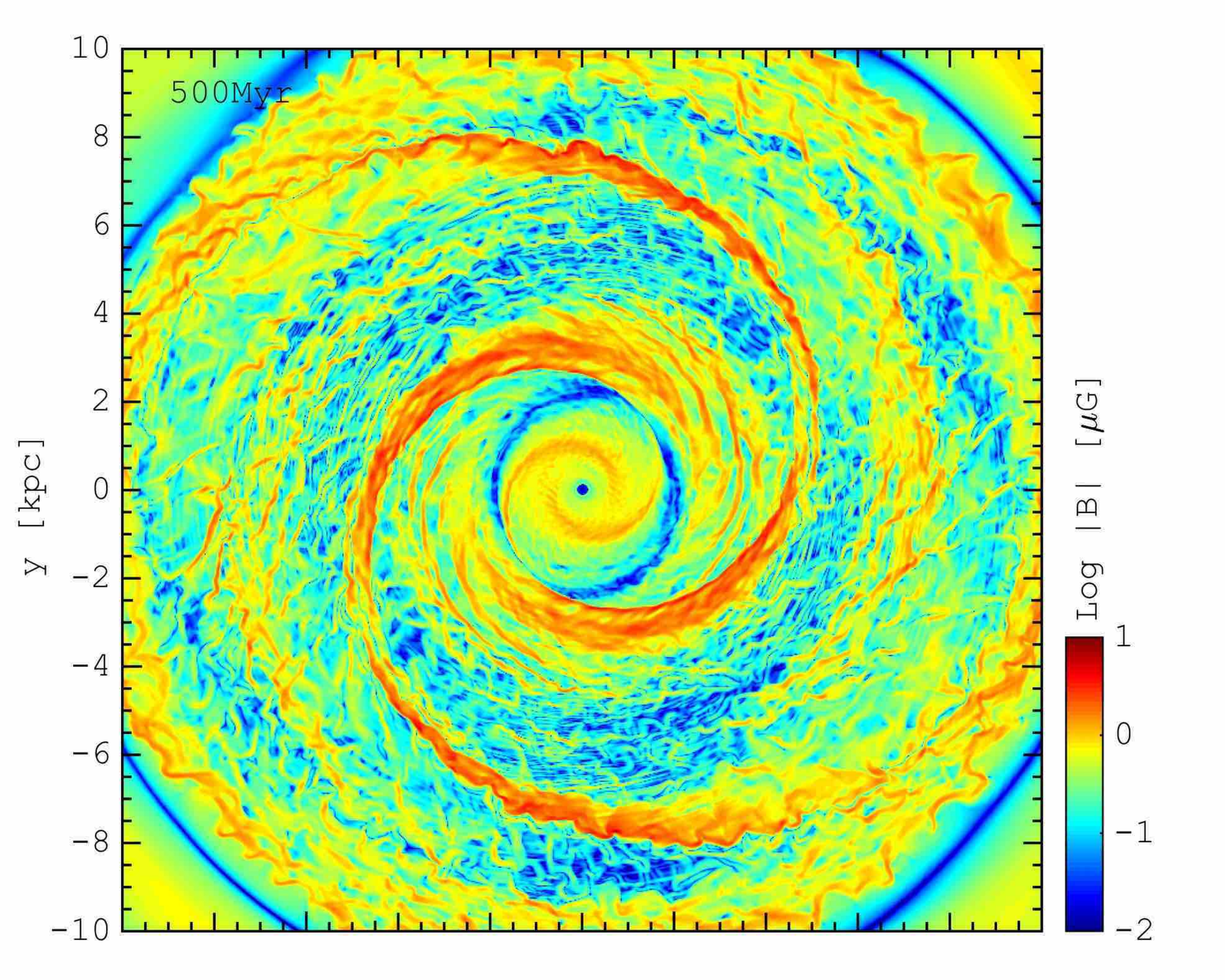}\caption{Model with turbulent initial magnetic field. On top: initial magnetic field components~($B_x$, $B_y$); on bottom: gas density distribution and the total field strength at $500$~Myr. The inter-arm structure becomes more distinct for initial turbulent field magnetic field strengths, while for initial toroidal field~(see Fig.~\ref{fig::evolution_fid}).}\label{fig::random_field1}
\end{figure*}

In this section we analyze the impact of the initial magnetic field structure on the gaseous disk structures formation and the magnetic field growth.  To mimic the turbulent structure of the magnetic field in the disk plane we set up its components as superposition of $10$ modes with (pseudo-) random location in the disk and various amplitude~(positive and negative). Similar to our basic approach we set up the the magnetic field through the vector potential which we introduce in the following form:
\begin{equation}
A_x(x,y,z) = z \sum^{10}_{i=1}  B^c_{i}   \sin( - \sqrt{ (x + x^c_{i})^2 + (y + y^c_{i})^2 + z^2 } )\\
\end{equation}
\begin{equation}
A_y(x,y,z) = z \sum^{10}_{i=1}   B^c_{i}  \cos( - \sqrt{ (x + x^c_{i})^2 + (y + y^c_{i})^2 + z^2 } )\\
\end{equation}
\begin{equation}
A_z(x,y,z) = 0\\
\end{equation}
where $B^c$-th is the amplitude of mode, $x^c_{i}, y^c_{i}$ are the coordinates of $i$-th mode. The initial magnetic field components were found according to Eq.~\ref{eq::ini_magn_f}. The initial magnetic field structure components in the disk plane are shown in Figure~\ref{fig::random_field1}. Initially the gas density profile is smoothly exponential as described in Section~\ref{sec::model} and the external potential follows Eqs.~\ref{eq::potential},\ref{eq::spirals} with parameters of our fiducial model. For the magnetic field amplitude we assume that the mean beta parameter in the disk is $\beta=10$.

As the simulation begins the gas and magnetic field feel the impact of the potential.  Since we drive the global spiral potential, the disk gas has a predominately two-arm spiral structure with a number of small scale structures which are formed in the galaxy disc through hydrodynamical and gravitational fragmentation. As it is clearly seen after $500$~Myr of evolution the density distribution is almost identical to those we observe in our fiducial model~(see right top frame in Fig.~\ref{fig::evolution_fid}). Magnetic field spatial distribution~(right frame in Fig.~\ref{eq::ini_magn_f}) is also very similar but it slightly enhanced in comparison to the purely toroidal initial field especially in the inter-arm region~(see right bottom frame in Fig.~\ref{fig::evolution_fid}). Comparison of the maps with the fiducial model suggests that initial magnetic field spatial structure is unlikely to be very important in the presence of the global spiral pattern. 

\begin{figure}
\includegraphics[width=1\hsize]{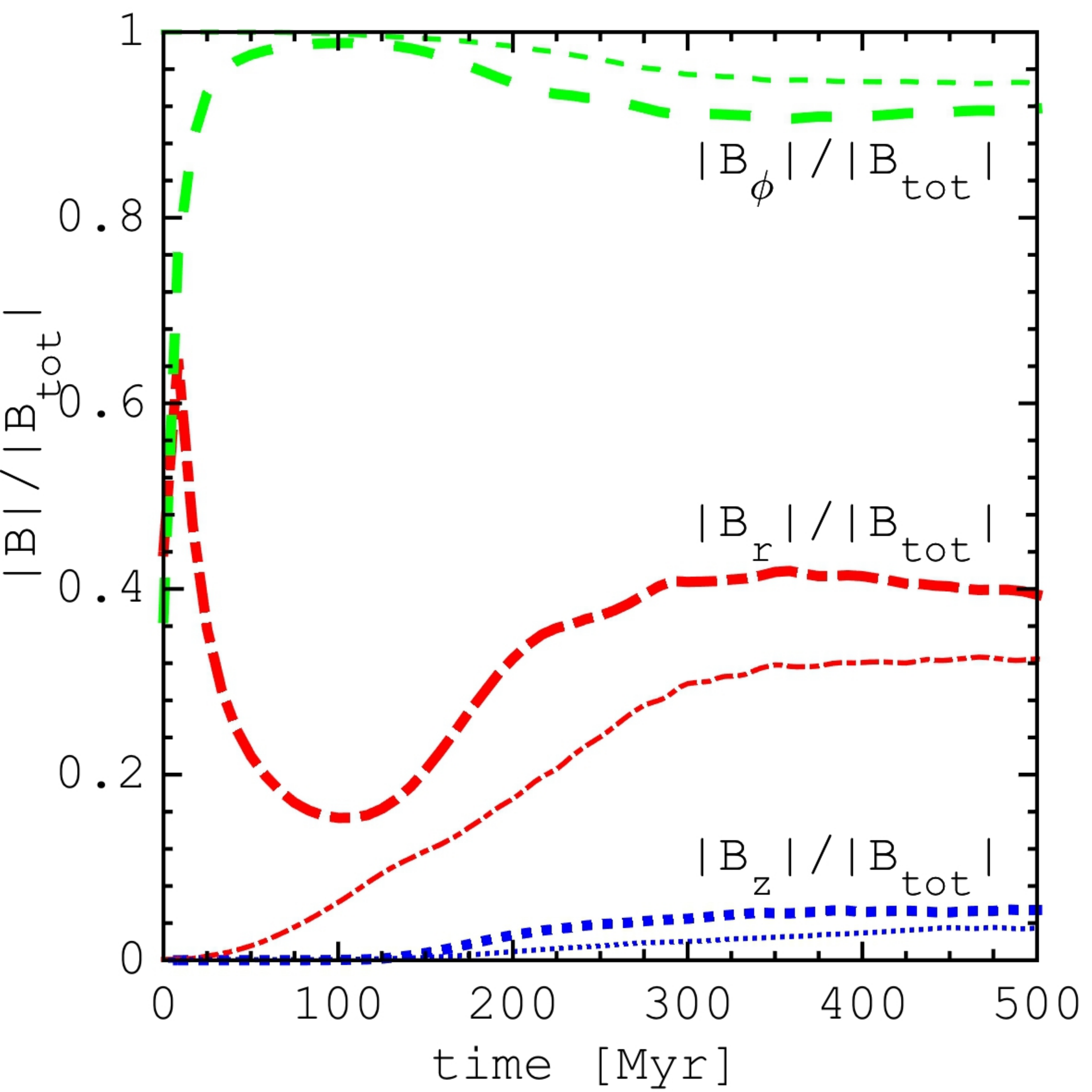}\caption{Evolution of the magnetic field components as a fraction of the total field is shown versus time for model with the initial turbulent magnetic field~(thick lines). For reference thin lines represent the same quantities evolution in our fiducial model. Total field strength is defined as $|B_{tot}|^2 = B^2_r + B^2_\phi + B^2_z$.}\label{fig::random_field2}
\end{figure}

In Fig.~\ref{fig::random_field2} we show the evolution of magnetic field components in the models with the random initial field. Since we impose random magnetic field at initial phase, we have almost equal amplitudes of radial and azimuthal fields. As it seen from the evolution of magnetic field components, toroidal field $B_\phi$ becomes dominant very rapidly meanwhile the radial component decreases. Fast initial transformation of turbulent magnetic field to the toroidal can be explained by differential rotation. Before the spiral pattern rises up~($t<100$~Myr), radial gas flow is negligible, but radial component of magnetic field is already presented. Since azimuthal velocity in the disk is the dominant velocity component, one would expect that toroidal field to be wound up by differential rotation $\Omega-$effect~\citep{1980opp..bookR....K,2002RvMP...74..775W}. Toroidal magnetic field amplification is saturated when all of the radial field is wound up. Transformation of the random magnetic field component to the toroidal field leads to the decrease of radial field strength~($t <100$~Myr). However, when angular momentum of gas is transported outward by spiral arms and the radial field is enhanced by a factor of $\approx 2$~($100-300$~Myr), but at the end of simulation its relative amplitude stays even smaller than the initial strength. Magnetic field strength across the disk plane~($B_z$) monotonically increases, but its strength stays relatively small. 

\subsection{Resolution study}

\begin{figure}
\includegraphics[width=1\hsize]{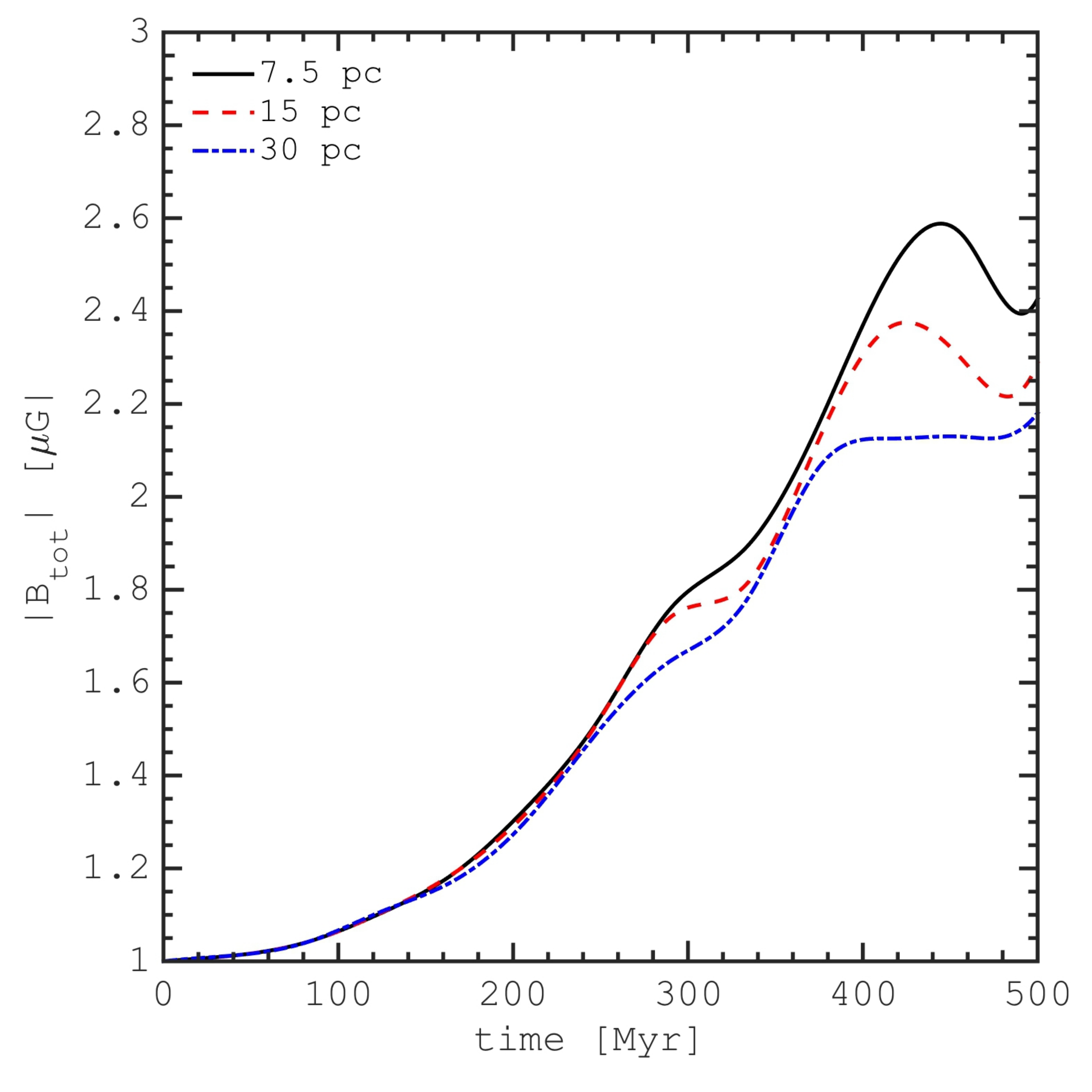}\caption{Time evolution of the mean value of the total magnetic field strength~($|B_{\rm tot}|$) for simulations with different spatial resolution: $7.5$~pc~(solid black line), $15$~pc~(fiducial model, dashed red line) and $30$~pc~(blue dash-dotted line). The evolution is shown for central region $<1$~kpc. }\label{fig::resolution_test}
\end{figure}

To check the effect of the grid resolution on the simulation results, we decrease the cell size to $7.5$~pc in the fiducial model and we also repeat the simulation with the lower resolution $30$~pc. Once we solve the equations are ideal MHD, the magnetic field is frozen into the plasma. In this conception the magnetic field moves together with gas, especially at higher plasma-beta. In Fig.~\ref{fig::resolution_test}, we show the evolution in the total  magnetic field strength in these simulations. The results obtained at the resolutions are evidently similar, although a slightly more rapid increase in magnetic-field strength is observed in simulations performed at the higher resolution, which can be explained by numerical dissipation. Nonetheless, the tendency will remain for the field strength to be strongly aligned with the gas density. As the disk rotates and the gas accumulates around the arms, the field will align with the arms and wind up. We found the saturation of the field correlates to the limit of the maximum gas density related to the model resolution.

\section{Conclusions}\label{sec::concl}

In our analysis, we rely on data from a suite of global spiral galaxy simulations that include the effects of self-gravity, magnetic fields and multiphase thermodynamics. Spiral structure in our simulations is the result of an externally imposed rotating spiral potential, representing the stellar spiral arms of a galaxy. MHD simulations was used to follow the evolution and magnetic field amplification in disks of spiral galaxies. We explore a range of values for the physical parameters describing the properties of the gas, magnetic field and the external potential. We come to the following conclusions.

\begin{itemize}
\item Similar to the previous studies we found that magnetic field does not change dramatically the morphology and physical structure of ISM in simulated spiral galaxies. The disk structure is very similar in the gas and magnetic field distribution. Magnetic fields suppress a purely hydrodynamic shear instability as a means of growing inter-arm spurs and small scale feathers. However, since our models include gas self-gravity and radiation cooling, we detect a number of filamentary structures which are more prominent in simulations with larger initial magnetic field strength. These gaseous filament are coherent with the large scale magnetic field structure.

\item We detect a sensible magnetic filed reversal at the outer edge of the large scale spurs. Meanwhile front edges of the spurs are characterized by the increase of the positive magnetic field component. Note, that amplitudes of positive and negative components are very similar and reach up to several units of $\Bunit$~(see Fig.~\ref{fig::shpurs}). The present treatment could be used for direct comparison with observations, coupled with radiative transfer calculations for various observational manifestations of magnetic field.

\item Gas in spiral arms is much more magnetized than it is in the inter-arm region. We found that the magnetic field in the spiral arms has a stronger mean-field component, and clear inverse correlation between the gas density and plasma-beta parameter $\beta\propto \rho^{-0.8}$, compared to the rest of the disk with more turbulent component to the field and an absence of correlation between gas density and plasma-beta~(see Fig.~\ref{fig::beta_dens}). In the inter-arm the relation is less significant due to the presence of large number of dense clumps and filaments which are surrounded by the warm low dense ISM. Moreover, magnetic filaments are seem to be disconnected from the gaseous structures exactly in the inter-arm region.

\item For various models we detect amplification of magnetic field by a factor of $2-2.5$ during $\approx 2$ rotation periods of the galaxy ($500$~Myr). Mean value of reversal magnetic field does not exceed 25\% of the total field strength. Amplification of ordered and random field is very similar, and since we start from the purely ordered state, at the end of simulation amplitude of ordered field component is three times higher than the random field~(see right frame in Fig.~\ref{fig::evolution_fid}).
 
\item Amplification of magnetic field strongly depends on the initial magnetic field value~(Fig~\ref{fig::evolution_different_models}). Large initial magnetic field suppress the formation of small scale structures and smooth large scale shocks which in turn are less dense and generate less magnetic field. We also found that magnetic field amplification action is a function of radius: in the very center it is very small, meanwhile at galactic outskirts~($\approx 3$ disk scales or $9$~kpc)  it can be larger by a factor of $\approx 10$ than the initial magnetic field strength. At the same time the simulation where the initial magnetic field is turbulent demonstrate generically similar field configuration at the end of the run, but with slightly enhanced field strength in the inter-arm region in comparison to fiducial model~(see Fig.~\ref{fig::random_field1}). Initially disordered magnetic field contributions from randomly oriented field are efficiently ordered during one rotational period. 

\item We found that radial component of Poynting flux peaks near the spiral arms which spread the magnetic field towards the outer disk over time~(see Figs.~\ref{fig::poynting_flux_2d} and~\ref{fig::poynting_flux_1d}). Hence the magnetic field strength is enhanced stronger at the galactic outskirts due to the radial transfer of magnetic energy by the spiral arms pushing the magnetic field outward.

\item  Our models support the presence of MRI at distances $ > 11$~kpc~(see Fig.~\ref{fig::mri_check}) where we detect that the gas velocity dispersion is lower than the critical values suggested in linear analysis by~\cite{2010AN....331...34K}.

\end{itemize}

There are some processes which are likely important for the amplification of magnetic fields in galaxies but which are not included in our models, e.g. star formation and supernova feedback. Therefore, we miss a possible turbulent amplification of the magnetic field on small scales~\citep{2011ApJ...731...62F,2012PhRvE..85b6303S,2012MNRAS.423.3148S}. Such subgrid physics should comprise the $\alpha$-effect although the mechanism remains poorly understood~\citep[see e.g.,][]{1992ApJ...393..165V,1996ARA&A..34..155B,2005PhR...417....1B}.  Because we evolve an isolated galaxy, it does not experience any large-scale shearing motions caused by infalling material or mergers with other galaxies that could lead to an additional amplification of the magnetic field~\citep{2010A&A...523A..72D,2012MNRAS.419.3571G,2017arXiv170107028P}. Nevertheless, in our limited model the magnetic field strength, its structure and parameters of gaseous structures are comparable to the observations. The next step is to use the work presented here as a basis to reproduce galaxies with various morphology. To better match real external galactic gas structure it will be prudent to include the effects of live stellar disk, star formation and feedback.

\begin{acknowledgements}
The authors thank an anonymous referee for his/her rapid reports and numerous very helpful comments and suggestions which improved this paper a lot. Authors also thank Dmitry Sokoloff and Eugeny Mikhailov for fruitful discussions of the paper. The authors are also grateful to Nicolas Leclerc and Yulia Venichenko for their assistance with the data processing. This work was supported by RFBR grant 16-32-60043. SSK has been supported by the Ministry of Education and Science of the Russian Federation (government task No.~2.852.2017/4.6). The software used in this work was developed with support of the Russian Foundation for Basic Research (grant 15-02-06204). The calculations for this paper were performed on Lomonosov supercomputer at the Research Computing Center~(Moscow State University).
\end{acknowledgements}

\bibliographystyle{aa} 
\bibliography{mhd_galaxies} 

\end{document}